%% file: main.tex
\RequirePackage{etoolbox}
\documentclass[camera,letterpaper,nomarginnotes,nonarrowgutter]{jpaper}

\include{macros}

\begin{document}
\urlstyle{tt}
\bstctlcite{IEEEexample:BSTcontrol}
\title{Hermes: Accelerating Long-Latency Load Requests \\via Perceptron-Based Off-Chip Load Prediction}

\author{
Rahul Bera$^1$ \hspace{1em} Konstantinos Kanellopoulos$^1$ \hspace{1em} Shankar Balachandran$^2$ \hspace{1em} David Novo$^{3}$ \vspace{0.3em} \\
Ataberk Olgun$^1$ \hspace{1em} Mohammad Sadrosadati$^1$ \hspace{1em} Onur Mutlu$^1$ \vspace{0.8em} \\
\normalsize{
    $^1$ETH Zürich \hspace{0.5em} $^2$Intel Processor Architecture Research Lab \hspace{0.5em} $^3$LIRMM, Univ. Montpellier, CNRS
}
}

\maketitle
\thispagestyle{firstpage}

\input{00abstract}

\input{01introduction}
\input{02motivation}
\input{03design_overview}
\input{04detailed_design}
\input{05methodology}
\input{06evaluation}

\input{07related_works}

\input{08conclusion}

\input{09ack}

\bibliographystyle{IEEEtranS}
\bibliography{refs}

\clearpage
\appendix
\input{10appendix}
\clearpage
\input{11appendix2}

\end{document}

%% file: macros.tex
\usepackage{xcolor}
\definecolor{freakishgreen}{HTML}{0A982B}
\definecolor{urlblue}{HTML}{319dd6}
\usepackage{hyperref}
\hypersetup{
    colorlinks=true,
    linkcolor=blue,
    filecolor=magenta,
    citecolor=violet,
    urlcolor=urlblue,
    pdfpagemode=FullScreen,
}
\usepackage{dblfloatfix}
\usepackage{multirow}
\usepackage{booktabs}
\usepackage{array}
\usepackage{algorithm}
\usepackage{eucal}
\usepackage{fancyhdr}
\usepackage{enumitem}%
\usepackage[noend]{algpseudocode}
\newcolumntype{L}[1]{>{\raggedright\let\newline\\\arraybackslash\hspace{0pt}}m{#1}}
\newcolumntype{C}[1]{>{\centering\let\newline\\\arraybackslash\hspace{0pt}}m{#1}}
\newcolumntype{R}[1]{>{\raggedleft\let\newline\\\arraybackslash\hspace{0pt}}m{#1}}
\usepackage{tikz}
\newcommand*\circled[1]{\tikz[baseline=(char.base)]{
            \node[shape=circle,fill,inner sep=1pt] (char) {\textcolor{white}{#1}};}}
\makeatletter
\def\thickhline{%
  \noalign{\ifnum0=`}\fi\hrule \@height \thickarrayrulewidth \futurelet
   \reserved@a\@xthickhline}
\def\@xthickhline{\ifx\reserved@a\thickhline
               \vskip\doublerulesep
               \vskip-\thickarrayrulewidth
             \fi
      \ifnum0=`{\fi}}
\makeatother
\newlength{\thickarrayrulewidth}
\usepackage{pifont}

\newcommand\Tstrut{\rule{0pt}{2ex}}         %
\newcommand\Bstrut{\rule[-1ex]{0pt}{0pt}}   %
\newcommand{\Tabval}[1]{{\Tstrut #1 \Bstrut}}   %

\usepackage{cleveref}
\crefname{section}{§\hspace{-2pt}}{§§}
\Crefname{section}{§}{§§}
\usepackage{datetime}

\usepackage[compress,sort]{cite}

\newcommand{\paraheading}[1]{\vspace{1em}\noindent \textbf{#1}}

\newcommand{\shellcmd}[1]{\\\indent\indent\texttt{\footnotesize\$ #1}}

\newif\ifsubmission
\submissiontrue

\ifsubmission
    \newcommand\kon[1]{}
    \newcommand{\rbc}[1]{{#1}}
    \newcommand{\rbd}[1]{{#1}}
    \newcommand{\rbe}[1]{{#1}}
    \newcommand{\rbf}[1]{{#1}}
    \newcommand{\rbg}[1]{{#1}}
    \newcommand{\rbh}[1]{{#1}}
    \newcommand{\rbi}[1]{{#1}}

    \fancypagestyle{firstpage}
    {
        
        \fancyhf{}
        \fancyfoot{}
        \fancyfoot[C]{\thepage}
    }
\else
    \newcommand\kon[1]{\noindent{\color{cyan}{\bf\textbf{}}~{\it Kon: #1}}~}
    \newcommand{\rbc}[1]{{#1}}
    \newcommand{\rbd}[1]{{#1}}
    \newcommand{\rbe}[1]{{#1}}
    \newcommand{\rbf}[1]{{#1}}
    \newcommand{\rbg}[1]{{\color{blue}#1}}
    \newcommand{\rbh}[1]{{\color{magenta}#1}}
    \newcommand{\rbi}[1]{{\color{brown}#1}}
    
    \newcommand{\versionnum}[0]{5.0}
    \fancypagestyle{firstpage}
    {
        \fancyhead[C]{\textcolor{blue}{\emph{Version \versionnum~---~\today, \xxivtime \ UTC}}}
        \fancyfoot[C]{\thepage}
    }
\fi

\newif\ifarxiv
\arxivtrue

\ifarxiv
    \fancypagestyle{firstpage}
    {
        \fancyhead{}
        \begin{tikzpicture}[remember picture,overlay]
        \node [xshift=153mm,yshift=-10mm]
        at (current page.north west) {\href{https://www.acm.org/publications/policies/artifact-review-and-badging-current}{\includegraphics[width=1.8cm]{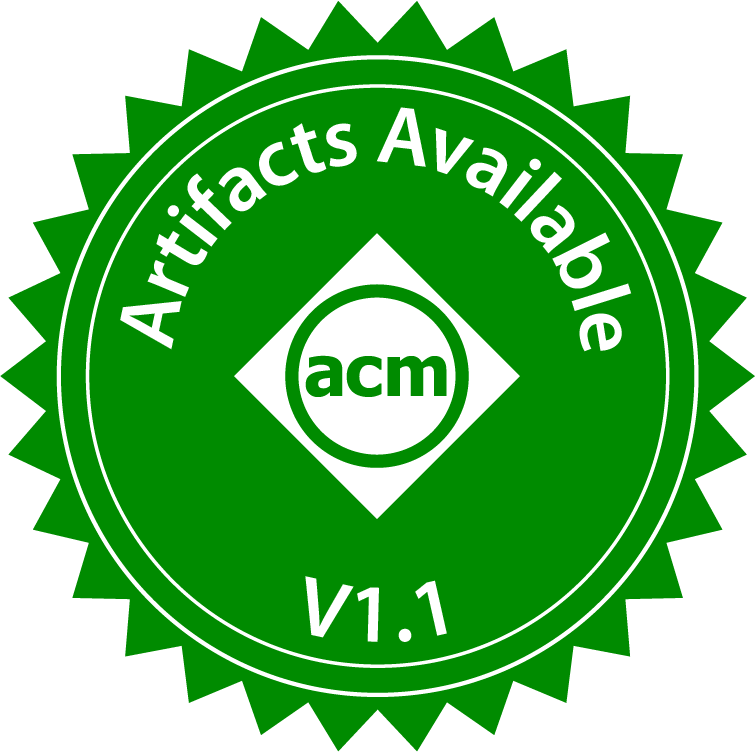}}} ;
        \node [xshift=172mm,yshift=-10mm]
        at (current page.north west) {\href{https://www.acm.org/publications/policies/artifact-review-and-badging-current}{\includegraphics[width=1.8cm]{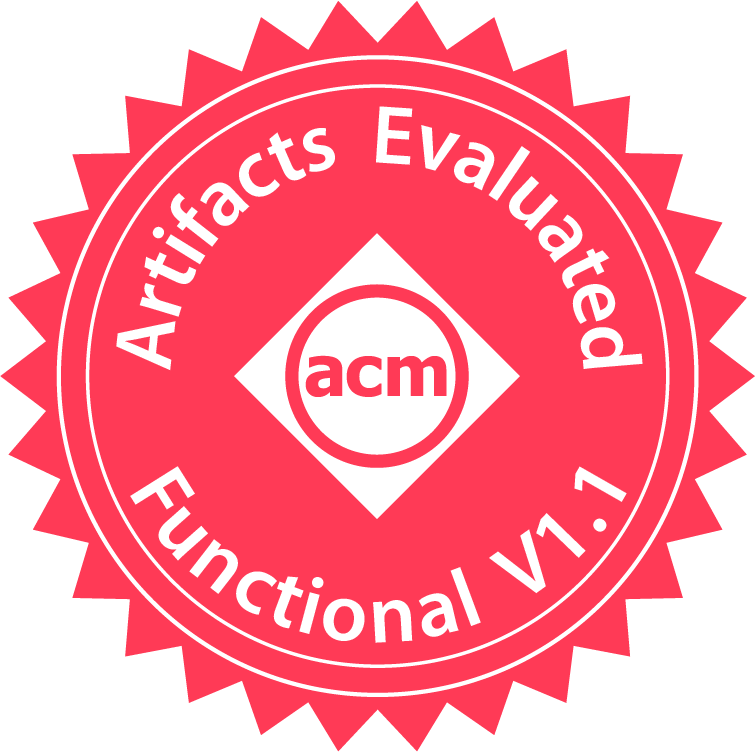}}} ;
        \node [xshift=191mm,yshift=-10mm]
        at (current page.north west) {\href{https://www.acm.org/publications/policies/artifact-review-and-badging-current}{\includegraphics[width=1.8cm]{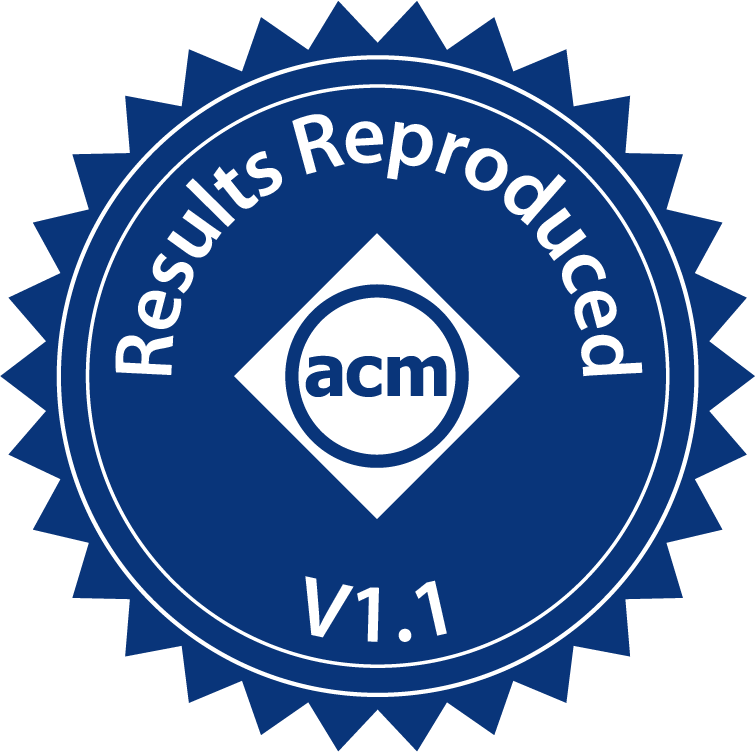}}} ;
        \end{tikzpicture}

      \pagenumbering{arabic}
      \fancyfoot[C]{\large\thepage}
    }
\else

\fi

\usepackage{flushend} 

\usepackage{xspace}
\newcommand{\pred}[0]{POPET\xspace}

%% file: 00abstract.tex
\begin{abstract}
Long-latency load requests continue to limit the performance of modern high-performance \rbc{processors}. To increase the latency tolerance of a \rbc{processor}, architects have primarily relied on two key techniques: \rbc{sophisticated data prefetchers and large on-chip caches}. In this work, we show that: 
(1) even a sophisticated state-of-the-art prefetcher can only predict half of the off-chip load requests on average \rbd{across a wide range of workloads}, and
(2) due to the \rbc{increasing} size and complexity of on-chip caches, a \rbc{large} fraction of 
the latency of an off-chip load request
is spent accessing the on-chip cache hierarchy \rbd{to solely determine that it needs to go off-chip}.

The goal of \rbc{this} work
is to accelerate off-chip load requests by removing the on-chip cache access latency from their critical path.
\rbc{To this end}, we propose \rbc{a new technique called} Hermes, \rbc{whose} key idea is to: (1) accurately predict which load requests might go off-chip, and (2) \rbc{speculatively fetch} the data required by the predicted off-chip loads directly from the main memory, \rbc{while also concurrently accessing the cache hierarchy for such loads}.

\begin{sloppypar}
\rbc{To enable Hermes, we develop}
a new lightweight, perceptron-based off-chip load prediction \rbc{technique} that learns to identify off-chip load requests 
using multiple program features (e.g., sequence of program counters, byte offset of a load request). 
For every load request generated by the \rbc{processor}, 
the predictor observes a set of program features to predict whether \rbc{or not} the load would go off-chip. 
If the load is predicted to go off-chip, Hermes issues a speculative load request directly to the main memory controller once the load's physical address is generated. 
If the prediction is correct, 
the load eventually misses the cache hierarchy and waits for the ongoing speculative load request to finish, 
\rbc{and} thus \rbc{Hermes} completely hides the on-chip cache \rbc{hierarchy} access latency from the critical path \rbc{of the \rbe{correctly-predicted off-chip} load}.
\rbc{Our} extensive evaluation using a wide range of workloads \rbc{shows} that Hermes provides consistent performance improvement on top of a \rbc{state-of-the-art} baseline system \rbc{across} a wide range of configurations with varying core count, main memory bandwidth, high-performance data prefetchers, and on-chip cache hierarchy access latencies, while incurring \rbc{only modest} storage overhead.
The source code of Hermes is freely available at: \url{https://github.com/CMU-SAFARI/Hermes}.
\end{sloppypar}
\end{abstract}

%% file: 01introduction.tex
\section{Introduction} \label{sec:introduction}

Long-latency load requests significantly limit the performance of high-performance out-of-order (OOO) \rbc{processors}. A load request that misses \rbc{in} the on-chip cache hierarchy and goes to the off-chip main memory \rbc{(\rbd{i.e., an} \emph{off-chip load})} often stalls the processor core by blocking the instruction retirement from the reorder buffer (ROB), thus limiting the core's performance~\cite{mutlu2005techniques,mutlu2003runahead,mutlu2003runahead2}.
To increase the latency tolerance of a core, computer architects primarily rely on two key techniques. 
First, they \rbc{employ} \rbd{increasingly} sophisticated hardware prefetchers \rbd{that} can learn complex memory address patterns and fetch data required by future load requests before the core demands them~\cite{pythia,dspatch,spp,ppf,bingo}.
Second, they significantly scale up the size of the \rbc{on-chip cache hierarchy with \rbd{each new} generation of processors}~\cite{goldencove_microarch,goldencove_microarch2,goldencove}. 

\textbf{Key problem.} 
Despite recent \rbc{advances} in processor core design, we observe two key trends in new processor designs that leave a significant opportunity for performance improvement on the table. 
First, even a sophisticated state-of-the-art prefetcher can only predict half of the long-latency off-chip load requests on average \rbd{across a wide range of workloads} (\rbc{see ~\cref{sec:motivation}}).
This is because 
even the most sophisticated prefetchers cannot easily learn the irregular access patterns in programs.

Second, a \rbc{large} fraction of the latency of an off-chip load request is spent \rbc{on accessing} the multi-level on-chip cache hierarchy. This is primarily due to the increasing size of the on-chip caches~\cite{llc_lat1,llc_lat2,llc_lat3}. To cater \rbc{to} workloads with ever increasing data \rbc{footprints}, on-chip caches in recent processors are growing in size and complexity~\cite{beckmann2004managing,hardavellas2009reactive,wang2021stream}. 
A larger on-chip cache, on \rbc{the} one hand, improves a core's performance by \rbc{reducing the fraction of} load requests \rbc{that go} off-chip~\cite{qureshi2007adaptive,jaleel2010high,ship}.
On the other hand, a larger cache comes with longer cache access latency, which increases
the latency of \rbc{each} off-chip load request~\cite{l3_lat_compare1}.

\begin{figure}[!h]
\centering
\includegraphics[width=3.3in]{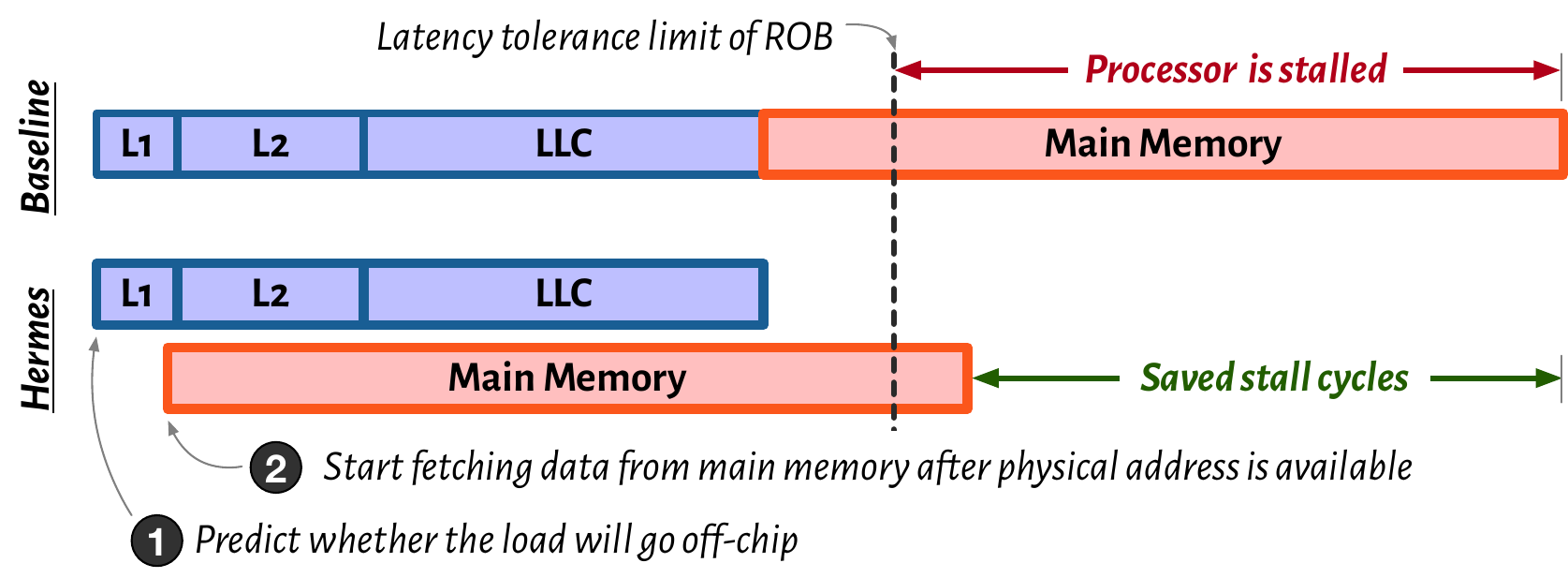}
\vspace{0.5em}
\caption{Comparison of the \rbc{execution timeline} of an off-chip load request in a \rbe{conventional processor} and in Hermes.}
\label{fig:hermes_overall_idea}
\end{figure}

\textit{\textbf{Our goal}} in this work is to 
accelerate long-latency off-chip load requests by removing on-chip cache access latency from their critical path.
\rbc{To this end}, we \rbd{introduce} \rbc{a new technique called} \emph{Hermes},
\rbc{whose} \textbf{\emph{key idea}} is to predict which load requests might go off-chip and start fetching their corresponding data \emph{directly} from the main memory, \rbc{while also concurrently accessing the cache hierarchy for such a load.}\footnote{Hence named after Hermes, the Olympian deity~\cite{hermes} who can quickly move between the realms of the divine (i.e., the \rbc{processor}) and the mortals (i.e., the main memory).}
\rbd{By doing so}, Hermes hides the on-chip cache access latency under the shadow of the main memory access latency (as \rbd{illustrated} in Fig.~\ref{fig:hermes_overall_idea}), \rbc{thereby} significantly reducing the overall latency of an off-chip load request.
Hermes works in tandem with any \rbd{hardware} data prefetcher and reduces the long memory access latency of \rbf{off-chip} load requests that otherwise could not have been prefetched by sophisticated \rbc{state-of-the-art} prefetchers.

\textbf{Key challenge.}
\rbc{Although} Hermes can potentially improve performance
\rbd{by removing the on-chip cache access latency from the critical path of a \rbf{correctly-predicted} off-chip load request,}
\rbd{its} performance gain significantly depends on how accurately it can identify \rbc{the} off-chip load requests. This is because \rbc{a} false-positive off-chip prediction (i.e., a load request \rbd{that} is predicted to go off-chip but hits in \rbc{the} cache) generates \rbe{an} \rbd{unnecessary} main memory request, \rbc{incurring} additional \rbc{main memory} bandwidth and latency overheads, which \rbc{\rbd{can easily} diminish the} performance benefit gained by the load latency reduction. 

We identify two key challenges in designing an accurate off-chip \rbf{load} prediction mechanism. First, in a system with \rbc{state-of-the-art} high-performance prefetchers, only $1$ out of $20$ load requests generated by a program on average eventually goes \rbd{off-chip (see~\cref{sec:key_challenge})}. \rbc{Such a} small fraction of off-chip loads makes it \rbc{difficult} for \rbd{an off-chip} \rbf{load} predictor to accurately and robustly learn from program behavior to produce highly-accurate predictions. Second, the \rbc{accuracy of the} off-chip \rbc{prediction} of a \rbc{load} \rbd{can change} in \rbd{the} presence of \rbc{sophisticated} prefetching techniques, making it even harder for an off-chip \rbf{load} predictor to learn from both the program's and the prefetcher's behavior.

\textbf{Limitations of prior works.}
\rbc{Several} prior works~\cite{yoaz1999speculation,lp,mnm,d2d} propose predicting the cache level that would serve a given load request to enable various performance optimizations (e.g., better instruction scheduling). However, most of these works suffer from two key limitations that make them unsuitable for off-chip load prediction. First, prior predictors 
suffer from \rbc{low} prediction accuracy
(i.e., \rbc{the fraction of predicted off-chip load requests that actually went off-chip})~\cite{yoaz1999speculation,mnm}, which \rbc{increases} the bandwidth overhead in main memory. \rbc{As a result, they often lose the performance benefit gained by the load latency reduction and might \rbe{lower performance than} the baseline system} \rbd{(see~\cref{sec:acc_cov_ocp} and \cref{sec:dmp_with_hmp_perf})}. Second, \rbd{prior off-chip \rbf{load} prediction mechanisms} often incur impractical metadata overhead (e.g., an operating-system-managed metadata storage inside the physical main memory~\cite{lp}, \rbc{extending each TLB and cache entry with additional metadata for tracking cache \rbe{residence} and \rbe{coherence of data}~\cite{d2d, d2m}}), which \rbc{hinders} adoption in commercial processors.

\begin{sloppypar}
\textbf{Key mechanism.}
\rbc{To enable Hermes,} we introduce a new lightweight \underline{p}erceptron-based \underline{o}ff-chip \underline{p}r\underline{e}dic\underline{t}or, called \emph{\pred}, that learns to identify off-chip load requests using multiple program features (e.g., sequence of program counters, byte offset of a load request). 
For every load generated by the \rbc{processor}, \pred observes a set of program features to predict whether or not the load would go off-chip. 
If the load is predicted to go off-chip, Hermes issues a speculative load request \rbd{(called a \emph{Hermes request})} directly to the main memory controller once the load's physical address is generated. 
\rbe{This Hermes request is serviced by the main memory controller concurrently with the \emph{regular load request} (i.e., the load issued by the processor that generated the Hermes request) that accesses the on-chip cache hierarchy.}
If the prediction is correct, the \rbd{regular} load request eventually misses the cache hierarchy and waits for the ongoing \rbd{Hermes request} to finish, \rbc{and} thus \rbc{Hermes} completely \rbc{hides} the on-chip cache \rbc{hierarchy} access latency from the critical path \rbc{of \rbd{a correctly-predicted} \rbe{off-chip} load}.
\end{sloppypar}

\textbf{Results summary.}
We evaluate Hermes with a diverse set of $110$ single-core and $220$ multi-core workloads spanning \texttt{SPEC CPU2006}~\cite{spec2006}, \texttt{SPEC CPU2017}~\cite{spec2017}, \texttt{PARSEC}~\cite{parsec}, \texttt{Ligra}~\cite{ligra} graph processing workloads, and commercial workloads~\cite{cvp2}. 
\rbe{Our evaluation yields} five key results \rbc{that \rbd{demonstrate} Hermes's effectiveness}.
First, \pred achieves on average $77.1\%$ accuracy \rbe{and} $74.3\%$ coverage \rbd{(i.e., the fraction of off-chip load requests of a workload that are successfully predicted)}, both \rbe{of which are significantly} higher ($1.6\times$ higher accuracy, $3.3\times$ higher coverage) than \rbe{that of} the \rbg{prior best-accuracy} off-chip predictor, HMP~\cite{yoaz1999speculation}.
\rbd{Second, Hermes \rbc{improves performance on average by (up to)} $5.4\%$ ($23.4\%$), $5.1\%$ ($25.7\%$), and $6.2\%$ ($32.2\%$) in single-core, eight-core, and bandwidth-constrained system \rbe{configurations}, \rbe{on top of} the best-performing state-of-the-art data prefetcher Pythia~\cite{pythia}.}
Third, Hermes consistently \rbc{improves} performance \rbc{when combined with} \emph{any} baseline \rbd{hardware data} prefetcher. 
\rbd{When implemented combined with four recently-proposed high-performance prefetchers (SPP~\cite{spp,ppf}, Bingo~\cite{bingo}, MLOP~\cite{mlop}, and SMS~\cite{sms}) in single-core system,} Hermes \rbc{improves performance on average by (up to)} $5.1\%$ ($27\%$), $6.2\%$ ($22.4\%$), $7.6\%$ ($26.7\%$), and $7.7\%$ ($25.7\%$).
Fourth, Hermes provides better performance-to-overhead benefit than traditional prefetchers due to \rbd{its} highly-accurate off-chip predictions. 
For every $1\%$ performance increase, \rbd{Hermes increases the main memory requests by only $0.5\%$, whereas Pythia increases them by $2\%$}.
Fifth, all of Hermes's benefits come at a \rbc{very modest} storage overhead of only $4$ KB per core, \rbe{while the state-of-the-art prefetcher Pythia consumes $25.5$~KB per core}.

\vspace{0.5em}
\noindent We make the following contributions in this paper:
\begin{itemize}
    \setlength\itemsep{0.1em}
    \item We identify two key opportunities for performance improvement \rbc{in modern processors}: (1) a significant fraction of the load requests continues to go off-chip even in \rbc {the} presence of sophisticated data prefetchers, and (2) an increasing fraction of the off-chip load latency is spent accessing on-chip caches due to the \rbc{increasing} size of \rbc{the} on-chip cache hierarchy.
    \item We introduce \emph{Hermes}, \rbc{a new technique} that reduces long memory access latency by predicting off-chip load requests and fetching their corresponding data \emph{directly} from the main memory, \rbc{while concurrently accessing the on-chip cache hierarchy for such loads.}
    \item We design a new perceptron-based off-chip \rbc{load} predictor, called \pred, that accurately identifies \rbc{and predicts} \rbc{the} off-chip load requests using multiple program features.
    \item \rbe{We show that Hermes significantly improves performance across a wide range of workloads and system configurations with varying core count, main memory bandwidth, high-performance data prefetchers, and on-chip cache access latencies.}
    \item We open-source Hermes and all necessary traces \rbc{and scripts} to reproduce results in \url{https://github.com/CMU-SAFARI/Hermes}.
\end{itemize}

%% file: 02motivation.tex
\section{Motivation} \label{sec:motivation}

High main memory access latency continues to limit the performance of modern out-of-order (OOO) \rbc{processors}. A load request that misses the on-chip cache hierarchy and goes to off-chip main memory often blocks instruction retirement from the reorder buffer (ROB), \rbc{preventing the processor} from allocating new instructions into the ROB~\cite{mutlu2003runahead,mutlu2003runahead2,mutlu2005techniques,hashemi2016continuous}, limiting performance.

\begin{sloppypar}
To tolerate long memory latency, recent high-performance OOO cores have primarily relied on two key techniques. First, modern cores have significantly scaled up their 
\rbc{on-chip cache size} (e.g., each Intel Alder Lake core~\cite{goldencove_microarch} employs $4.3$MB on-chip cache (including L1, L2 and a per-core last-level cache (LLC) slice), which is $1.88\times$ larger than the \rbd{on-chip cache in the} previous-generation Skylake core~\cite{skylake}). Second, modern cores \rbc{employ} increasingly sophisticated hardware prefetchers~\cite{goldencove,goldencove_microarch2} that can \rbc{more effectively} predict the addresses of load requests in advance and fetch their corresponding data to on-chip caches before the program demands it, \rbd{thereby} completely or partially hiding the long off-chip load latency \rbc{for a fraction of off-chip loads}~\cite{goldencove,ppf,dspatch,pythia}.
\end{sloppypar}

Despite \rbc{these advances}, we observe two key trends \rbc{in} processor \rbc{design} that leave a significant performance improvement opportunity on the table: (1) a large fraction of load requests continues to go \rbd{off-chip} even in \rbd{the} presence of state-of-the-art prefetchers, and (2) an increasing fraction of the latency of an off-chip load request is spent accessing \rbc{the increasingly larger} on-chip caches.
 
\begin{figure}[!h]
\centering
\includegraphics[width=3.3in]{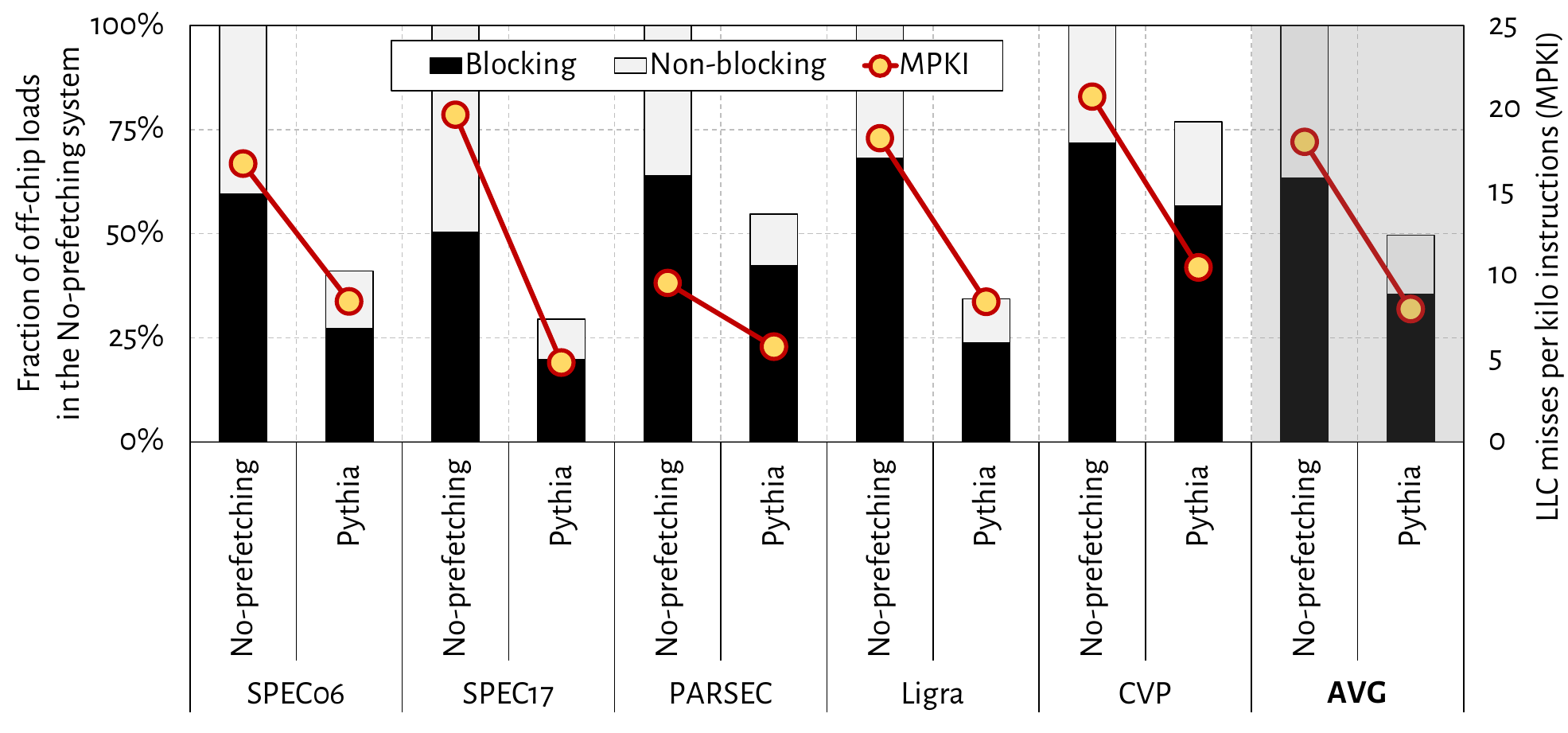}
\vspace{0.5em}
\caption{The distribution of ROB-blocking and non-blocking load requests (on \rbe{the left} y-axis), \rbd{and LLC misses per kilo instructions} (on \rbe{the right} y-axis) in \rbd{the} absence and presence of \rbd{a state-of-the-art hardware} data prefetcher~\cite{pythia}.}
\label{fig:impact_of_pref}
\end{figure}

\vspace{2pt}
\noindent \textit{\textbf{A large fraction of loads is still uncovered by state-of-the-art prefetchers.}} Over the past decades, researchers have proposed \rbc{many} hardware prefetching techniques that have consistently pushed the limits of performance improvement (e.g.,~\cite{stride,streamer,baer2,jouppi_prefetch,ampm,fdp,footprint,sms,sms_mod,spp,vldp,sandbox,bop,dol,dspatch,mlop,ppf,ipcp,pythia,litz2022crisp}). We observe that state-of-the-art prefetchers provide a \rbc{large} performance gain by accurately predicting future load addresses.
Yet, a large fraction of \rbc{off-chip} load requests cannot be predicted even by the most advanced prefetchers. 
These uncovered requests limit the \rbd{processor's} performance by blocking instruction retirement in \rbd{the} ROB. 
Fig.~\ref{fig:impact_of_pref} shows a stacked graph of total number of off-chip load requests in a no-prefetching system and a system with the recently-proposed hardware \rbf{data} prefetcher Pythia~\cite{pythia}, normalized to the no-prefetching system, \rbe{across $110$ workload traces categorized into five workload categories}.\footnote{We select Pythia as the baseline prefetcher as it provides the highest prefetch coverage and performance benefit among the five contemporary prefetchers considered in this paper (see \cref{sec:dmf_configuration} and \cref{sec:prefetchers}). Nonetheless, \rbc{our qualitative} observation holds equally true for other prefetchers \rbc{considered in this work} (see~\cref{sec:dmf_configuration}).} Each bar further categorizes load requests into two classes: loads that block instruction retirement from \rbc{the} ROB (called \emph{blocking}) and loads that do not (called \emph{non-blocking}).
\cref{sec:methodology} discusses our evaluation methodology. 

We make two key observations from Fig.~\ref{fig:impact_of_pref}. 
First, on average, Pythia \rbc{accurately prefetches} nearly half of \rbc{all} off-chip load requests in the no-prefetching system, \rbd{thereby} improving \rbf{the} overall performance (not shown here; see~\cref{sec:perf_1c}).
Second, the remaining half of the off-chip loads \rbc{are not prefetched even by} a sophisticated prefetcher like Pythia. $71.4\%$ of these \rbd{non-prefetched} off-chip loads block instruction retirement from \rbc{the} ROB, \rbc{significantly limiting performance}. \rbc{We} conclude that, \rbc{state-of-the-art prefetchers, while effective at improving performance,} still leave a significant performance \rbd{improvement} opportunity on the table.

\vspace{2pt}
\noindent \textit{\textbf{An increasing fraction of off-chip load latency is spent \rbd{accessing the} on-chip cache \rbd{hierarchy}.}}
We observe that the on-chip cache hierarchy has not only grown tremendously in size but also in design complexity (e.g., sliced last-level cache organization~\cite{beckmann2004managing,kim2002adaptive,hardavellas2009reactive}) in recent processors, \rbd{in order} to cater \rbc{to} workloads with \rbc{large} data \rbc{footprints}. A larger on-chip cache hierarchy, on \rbc{the} one hand, improves a core's performance by \rbc{preventing more} load requests from going off-chip. 
\rbd{On the other hand, all on-chip caches need to be accessed to determine if a load request should be sent off-chip. As a result, on-chip cache access latency significantly contributes to the total latency of an off-chip load. With increasing on-chip cache \rbe{sizes}, \rbe{and the complexity of the} cache hierarchy design and \rbf{the} on-chip network~\cite{wang2021stream,besta2018slim}, the on-chip cache access latency is increasing in processors~\cite{l3_lat_compare1,llc_lat3}.}
An analysis of the Intel \rbe{Alder Lake} core suggests that the
load-to-use latency of an LLC access
has increased to $14$~ns (which \rbe{is} equivalent to $55$ cycles for a core running at $4$~GHz)~\cite{llc_lat1,llc_lat2,llc_lat3}. 

\begin{figure}[!h]
\centering
\includegraphics[scale=0.3]{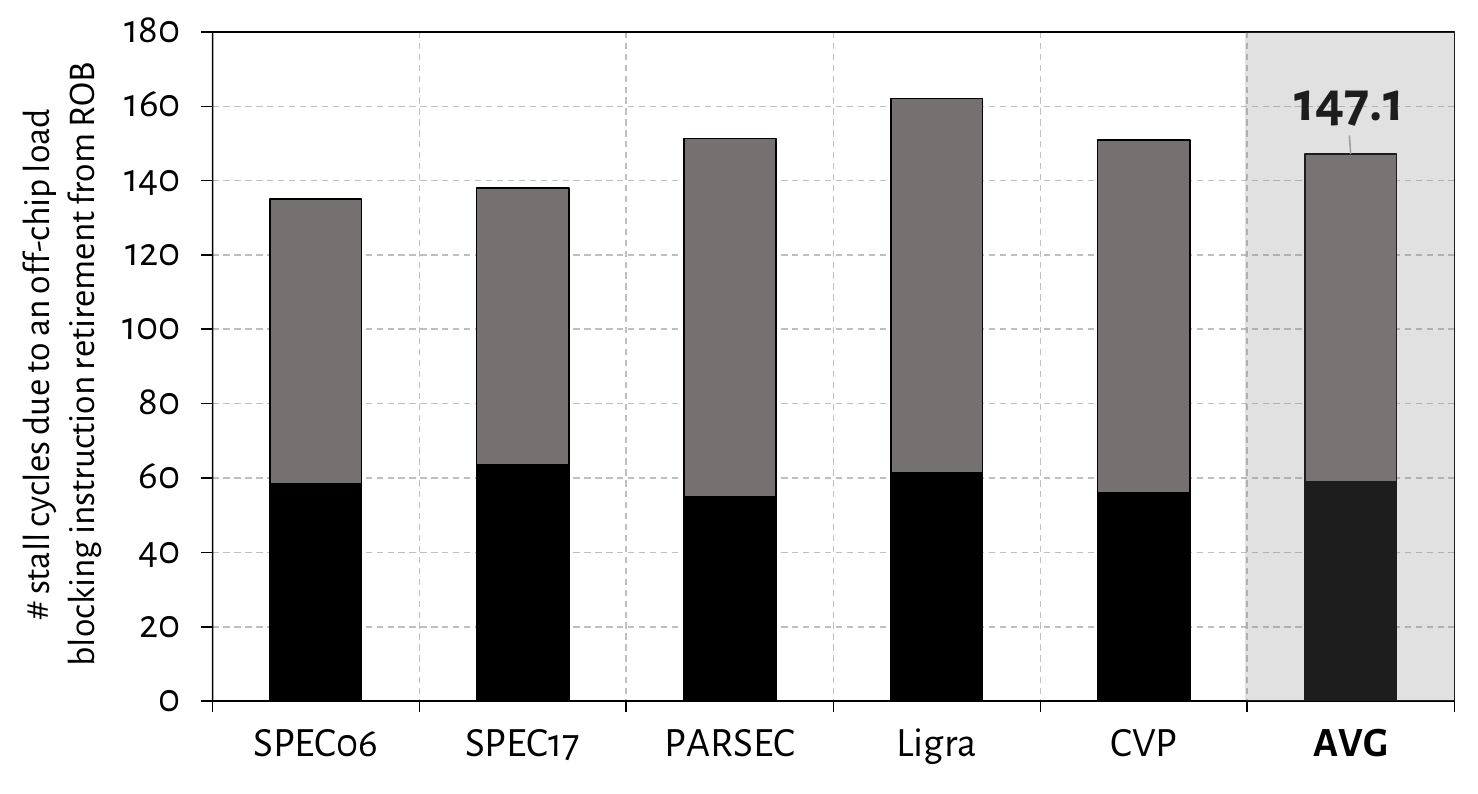}
\vspace{0.5em}
\caption{\rbd{The average number of cycles a core stalls due to an off-chip load blocking any instruction from retiring from the ROB} across all workload categories. \rbc{The dark portion in each bar shows the cycles that can be \rbe{completely} \rbd{eliminated} by removing the on-chip cache access latency from an off-chip load's critical path.}}
\label{fig:load_lat}
\end{figure}

To demonstrate the effect of long on-chip cache access latency on the \rbd{total latency of an off-chip load}, Fig.~\ref{fig:load_lat} plots the 
average number of cycles a core stalls due to an off-chip load blocking any instruction from retiring from the ROB, averaged across each workload category in our baseline system with Pythia. 
\rbd{Each bar further shows the \rbe{average} number of cycles an off-chip load spends for accessing the on-chip cache hierarchy.}
Our simulation configuration faithfully models an Intel Alder Lake performance-core with \rbc{a large} ROB, large on-chip caches \rbc{and} publicly-reported cache access latencies (see \cref{sec:methodology}). As Fig.~\ref{fig:load_lat} shows, an off-chip load stalls the core for \rbc{an average of} $147.1$ cycles. $40.1\%$ of these stall cycles (i.e., $58.9$ cycles) can be \rbe{completely} \emph{eliminated} by removing the on-chip cache access latency from \rbe{the off-chip load's} critical path.
We conclude that a large \rbc{and complex} on-chip cache hierarchy \rbd{is directly responsible for a large fraction of} the overall stall cycles \rbc{caused by} an off-chip load request. We envision that this problem will only get exacerbated with new \rbe{processor} designs \rbc{as on-chip caches \rbe{continue to} grow in size and complexity}~\cite{l3_lat_compare1}.

\section{Our Goal and Key Idea}

\textit{\textbf{Our goal}} is to improve processor performance by removing the on-chip cache access latency from the critical path of off-chip load requests.

\subsection{The Key Idea and Potential Benefits} \label{sec:headroom_study}

\rbd{To this end}, we propose \rbd{a new technique called} \textit{\textbf{Hermes}},
\rbc{whose} \textbf{\emph{key idea}} is \rbd{to predict which load requests might go off-chip and start fetching their corresponding data \emph{directly} from the main memory, while also concurrently accessing the cache hierarchy for such a load.}

\begin{figure}[!h]
\centering
\includegraphics[width=3.3in]{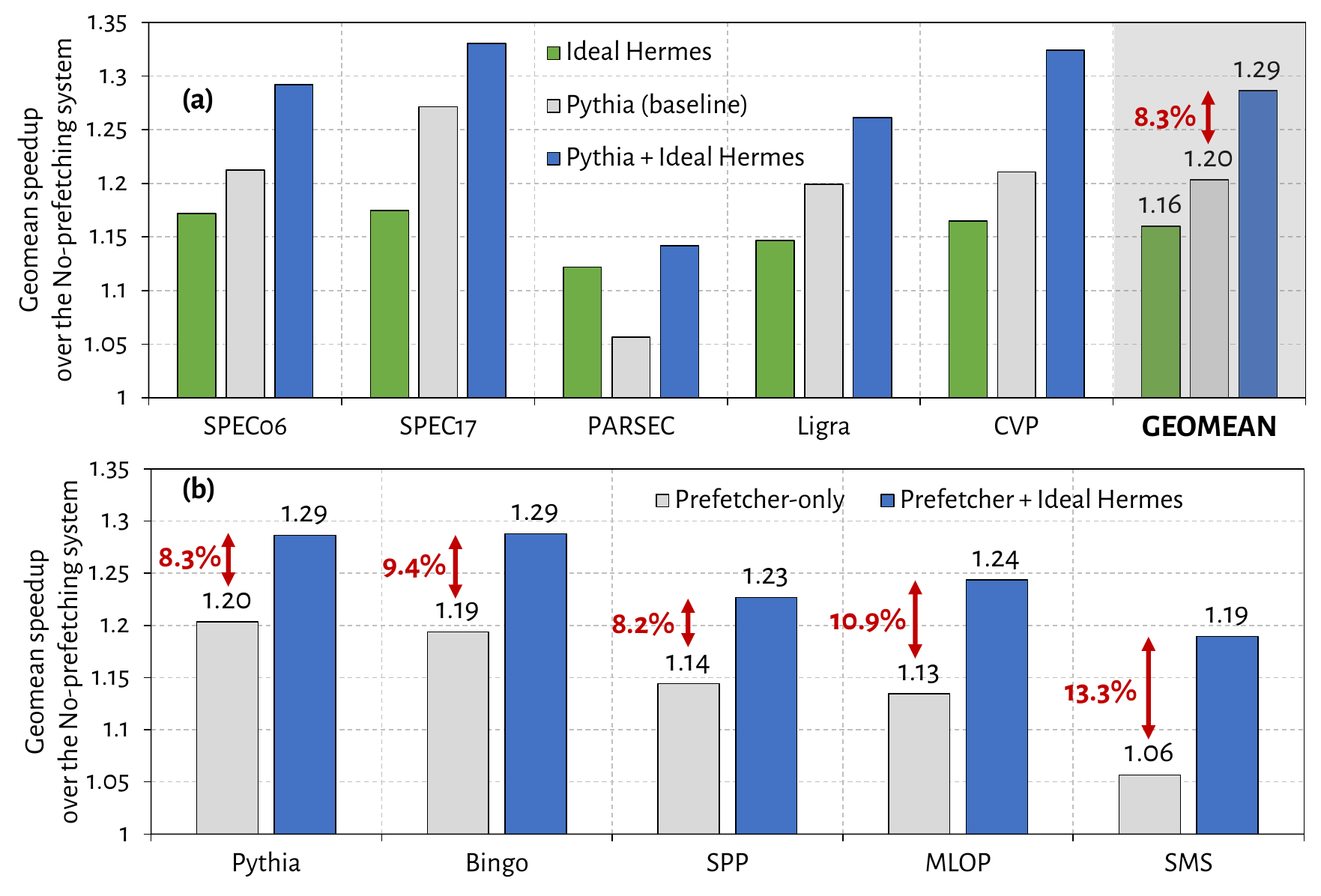}
\vspace{0.5em}
\caption{(a) \rbd{\rbe{Speedup of} Ideal Hermes \rbd{by itself and when combined with} Pythia in single-core workloads}. (b) \rbe{Speedup of} Ideal Hermes \rbd{when combined with} four recently-proposed prefetchers: Bingo~\cite{bingo}, SPP~\cite{spp,ppf}, MLOP~\cite{mlop}, and SMS~\cite{sms}.}
\label{fig:Ideal_dmf_master}
\end{figure}

To understand the potential performance benefits of Hermes, we model an \emph{Ideal Hermes} system \rbd{in simulation} where we reduce the main memory access latency of \emph{every} off-chip load request by the \rbe{post-L1 on-chip cache hierarchy access latency (which includes L2 and LLC access, and interconnect latency)}. 
\rbd{In other words, in \rbe{the} Ideal Hermes system, we \rbe{(1)} magically and perfectly know if a load request would go off-chip after its physical address is available (i.e., after the translation lookaside buffer access, which happens in parallel with the L1 \rbf{data} cache access in modern processors~\cite{wood1986cache,cekleov1997virtual,patterson2016computer,basu2012reducing}), and \rbe{(2)} directly access the off-chip main memory for such a load, eliminating the non-L1-cache related on-chip cache \rbf{hierarchy} access latency from such a load’s total latency.}
Fig.~\ref{fig:Ideal_dmf_master}(a) shows \rbe{the speedup} of Ideal Hermes \rbd{by itself} and \rbd{when combined with} Pythia \rbe{normalized to the no-prefetching system in single-core workloads.} 
We make two key observations from Fig.~\ref{fig:Ideal_dmf_master}(a). 
First, Ideal Hermes \rbd{combined with} Pythia outperforms Pythia \rbd{alone} by $8.3\%$ on average across \rbd{all} workloads. 
Second, Ideal Hermes \rbd{by itself} provides nearly $80\%$ \rbd{of the} performance improvement \rbd{that} Pythia \rbd{provides}.
Fig.~\ref{fig:Ideal_dmf_master}(b) shows the \rbe{speedup} of Ideal Hermes \rbd{when combined with} four other recently-proposed high-performance prefetchers: Bingo~\cite{bingo}, SPP~\cite{spp} (with perceptron filter~\cite{ppf}), MLOP~\cite{mlop}, and SMS~\cite{sms}.
\rbd{Ideal Hermes improves performance by $9.4\%$, $8.2\%$, $10.9\%$, and $13.3\%$ on top of four state-of-the-art prefetchers Bingo, SPP, MLOP, and SMS, respectively.}
Based on these results, we conclude that Hermes has high potential performance benefit not only \rbd{when implemented alone} but also \rbd{when combined with} a wide variety of high-performance prefetchers.

\subsection{Key Challenge} \label{sec:key_challenge}

Even though Hermes has a significant potential \rbd{to improve} performance, Hermes's performance gain heavily depends on the accuracy (i.e., the fraction of predicted off-chip loads that actually go off-chip) \rbd{and the coverage (i.e., the fraction of off-chip loads that are successfully predicted)} of the off-chip load prediction. 
\rbe{A low-accuracy off-chip load predictor}
generates \rbd{useless} main memory requests, which incur both latency and bandwidth overheads, and causes interference to the useful requests in the main memory.
\rbe{A low-coverage predictor}
loses opportunity \rbe{to improve} performance.

\begin{figure}[!h]
\centering
\includegraphics[width=3.3in]{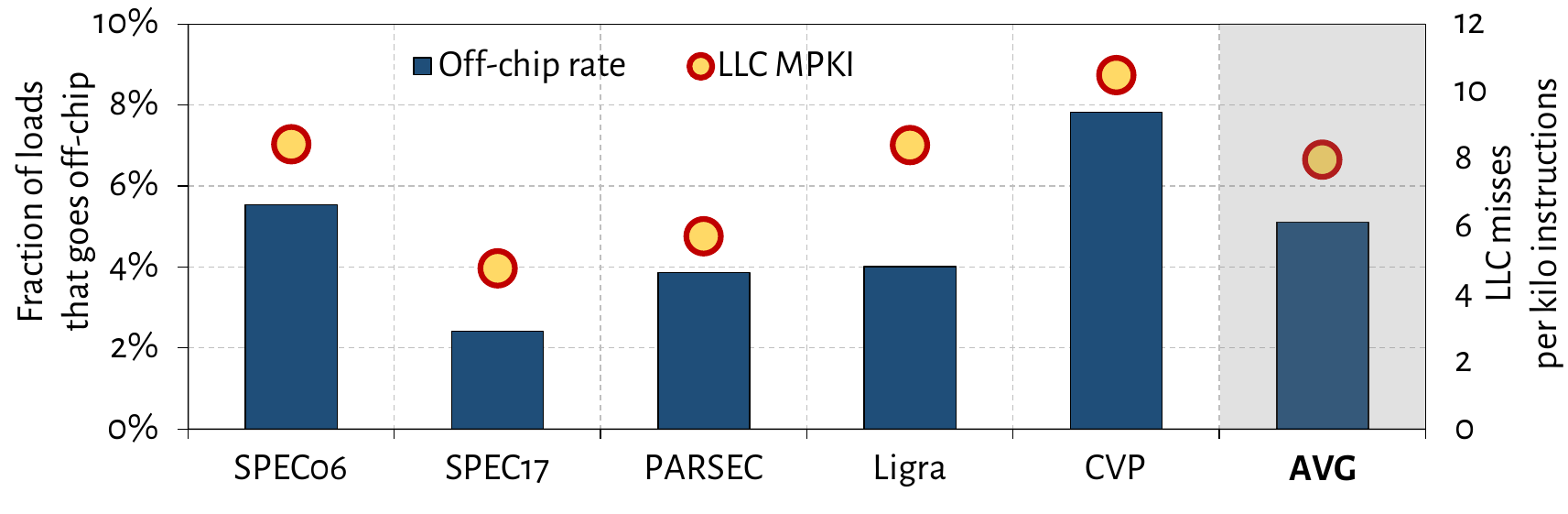}
\vspace{0.5em}
\caption{\rbd{\rbe{Percentage of} loads that miss the LLC and goes off-chip (on \rbe{the left} y-axis) and the LLC MPKI (on \rbe{the right} y-axis) in \rbe{the} baseline system with Pythia.}}
\label{fig:off_chip_rate}
\end{figure}

We identify two key challenges in designing an off-chip \rbf{load} predictor with high accuracy and high coverage.
\rbd{First, only a small fraction of the total loads generated by a workload goes off-chip in presence of a sophisticated data prefetcher. As shown in Figure~\ref{fig:off_chip_rate}, on average $7.9$ loads per kilo instructions miss the LLC and go off-chip in our baseline system with Pythia. However, these loads \rbe{constitute} only $5.1\%$ of the total loads generated by a workload.}
This small fraction of off-chip loads makes it difficult for an off-chip load predictor to accurately learn from the workload behavior to produce highly-accurate predictions.

Second, the off-chip predictability of a workload \rbd{can change} in \rbd{the} presence of modern sophisticated data prefetchers. This is because in \rbe{the} presence of a sophisticated  prefetcher, the likelihood of a load request going off-chip not only depends on the program behavior but also on the prefetcher's ability to successfully prefetch for the load.

In this work, we \rbd{overcome these two key challenges by} \rbe{designing} a new off-chip \rbf{load} prediction technique, called \pred, based on perceptron learning~\cite{mcculloch1943logical,rosenblatt1958perceptron,jimenez2002neural}.
\rbd{By learning to identify off-chip loads using multiple program features (e.g., sequence of program counters, byte offset of a load request, \rbe{page number of the load address}), \pred provides both higher accuracy and coverage than \rbi{a} prior cache hit-miss prediction \rbi{technique~\cite{yoaz1999speculation} and \rbi{higher accuracy than} another off-chip load prediction technique that we develop (see~\cref{sec:dmf_configuration}),} in \rbf{the} presence of modern sophisticated prefetchers, without requiring large metadata storage overhead.}
With \rbd{small} changes to the existing on-chip datapath design, we demonstrate that Hermes with \pred significantly outperforms the baseline system with \rbf{a} state-of-the-art prefetcher \rbe{across} a wide range of workloads and system configurations.

\vspace{-0.5em}
\section{Key Related Works} \label{sec:differences_from_prior_work}

The key idea of load hit-miss prediction (HMP) was proposed in~\cite{yoaz1999speculation} and demonstrated as a method to improve the instruction \rbd{scheduler} efficiency \rbd{and performance}. By predicting which load instructions will miss the \rbd{L1 data cache},\footnote{\rbd{Even though HMP was originally proposed to predict loads that miss L1 data cache, it can be extended to predict loads that miss the entire multi-level on-chip cache hierarchy.}} HMP enables the instruction scheduler to delay the scheduling of dependent instructions until the data is fetched. 
This scheduler optimization improves \rbd{processor} performance by \rbd{scheduling a load-dependent instruction to execute at the time when \rbe{the data in available}}. 
Even though Hermes leverages off-chip \rbf{load} prediction in a \rbe{very} different way than HMP, we compare the accuracy and coverage of our perceptron-based off-chip load predictor \pred, against \rbe{those of} HMP in \cref{sec:acc_cov_ocp} and show that Hermes with \pred greatly outperforms Hermes with HMP in \cref{sec:dmp_with_hmp_perf}.

Cache-level \rbe{hit/miss} prediction \rbd{(i.e., predicting which cache level a load might hit)} has also been explored in three works: Direct-to-Data cache (D2D~\cite{d2d}), Direct-to-Master cache (D2M~\cite{d2m}), and Level Prediction (LP~\cite{lp}). 
\rbg{All three works employ \rbh{different} mechanisms to track cacheline \rbi{addresses} present in the cache hierarchy along with the \rbi{cache-level(s) a cacheline is present in}.}
\rbg{For a given load, these works predict which cache-level the load would likely hit.}
\rbf{If the cache-level hit/miss prediction is correct, the processor fetches the data of the correctly-predicted load by only accessing the predicted memory level (L1, L2, LLC, or the main memory) and bypassing all other memory levels. By doing so, a cache-level hit/miss predictor reduces the latency of a correctly-predicted load and improves the processor's energy efficiency. However, if the predictor incorrectly bypasses a memory level that has more up to date data (e.g., if the data is present in L2, but the predictor suggests fetching the data from LLC), the processor needs to detect such \emph{mispredictions} and access the correct memory level to maintain correct execution, which comes with performance overhead \rbh{and additional complexity}.}
Hermes differs from all these prior works in \rbg{three} major ways:

\rbg{\textbf{Prediction via tracking addresses vs. learning from program \rbh{context}.}
To make accurate cache-level prediction, D2D, D2M, and LP rely on tracking \rbh{cacheline addresses} present in the \rbh{on-chip caches} \rbh{by} using efficiently-managed metadata structures. For example, D2D and D2M \rbh{extend} each translation look-aside buffer (TLB) entry (called \emph{eTLB}) to keep additional cache-level-related metadata. LP manages the cache-level metadata as an in-memory table and caches the metadata on-chip using \rbh{a metadata} prefetching mechanism. These metadata structures need to be updated for \rbi{almost all} cache \rbi{operations} (e.g., cache \rbi{insertions}, \rbi{evictions}) in order to faithfully track the cache contents and to provide accurate cache-level predictions. \rbh{In contrast,} \pred is built on the insight that 
\rbh{the correlation between different types of program context information and off-chip loads can be accurately \emph{learned without tracking cache contents}.}
As a result, \pred does \rbh{\emph{not}} explicitly track addresses present in the cache hierarchy, but learns to predict \rbh{off-chip loads by aggregating various program context information}.}

\textbf{\rbe{Lower} complexity and hardware overhead.}
To enable accurate cache-level hit/miss prediction and \rbe{to recover from a misprediction}, prior works need \rbh{relatively intrusive} changes to multiple components on the on-chip datapath.
Both D2D and D2M cache \rbd{designs} require extending the TLB and operating system support to store the way and cache-level information for every cacheline within the reach of eTLB. 
LP requires $2$-bit metadata for every $64$B chunk of \emph{the physical main memory}. LP manages this metadata as an in-memory table in the system-reserved physical main memory space~\cite{lp}, which necessitates support from the operating system. 
In \rbe{the} worst case, to make a cache-level hit/miss prediction, LP needs to fetch the metadata from main memory, which \rbh{can take} longer than the cache lookup \rbd{LP is designed to avoid}. 
\rbg{\rbh{In contrast, \pred} \rbd{incurs a very modest \rbe{(4~KB)} \rbh{total} storage overhead and \rbe{small} changes to the existing datapath}.}

\textbf{No misprediction detection and recovery.} 
\rbd{LP \rbh{requires} cache level misprediction detection and recovery to prevent the program from using any stale, incoherent data from the memory subsystem. On the other hand, Hermes \emph{never} brings any data to the on-chip cache hierarchy unless the data is required by an LLC miss request (see~\cref{sec:returning_data_to_core}). Hence, Hermes \emph{does not} require any misprediction detection and recovery mechanism to maintain \rbe{correct execution}.}

\rbg{
\rbi{We design an address tag-tracking based off-chip load predictor (called \emph{TTP}; see~\cref{sec:dmf_configuration}) inspired by these prior works~\cite{loh_miss_map,lp,d2d} and evaluate it against \pred (see ~\cref{sec:evaluation})}.
\rbh{TTP} tracks address tags present in the \emph{entire} cache hierarchy \rbh{in its metadata structure (see~\cref{sec:dmf_configuration})} \rbh{and predicts that a load would go off-chip if the tag of the load address is not present in its metadata structure}.
\rbh{We open-source the implementation of TTP in our repository~\cite{hermes_github}}. \rbh{Our results show} that \pred provides both higher accuracy and \rbi{higher} performance than TTP (see~\cref{sec:acc_cov_ocp} and~\cref{sec:perf_1c}).}

%% file: 03design_overview.tex
\section{Hermes: Design Overview} \label{sec:design_overview}

\begin{figure}[!h]
\centering
\includegraphics[scale=0.4]{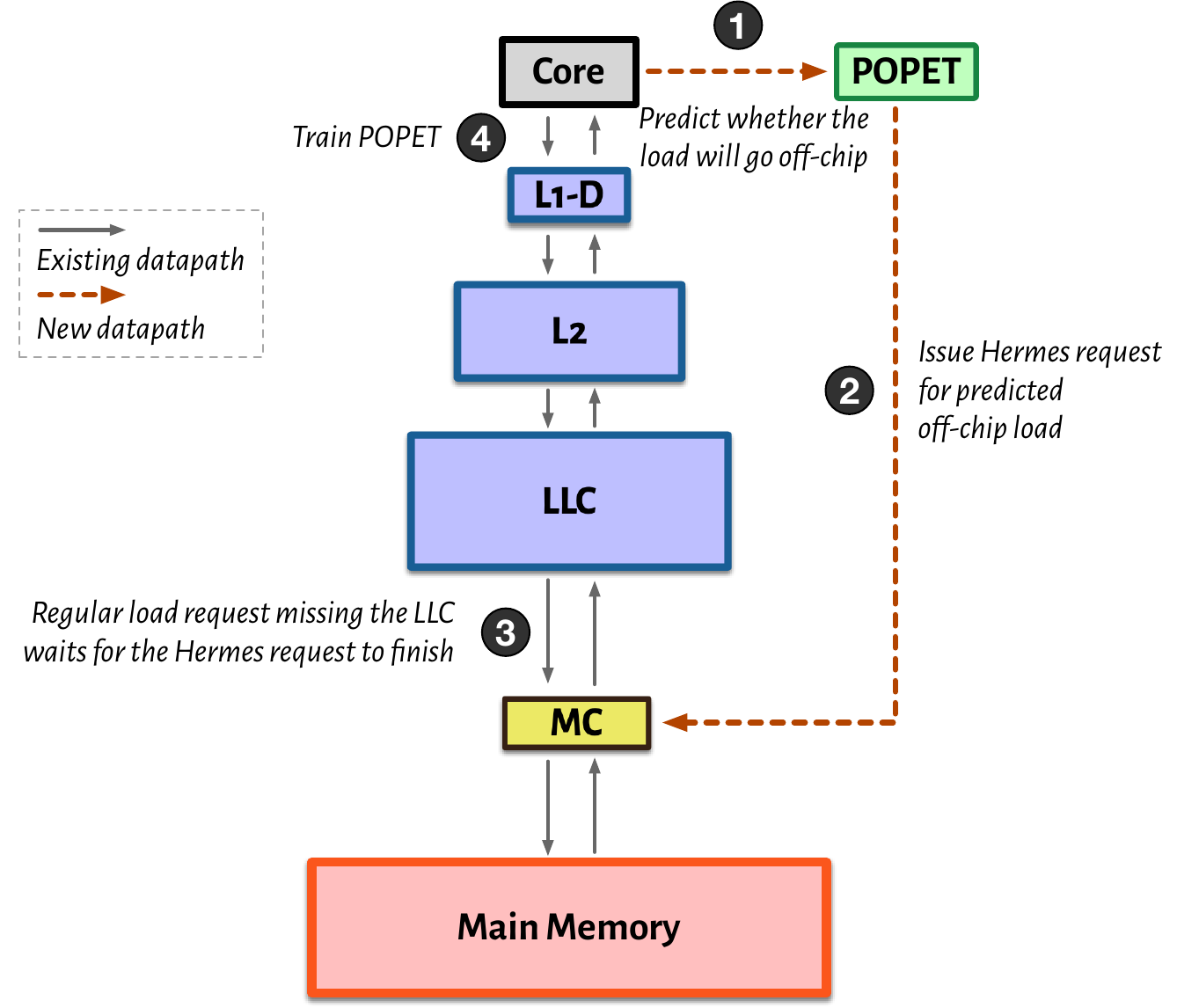}
\caption{Overview of Hermes}
\label{fig:dmf_overview}
\end{figure}

Fig.~\ref{fig:dmf_overview} shows a high-level overview of Hermes. 
\pred is the key component of Hermes that is responsible for making highly-accurate off-chip load predictions.  
For every demand load request generated by the processor, \pred predicts whether or not the load request would go off-chip (\circled{1}). 
If the load is predicted to go off-chip, Hermes issues a speculative \rbd{memory} request \rbd{(called a \emph{Hermes request})} \emph{directly} to the main memory controller once the \rbd{load's} physical address is generated to start fetching the corresponding data from the main memory (\circled{2}). 
\rbe{This Hermes request is serviced by the main memory controller concurrently with the \emph{regular load request} (i.e., the load issued by the processor that generated the Hermes request) that accesses the on-chip cache hierarchy.}
If the prediction is correct, the \rbd{regular} load request to the same address
eventually misses the LLC and waits for  the ongoing \rbd{Hermes request} to finish, \rbd{thereby completely hiding the on-chip cache hierarchy access latency from the critical path of the correctly-predicted \rbe{off-chip} load} (\circled{3}).
\rbd{\rbe{If a} Hermes request \rbe{returns} from the main memory \rbe{but there has been no} regular load request to the same address, Hermes drops the request and does not fill the data into the cache hierarchy. \rbe{By doing so,} Hermes keeps the on-chip cache hierarchy fully coherent even in case of a misprediction.}
For every regular load request returning to the core, Hermes trains \pred based on whether or not this load has actually gone off-chip (\circled{4}).

%% file: 04detailed_design.tex
\section{Hermes: Detailed Design} \label{sec:detailed_design}

We first describe the design of \pred in \cref{sec:ocp_design}, followed by the changes introduced by Hermes to the on-chip cache access datapath in \cref{sec:data_path_design}.

\subsection{\pred Design} \label{sec:ocp_design}

The purpose of \pred is to accurately predict whether or not a load request generated by the processor will go off-chip. We design \pred using \rbe{the} multi-feature perceptron learning mechanism~\cite{rosenblatt1958perceptron, perceptron,jimenez2002neural,jimenez2003fast,jimenez2017multiperspective,teran2016perceptron,ppf,garza2019bit}. 

\textbf{What is perceptron learning?}
\rbd{Perceptron learning, \rbe{whose roots are in~\cite{mcculloch1943logical} and was} demonstrated by Rosenblatt~\cite{rosenblatt1958perceptron}, is a simplified learning model to mimic biological neurons.
Fig.~\ref{fig:perc_basic} shows a \emph{single-layer perceptron} network where each \emph{input} is connected to the \emph{output} \rbe{via an} \emph{artificial neuron}. Each artificial neuron is represented by a numeric value, called \emph{weight}. 
\rbe{The perceptron network as a whole iteratively learns a binary classification function $f(x)$ (shown in Eq.~\ref{eq:perceptron}), a function that maps the input $X$ (a vector of $n$ values) to a single binary output.}

\begin{equation} \label{eq:perceptron}
f(x)=\left\{\begin{array}{ll}
1 & \text { if } w_{0}+\sum_{i=1}^{n} w_{i} x_{i}>0 \\
0 & \text { otherwise }
\end{array}\right.
\end{equation}

\rbe{The perceptron learning algorithm starts by initializing the weight of each neuron and iteratively trains the weights using each input vector from the training dataset in two steps.
First, for an input vector $X$, the perceptron network computes a binary output using Eq.~\ref{eq:perceptron} and the current weight values of its neurons. Second, if the computed output differs from the desired output for that input vector provided by the dataset, the weight of each neuron is updated~\cite{rosenblatt1958perceptron}. This iterative process is repeated until the error between the computed and desired output falls below a user-specified threshold.}

\begin{figure}[!h]
\vspace{0.5em}
\centering
\includegraphics[scale=0.3]{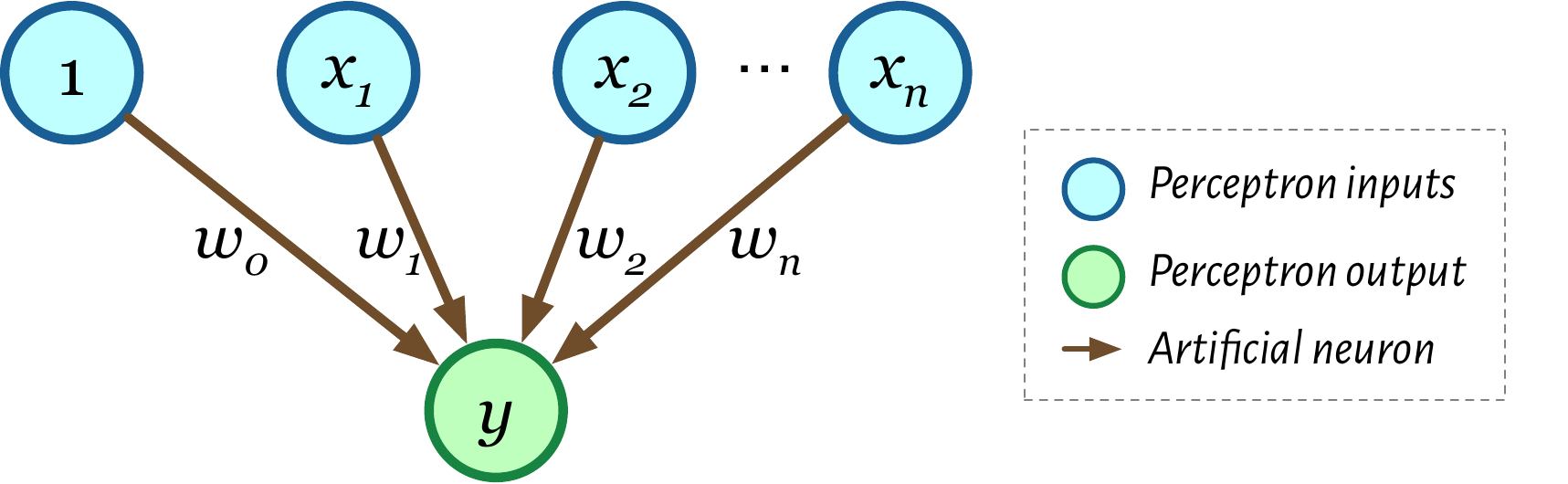}
\caption{Overview of a single-layer perceptron model. Each blue circle denotes an input and the green circle denotes the output of the perceptron.}
\label{fig:perc_basic}
\end{figure}

Jimenez et al.~\cite{jimenez2002neural} have applied the perceptron learning algorithm to design a lightweight, high-performance branch predictor in a processor core. Today, \rbe{multiple} commercial processors also use perceptron learning for making various microarchitectural predictions (e.g., Samsung Exynos~\cite{samsung_exynos}, and AMD Ryzen~\cite{ryzen_perceptron}).

We design \pred using \rbe{an existing} microarchitectural perceptron model, known as the \emph{hashed perceptron}~\cite{tarjan2005merging}. A hashed perceptron model hashes multiple feature values to retrieve weights of each feature from small tables. If the sum of these weights exceeds a threshold, the model makes a positive prediction. Hashed perceptron, as compared to other perceptron models, is lightweight and easy to implement in hardware. Prior works  successfully \rbe{apply} hashed perceptron for various microarchitectural predictions, e.g., branch outcome~\cite{jimenez2002neural,jimenez2003fast,garza2019bit}, LLC reuse~\cite{jimenez2017multiperspective,teran2016perceptron}, prefetch usefulness~\cite{ppf}. This is the first work that applies hashed perceptron \rbe{to} off-chip load prediction.
}

\textbf{Why perceptron?}
We choose to design \pred based on perceptron learning for two key reasons. First, by learning using multiple program features, perceptron learning can provide highly accurate predictions that could not be otherwise provided by simple history-based learning prediction (e.g., HMP~\cite{yoaz1999speculation}, described in~\cref{sec:differences_from_prior_work}). Second, perceptron learning can be implemented with low storage overhead, without requiring any impractical metadata support (e.g., extending TLB~\cite{d2d,d2m} or in-memory metadata storage~\cite{lp}, described in~\cref{sec:differences_from_prior_work}).

\textbf{\pred overview.}
\pred is organized as a collection of one-dimensional tables (\rbd{each} called a \emph{weight table}), where each table corresponds to a single program feature. Each table entry stores a \emph{weight} value, implemented using a $5$-bit saturating signed integer, that represents the correlation between the corresponding program feature value and the true outcome \rbd{(i.e., whether a given load actually went off-chip)}. A weight value saturated near the maximum (i.e., $+15$) or the minimum (i.e., $-16$) value represents a strong positive or negative correlation between the program feature value and the true outcome, respectively. A weight value closer to zero signifies a weak correlation. The weights are adjusted \rbd{during training (step \circled{4} in Fig.~\ref{fig:dmf_overview})} to \rbe{update} \pred's prediction with the true outcome. Each weight table is sized differently based on its corresponding program feature (see Table~\ref{table:overhead}).

\subsubsection{Making a Prediction} \label{sec:inference}

\begin{figure}[!h]
\centering
\includegraphics[width=3.3in]{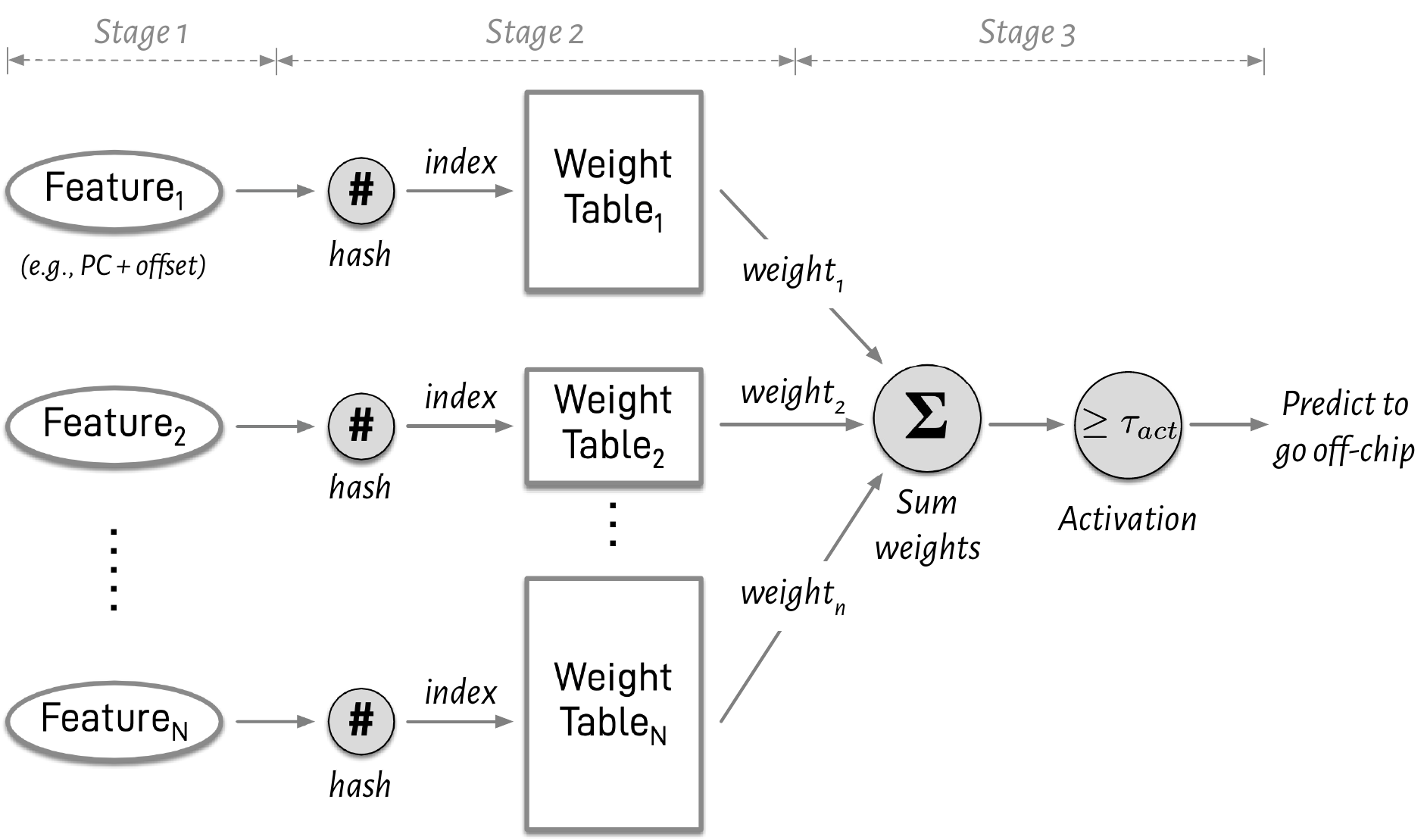}
\vspace{0.5em}
\caption{Stages to make a prediction by \pred}
\label{fig:perc_design}
\end{figure}

During load queue (LQ) allocation for a load generated by the core (step \circled{1} in Fig.~\ref{fig:dmf_overview}), \pred makes a binary prediction on whether or not the load request would go off-chip.
The prediction happens in three stages \rbd{as shown in Fig.~\ref{fig:perc_design}}.
In the first stage, \pred extracts a set of program features from the current load request and a history of prior requests (\cref{sec:feature_selection} shows the list of program features used by \pred). 
In the second stage, each feature value is hashed and used as an index to retrieve a weight value from the weight table of the corresponding feature.
In the third stage, all weight values from individual features are accumulated to generate the \emph{cumulative perceptron weight} ($W_{\sigma}$).
If $W_{\sigma}$ exceeds a predefined threshold (called the \emph{activation threshold}, $\tau_{act}$), \pred
makes a positive prediction (\rbe{i.e., it predicts that} the current load request would go off-chip). Otherwise, \pred makes a negative prediction.
The hashed feature values, the cumulative perceptron weight $W_{\sigma}$, and the predicted outcome are stored in the LQ entry to be reused \rbe{to train} \pred when the load request returns to the processor core (step \circled{4} in Fig.~\ref{fig:dmf_overview}).

\subsubsection{Training the Predictor} \label{sec:training}

\pred training is invoked when a demand load request returns to the core and prepares to release its corresponding LQ entry (step \circled{4} in Fig.~\ref{fig:dmf_overview}). Every demand load that misses the LLC and goes to the main memory controller is marked as a \emph{true} off-chip load request. This true off-chip outcome, along with the predicted outcome stored in the LQ entry of the demand load, are used to appropriately train the feature weights of \pred. 
The training happens in two stages. 
In the first stage, the $W_{\sigma}$ (computed during prediction) is retrieved from the LQ entry. If $W_{\sigma}$ is neither positively nor negatively saturated (i.e., $W_{\sigma}$ lies within a negative and a positive training threshold, $T_N$ and $T_P$, respectively), the weight training is triggered. This saturation check prevents the individual feature weight values from getting over-saturated, \rbd{thereby} helping \pred to quickly adapt its learning to program phase changes. In the second stage, if the weight training is triggered, the weights for each individual program feature are retrieved from their corresponding weight table using the hashed feature indices stored in the LQ entry. If the true outcome is positive (meaning the load actually went off-chip), the weight value for each feature is incremented by one. If the true outcome is negative, the weight values are decremented by one. This simple weight update mechanism moves each individual feature weight \rbe{towards} the direction of the true outcome, thus gradually increasing the prediction accuracy.

\subsubsection{Automated Feature Selection} \label{sec:feature_selection}

The selection of the program features used to make the off-chip \rbf{load} prediction is critical to \pred's performance. 
A \rbd{carefully-crafted and selected} set of features can significantly improve the accuracy and the coverage of \pred. In this section we propose an automated, offline, performance-driven methodology to find a set of program features for \pred.

Using our domain expertise, we initially select a set of $16$ individual program features that can correlate well with \rbd{a load going off-chip}. Table~\ref{table:initial_feature_set} shows the initial feature set.

\begin{table}[htbp]
  \vspace{0.5em}
  \centering
  \footnotesize
  \caption{The initial \rbd{set} of program features used for automated feature selection. $\oplus$ represents a \rbd{bitwise} XOR operation.}
  \small
    \begin{tabular}{m{12em}m{11em}}
    \toprule
    \textbf{\footnotesize Features without control-flow information} & \textbf{\footnotesize Features with control-flow information} \\
    \midrule
    \begin{minipage}{12em}
      \footnotesize
      \vskip 7pt
      \begin{enumerate}[leftmargin=1em]
        \setlength\itemsep{0.001em}
        \item Load virtual address
        \item Virtual page number
        \item Cacheline offset in page
        \item First access
        \item Cacheline offset + first access
        \item Byte offset in cacheline
        \item Word offset in cacheline
      \end{enumerate}
      \vskip 4pt
    \end{minipage} & 
    \begin{minipage}{11em}
      \footnotesize
      \vskip 1pt
      \begin{enumerate}[leftmargin=2em]
        \setcounter{enumi}{7}
        \setlength\itemsep{0.001em}
        \item Load PC
        \item PC $\oplus$ load virtual address
        \item PC $\oplus$ virtual page number
        \item PC $\oplus$ cacheline offset
        \item PC + first access
        \item PC $\oplus$ byte offset
        \item PC $\oplus$ word offset
        \item Last-$4$ load PCs
        \item Last-$4$ PCs
      \end{enumerate}
      \vskip 4pt
    \end{minipage}\\
    \bottomrule
    \end{tabular}%
  \label{table:initial_feature_set}%
\end{table}%

The automated feature selection process happens offline during the design time of \pred. The process starts with the initial set of 16 individual program features and iteratively creates a list of \emph{feature sets}, each containing $n$ features, at every iteration $n$ in the following way. In the first iteration, we design \pred with each \rbe{of the} 16 initial program features and test its prediction accuracy in $10$ randomly-selected workload traces (called \emph{testing workloads}). We select the top-$10$ features that produce the highest prediction accuracy for the second iteration. In the second iteration, we create $160$ two-combination feature sets (meaning, each feature set contains two initial features from Table~\mbox{\ref{table:initial_feature_set}}) by \rbe{combining} each \rbd{of the} $16$ initial features with each of the $10$ winning feature sets from the last iteration, and test the prediction accuracy on the testing workloads. We select the top-$10$ two-combination feature sets that produce the highest prediction accuracy for the third iteration. This iterative process repeats until the maximum prediction accuracy gets saturated \rbd{(i.e., the difference in accuracy of two successive iterations is less than $3\%$)}.\footnote{\rbe{For simplicity, our automated feature selection process optimizes for accuracy. A more comprehensive feature selection process can also include coverage or directly optimize for performance (i.e., execution time).}}
Table~\ref{table:ocp_config} shows the final list of program features selected by the automated feature selection process.

\begin{table}[htbp]
  \vspace{0.5em}
  \centering
  \footnotesize
  \caption{\pred configuration parameters}
  \small
    \begin{tabular}{m{9em}m{15em}}
    \toprule
    \vskip 15pt \textit{\textbf{Selected features}} & 
    \begin{minipage}{15em}
      \footnotesize
      \begin{itemize}[leftmargin=1em]
        \setlength\itemsep{0.001em}
        \item PC $\oplus$ cacheline offset
        \item PC $\oplus$ byte offset
        \item PC + first access
        \item Cacheline offset + first access
        \item Last-$4$ load PCs
      \end{itemize}
      \vskip 4pt
    \end{minipage} \\
    \midrule
    \textit{\textbf{Threshold values}} & $\tau_{act} = -18$, $T_{N} = -35$, $T_{P} = 40$ \\
    \bottomrule
    \end{tabular}%
  \label{table:ocp_config}%
\end{table}%

\vspace{1pt}
\noindent \textbf{Rationale \rbd{for selected features}.} Each selected feature correlates with the likelihood of observing an off-chip load request with a different program context information. \rbd{We explain the rationale for each selected feature below.}

\textbf{(1) PC $\oplus$ cacheline offset.} This feature is computed by XOR-ing the load PC value with the cacheline offset \rbd{of the load address} in the virtual page of the load request. 
The goal of this feature is to learn the likelihood of a load request going off-chip when a given load PC touches a certain cacheline offset in a virtual page.
The use of cacheline offset information, instead of load virtual address or virtual page number, enables this feature to apply the learning across different virtual pages.

\textbf{(2) PC $\oplus$ cacheline byte offset.} This feature is computed by XOR-ing the load PC with the byte offset of the load cacheline address. This feature is particularly useful in accurately predicting off-chip load requests when a program has a streaming access pattern over a linearly allocated data structure. For example, when a program streams through a large array of $4$B integers, every $16^{th}$ load (as a $64$B cacheline stores $16$ integers) generated by a load PC that is iterating over the array will go off-chip, and the remaining loads will hit in on-chip caches. In this case, this feature learns to identify only those loads \rbd{that have a byte offset of $0$ to go off-chip}.

\textbf{(3) PC + first access.} This feature is computed by left-shifting the load PC and adding the \emph{first access} hint at the most-significant bit position. \rbd{The first access hint is a binary value that represents whether or not a cacheline has been recently touched by the program}. The hint is computed using a small $64$-entry buffer (called the \emph{page buffer}) that tracks the demanded cachelines from last $64$ virtual pages. Each page buffer entry holds two \rbd{pieces of} information: a virtual page tag, and a $64$-bit bitmap, where each bit represents one cacheline in the virtual page. During every load request generation, \pred searches the page buffer with the virtual page number \rbd{of the load address}. \rbd{If a matching entry is found, \pred} uses the value of the bit corresponding to the cacheline offset \rbd{in the matching page buffer entry's bitmap} as the first access hint.
\rbd{If the bit is set (or unset), it signifies that the corresponding cacheline has (not) been recently accessed by the program. If the bit is unset, \pred sets the bit in the page buffer entry's bitmap.}
The first access hint provides a crude estimate of a cacheline's reuse in a short temporal window. However, it alone cannot determine the cacheline's residency in on-chip caches, as the memory footprint tracked by the page buffer is much smaller than the total cache size.

\textbf{(4) Cacheline offset + first access.} This feature is similar to the PC + first access feature, except that it learns the likelihood of a load request going off-chip when a given cacheline offset is recently touched by the program.

\textbf{(5) Last-4 load PCs.} This feature value is computed as a shifted-XOR of last four load PCs. \rbe{It} represents the execution path of a program and correlates it with the likelihood of observing an off-chip load request whenever the program follows the same execution path.

\subsubsection{Parameter Threshold Tuning} \label{sec:threshold_selection}

\rbd{\pred has three tunable parameters: negative and positive training thresholds ($T_{N}$ and $T_{P}$, respectively), and the activation threshold ($\tau_{act}$). Properly tuning the values of all these three parameters}
is also critical to \pred's performance, since both \pred's accuracy and coverage are sensitive to parameter values. 

We employ a three-step grid search technique to tune each of the three parameters separately. 
In the first stage, we uniformly sample values from \rbd{a parameter's} range. For example, $\tau_{act}$ can take values in the range $[-80,75]$.\footnote{\rbd{As \pred uses five program features (see~\cref{sec:feature_selection}), the sum of all five weights (each represented by a 5-bit saturating signed integer as described in ~\cref{sec:ocp_design}) can take a maximum and minimum value of $75$ and $-80$, respectively.}} 
We uniformly sample values from this range with a grid size of $5$. In the second stage, we run Hermes with the randomly-selected 10 test workloads (as mentioned in~\mbox{\cref{sec:feature_selection}}) for each of the sampled values and pick the top-10 values that provide the highest performance gain. In the third stage, we run Hermes with all single-core workload traces using the selected 10 parameter values from the second stage. We finally select the value that provides the highest average performance gain. Table~\mbox{\ref{table:ocp_config}} shows the \rbe{selected} threshold values of each parameter. 

\subsection{Hermes Datapath Design} \label{sec:data_path_design}

In this section, we describe the key changes introduced to the existing well-optimized on-chip cache access datapath to incorporate Hermes.
First, we show how the core issues \rbd{a Hermes request} directly to the main memory controller \rbe{if \pred predicts the load would go off-chip} and how a \rbd{regular} load request that misses the LLC waits for an ongoing \rbd{Hermes} request (see~\cref{sec:spec_load_issue}). Second, we discuss \rbd{how the data fetched from main memory is properly sent back to the core in presence of Hermes} while maintaining cache coherence (see~\cref{sec:returning_data_to_core}).

\subsubsection{Issuing a \rbd{Hermes} Request} \label{sec:spec_load_issue}

For every load request predicted to go off-chip, Hermes issues a Hermes request directly to the main memory controller (step~\circled{2} in Fig.~\ref{fig:dmf_overview}) once the load's physical address is generated.
The main memory controller enqueues the \rbd{Hermes request} in its read queue (RQ) and starts fetching the corresponding data from the main memory as dictated by its scheduling policy, \rbe{while the regular load request is concurrently accessing the on-chip cache hierarchy}. 
\rbd{If the off-chip prediction is correct}, the \rbd{regular} load request eventually misses the LLC and checks the main memory controller's RQ for any ongoing main memory access to the same load address (step~\circled{3}). If the address is found, the \rbd{regular} load request waits for the ongoing \rbd{Hermes} request to finish before sending the Hermes-fetched data back to the core.

Hermes's performance gain depends on the latency to directly issue \rbd{a Hermes} request to the main memory controller \rbd{(called \emph{Hermes request issue latency})}.
\rbe{Although} a \rbd{Hermes request} experiences a significantly shorter latency to arrive at the main memory controller than its corresponding \rbd{regular load request} \rbe{because a Hermes request bypasses the cache hierarchy and on-chip queueing delays}, \rbd{a Hermes request} nonetheless pays for a latency to route through the on-chip network. We model two variants of Hermes using an optimistic and a pessimistic estimate of \rbd{Hermes request} issue latency \rbe{to take into account a wide range of potential differences in on-chip interconnect designs} (see \cref{sec:dmf_configuration}). \rbd{In~\cref{sec:load_issue_latency}, we also evaluate Hermes with a wide range of Hermes request issue latencies (from $0$ cycle to $24$ cycles)} and show that Hermes consistently provides performance benefit even with the most pessimistic Hermes request issue latency.

\subsubsection{Returning Data to the Core} \label{sec:returning_data_to_core}

For every \rbd{Hermes request} returning from main memory, Hermes checks the RQ of the main memory controller and returns the fetched data back to the LLC if there is a \rbd{regular} load request already waiting for the same load address. If there is no \rbd{regular} load request waiting for the completed \rbd{Hermes} \rbd{request}, \rbd{Hermes drops the request and does \emph{not} fill the data into the cache hierarchy, \rbe{which} keeps the on-chip cache hierarchy \rbe{internally} coherent.}

\subsection{Storage Overhead} \label{sec:storage_overhead}

Table~\ref{table:overhead} shows the total storage overhead of Hermes. Hermes requires only $4$~KB of metadata storage \rbd{per processor core}. \pred 
consumes $3.2$~KB, whereas the metadata stored in LQ for \pred training consumes $0.8$~KB.

\begin{table}[htbp]
  \vspace{0.5em}
  \centering
  \footnotesize
  \caption{Storage overhead of Hermes}
    \begin{tabular}{|L{4.3em}||m{18em}||C{3em}|}
    \hline
    \textbf{Structure} & \textbf{Description} & \textbf{Size}\\
    \hline
    \hline
    \vskip 20pt \pred &
    \begin{minipage}{18em}
      \footnotesize
      \vskip 2pt
      \begin{itemize}[leftmargin=1em]
        \setlength\itemsep{0.01em}
        \item Perceptron weight tables
        \begin{itemize}[leftmargin=1em]
            \setlength\itemsep{0.01em}
            \item PC $\oplus$ cacheline offset: $1024\times5b$
            \item PC $\oplus$ byte offset: $1024\times5b$
            \item PC + first access: $1024\times5b$
            \item Cacheline offset + first access: $128\times5b$
            \item Last-$4$ load PCs: $1024\times5b$
        \end{itemize}
        \item Page buffer: $64\times80b$
      \end{itemize}
      \vskip 2pt
    \end{minipage} &
    \vskip 20pt \textbf{3.2 KB} \\
    \hline
    \hline
    LQ Metadata & Hashed PC: $128\times32b$; Last-4 PC: $128\times10b$; First access: $128\times1b$; perceptron weight: $128\times5b$; prediction: $128\times1b$ & \textbf{0.8 KB} \\
    \hline
    \hline
    Total &
    & \textbf{4.0 KB} \\
    \hline
    \end{tabular}%
  \label{table:overhead}%
\end{table}%

%% file: 05methodology.tex
\section{Methodology} \label{sec:methodology}

We use the ChampSim trace-driven simulator~\cite{champsim} to evaluate Hermes. We faithfully model the latest-generation Intel Alder Lake performance-core~\cite{goldencove} with its \rbd{large} ROB, large caches with publicly-reported on-chip cache access latencies~\cite{llc_lat1,llc_lat2,l3_lat_compare1}, and 
the state-of-the-art prefetcher Pythia~\cite{pythia} at the LLC. 
Table~\ref{table:sim_params} shows the key microarchitectural parameters. For single-core simulations, we \rbd{warm up} the core using $100$M instructions and simulate the next $500$M instructions. For multi-programmed simulations, we use $50$M and $100$M instructions from each workload for warmup and simulation, respectively. If a core finishes early, the workload is replayed \rbd{until} every core has finished executing at least $100$M instructions. 
\rbd{The source code of Hermes, along with all workload traces and scripts to reproduce \rbe{our} results are \rbe{freely} available at~\cite{hermes_github}.}

\begin{table}[h]
    \centering
    \vspace{0.5em}
    \caption{Simulated system parameters}
    \footnotesize
    \begin{tabular}{m{4em}m{23.5em}}
         \toprule
         \textbf{Core} & { 1 and 8 cores, 6-wide fetch/execute/commit, 512-entry ROB, 128/72-entry LQ/SQ, Perceptron branch predictor~\cite{perceptron} with 17-cycle misprediction penalty}\\
         \midrule
         \textbf{L1/L2 Caches} & Private, 48KB/1.25MB, 64B line, 12/20-way, 16/48 MSHRs, LRU, 5/15-cycle round-trip latency~\cite{llc_lat2} \\
         \midrule
         \textbf{LLC} & 3MB/core, 64B line, 12 way, 64 MSHRs/slice, SHiP~\cite{ship}, 55-cycle round-trip latency~\cite{llc_lat1,llc_lat2}, \textbf{Pythia} prefetcher~\cite{pythia}\\
         \midrule
         \textbf{Main Memory} & \textbf{1C:} 1 channel, 1 rank per channel; \textbf{8C:} 4 channels, 2 ranks per channel; 8 banks per rank, DDR4-3200 MTPS, 64b data-bus per channel, 2KB row buffer per bank, tRCD=12.5ns, tRP=12.5ns, tCAS=12.5ns \\
        \midrule
        \textbf{Hermes} & \textbf{Hermes-O/P}: $6$/$18$-cycle \rbd{Hermes request} issue latency\\
        \bottomrule
    \end{tabular}
    \label{table:sim_params}
\end{table}

\subsection{Workloads} \label{sec:workloads}

\begin{sloppypar}
We evaluate Hermes using a wide range of memory-intensive workloads spanning \texttt{SPEC CPU2006}~\cite{spec2006}, \texttt{SPEC CPU2017}~\cite{spec2017}, \texttt{PARSEC}~\cite{parsec}, \texttt{Ligra} graph processing workload suite~\cite{ligra}, and commercial workloads from \rbd{the} 2nd data value prediction championship (\texttt{CVP}~\cite{cvp2}). For \texttt{SPEC CPU2006} and \texttt{SPEC CPU2017} workloads, we reuse the instruction traces provided by the 2nd and the 3rd data prefetching championships (DPC~\cite{dpc2,dpc3}). For \texttt{PARSEC} and \texttt{Ligra} workloads, we reuse the instruction traces open-sourced by Pythia~\cite{pythia}. 
The \texttt{CVP} workload traces are collected by the Qualcomm Datacenter Technologies and capture complex program behavior from various integer, floating-point, cryptographic, and server applications in the \rbd{field}. 
We only consider workload traces in our evaluation that have at least $3$ LLC misses per kilo instructions (MPKI) in the no-prefetching system. In total, we evaluate Hermes using $110$ single-core workload traces \rbd{from} $73$ workloads, which are summarized in Table~\ref{table:workloads}.
\rbe{All these traces can be freely downloaded using a script as mentioned in Appendix~\ref{sec:preparing_traces}.}
For multi-programmed simulations, we create both homogeneous and heterogeneous trace mixes. For an eight-core homogeneous multi-programmed simulation, we run eight copies of each trace from our single-core trace list, one trace in each core. For heterogeneous multi-programmed simulation, we \emph{randomly} select any eight traces from our single-core trace list and run one trace in each core. In total, we evaluate Hermes using $110$ homogeneous and $110$ heterogeneous eight-core workloads.
\end{sloppypar}

\begin{table}[htbp]
  \vspace{0.5em}
  \centering
  \footnotesize 
  \caption{Workloads used for evaluation}
    \begin{tabular}{m{2.5em}C{5.5em}C{3.5em}m{13em}}
    \toprule
    \textbf{Suite} & \textbf{\#Workloads} & \textbf{\#Traces} & \textbf{Example Workloads} \\
    \midrule
    \texttt{SPEC06} & 14    & 22    & gcc, mcf, cactusADM, lbm, ... \\
    \texttt{SPEC17} & 11    & 23    & gcc, mcf, pop2, fotonik3d, ... \\
    \texttt{PARSEC} & 4     & 12    & canneal, facesim, raytrace, ... \\
    \texttt{Ligra}  & 11    & 20    & BFS, PageRank, Radii, ... \\
    \texttt{CVP}    & 33    & 33    & integer, floating-point, server, ... \\ 
    \bottomrule
    \end{tabular}%
  \label{table:workloads}%
\end{table}%

\subsection{Evaluated System Configurations} \label{sec:dmf_configuration}

For a comprehensive analysis, we compare Hermes with various off-chip load prediction mechanisms, as well as \rbd{in combination with} various recently proposed prefetchers. Table~\ref{table:overhead_comparison} compares the storage overhead of all evaluated mechanisms.

\textbf{\textit{(1) Various off-chip prediction mechanisms.}}
\rbg{We compare \pred against two cache hit/miss prediction techniques: (1) HMP, proposed by Yoaz et al.~\cite{yoaz1999speculation}, and (2) a simple cacheline tag-tracking based predictor, called \emph{TTP}, \rbh{which we design}.}
\rbh{HMP uses three predictors similar to a hybrid branch predictor: local~\cite{yeh1991two}, gshare~\cite{mcfarling1993combining,yeh1991two}, and gskew~\cite{michaud1997trading}, each of which individually \rbi{predicts} off-chip loads using a different \rbi{prediction} mechanism. For a given load, HMP consults each individual predictor and selects the majority prediction.}
\rbg{\rbh{We design} TTP \rbh{by taking inspiration from} prior cacheline address tracking-based mechanisms~\cite{loh_miss_map,lp,d2d} \rbh{(see~\cref{sec:differences_from_prior_work})}. \rbh{TTP} tracks partial tags of cacheline addresses that are likely to be present in the \rbh{entire on-chip} cache hierarchy in a separate metadata structure. For every cache fill (LLC eviction), the partial tag of the filled (evicted) cacheline address is inserted \rbi{into} (evicted \rbi{from}) TTP's metadata. To predict whether or not a given load would go off-chip, TTP searches the metadata \rbi{structure} with the partial tag of the load address. If the tag is not present in the metadata \rbi{structure}, TTP predicts the load would go off-chip. \rbh{We open-source TTP in our repository~\cite{hermes_github}.}}

\textbf{\textit{(2) Various data prefetchers.}}
We \rbd{evaluate} Hermes \rbd{combined with} five recently-proposed high-performance prefetching techniques: Pythia~\cite{pythia}, Bingo~\cite{bingo}, SPP~\cite{spp} (with perceptron filter~\cite{ppf}), MLOP~\cite{mlop}, and SMS~\cite{sms}. As mentioned in Table~\ref{table:sim_params}, Pythia is incorporated in our baseline system.

\begin{table}[h]
    \vspace{0.5em}
    \centering
    \footnotesize 
    \caption{Storage overhead of all evaluated mechanisms}
    \begin{tabular}{L{23.6em}||L{3.6em}}
    \thickhline
    \textbf{HMP}~\cite{yoaz1999speculation} with local, gshare, and gskew predictors & \textbf{11 KB}\\
    \rbg{\textbf{TTP} with a metadata budget similar to the L2 cache} & \textbf{1536 KB} \\
    \hline
    \hline
    \textbf{Pythia}~\cite{pythia} with the same configuration in~\cite{pythia} & \textbf{25.5 KB} \\
    \textbf{Bingo}~\cite{bingo} with the same configuration in~\cite{bingo} & \textbf{46 KB} \\
    \textbf{SPP}~\cite{spp} with perceptron-based prefetch filter~\cite{ppf} & \textbf{39.3 KB} \\
    \textbf{MLOP}~\cite{mlop} with the same configuration in~\cite{mlop} & \textbf{8 KB} \\
    \textbf{SMS}~\cite{sms} with the same configuration in~\cite{sms} & \textbf{20 KB} \\
    \hline
    \hline
    \textit{\textbf{Hermes with \pred}} \textit{(this work)} & \textbf{4 KB} \\
    \thickhline
    \end{tabular}
    \label{table:overhead_comparison}
\end{table}

We evaluate two variants of Hermes: \textbf{Hermes-O} and \textbf{Hermes-P}. These two variants differ only in Hermes request issue latency. \emph{Hermes-O} (i.e., the optimistic Hermes) and \emph{Hermes-P} (i.e., the pessimistic Hermes) use a request issue latency of $6$ cycles and $18$ cycles, respectively. Unless stated otherwise, \emph{Hermes} represents the optimistic variant \emph{Hermes-O}.

%% file: 06evaluation.tex
\section{Evaluation} \label{sec:evaluation}

\subsection{\pred Prediction Analysis} \label{sec:ocp_pred}

\subsubsection{Accuracy and Coverage of \pred} \label{sec:acc_cov_ocp}

\begin{figure}[!h]
\centering
\includegraphics[width=3.3in]{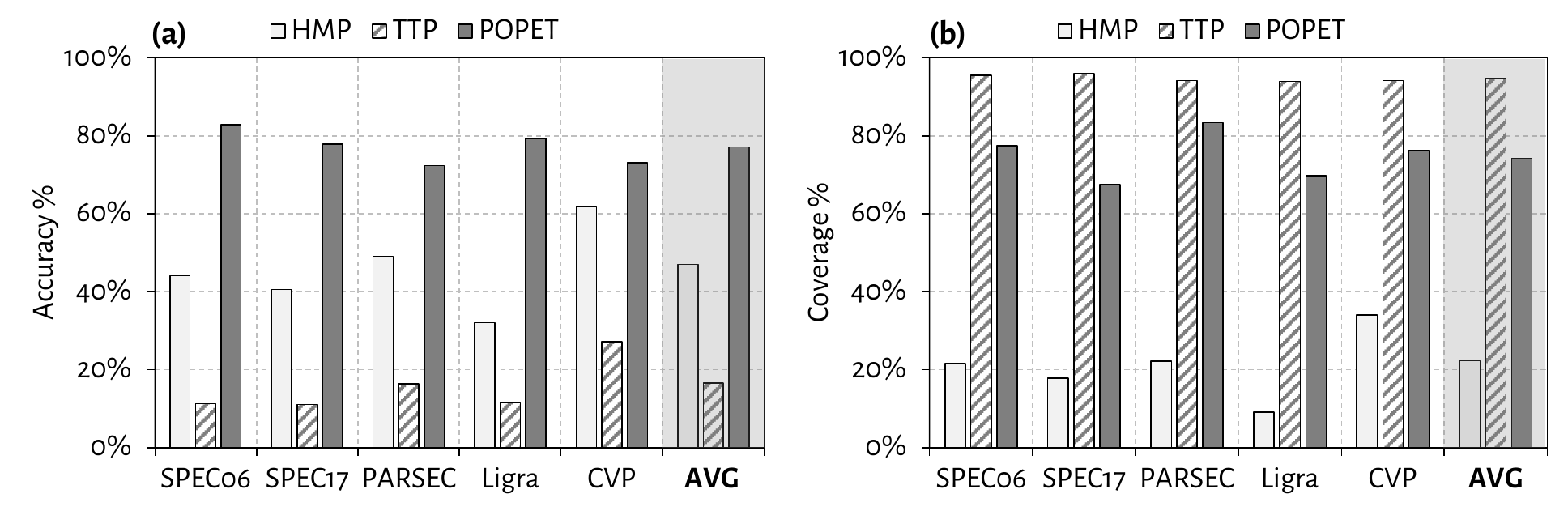}
\vspace{0.2em}
\caption{Comparison of (a) accuracy and (b) coverage of \pred against those of HMP~\cite{yoaz1999speculation} and TTP.
}
\label{fig:ocp_acc_cov}
\end{figure}

Fig.~\ref{fig:ocp_acc_cov} shows the comparison of \rbh{\pred's} \rbe{off-chip load prediction} accuracy and coverage  against \rbh{those of} HMP and TTP in the baseline system. The key takeaway is that \pred has significantly higher accuracy \emph{and} coverage than HMP. \rbh{\pred provides} $77.1\%$ accuracy with $74.3\%$ coverage \rbd{on average across all single-core workloads}, whereas HMP \rbh{provides} $47\%$ accuracy with $22.3\%$ coverage. 
\rbg{TTP, with a metadata budget of $1.5$~MB, \rbh{provides the highest} coverage \rbh{($94.8\%$)} but with a significantly lower accuracy \rbh{($16.6\%$)}.}
\pred's superior accuracy \emph{and} coverage directly translates to performance benefits both in single-core and eight-core system configuration (see ~\cref{sec:perf_1c_main} and~\cref{sec:perf_mc}).

\subsubsection{\rbd{Effect of} Different \pred Features} \label{sec:acc_cov_ocp_contri}

\begin{figure}[!h]
\centering
\includegraphics[width=3.3in]{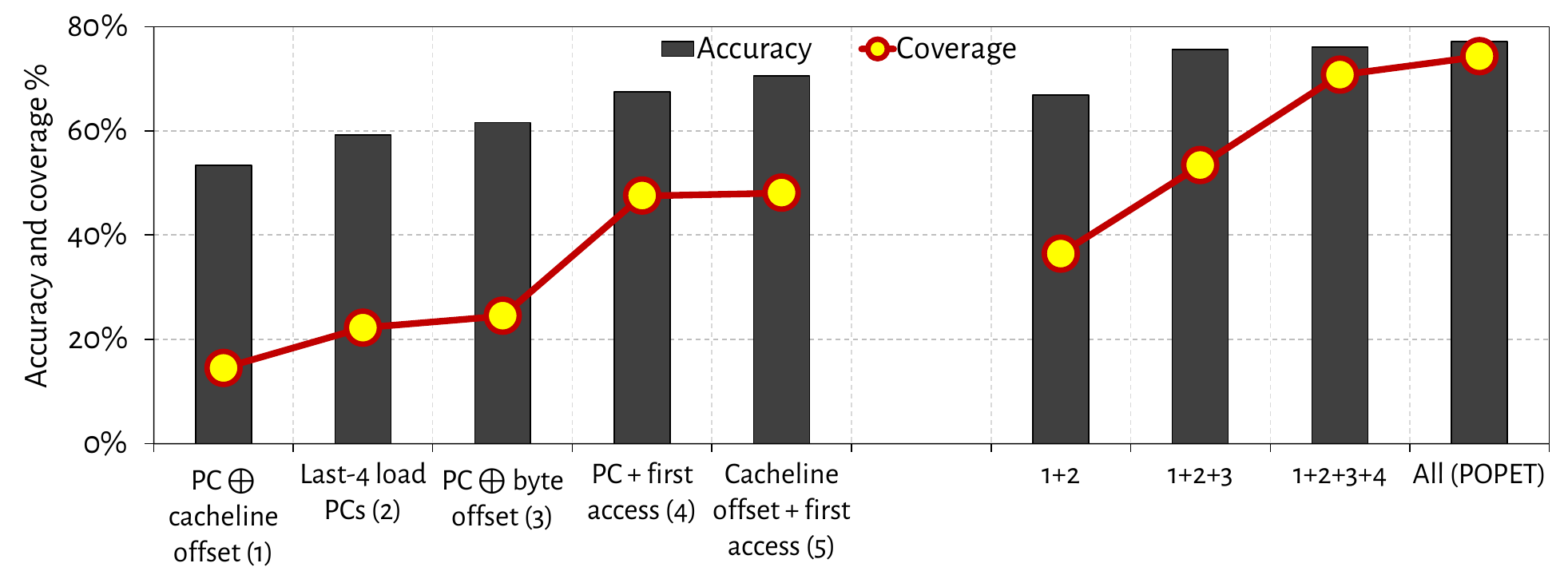}
\caption{The accuracy and coverage of \pred using each program feature individually and in various combinations.}
\label{fig:acc_cov_contri}
\end{figure}

Fig.~\ref{fig:acc_cov_contri} shows the accuracy and coverage of \pred using the five selected program features used individually and in \rbe{various} combinations. We make two key observations.
First, each program feature individually produces predictions with a wide range of accuracy and coverage.
The PC $\oplus$ cacheline offset feature produces the lowest-quality predictions with only $53.4\%$ accuracy and $14.5\%$ coverage,
whereas the cacheline offset + first access feature produces the highest-quality predictions with $70.6\%$ accuracy and $48.1\%$ coverage.
Second, by stacking multiple features together, the final \pred design achieves \emph{both} higher accuracy and coverage than \rbd{those provided by} any \rbd{single} individual program feature.
We conclude that \pred is capable of learning from multiple program features to achieve both higher off-chip load prediction accuracy and coverage than any individual program feature can provide.

\subsubsection{Usefulness of all features} \label{sec:feature_necessity}
\begin{figure}[!h]
\centering
\includegraphics[width=3.3in]{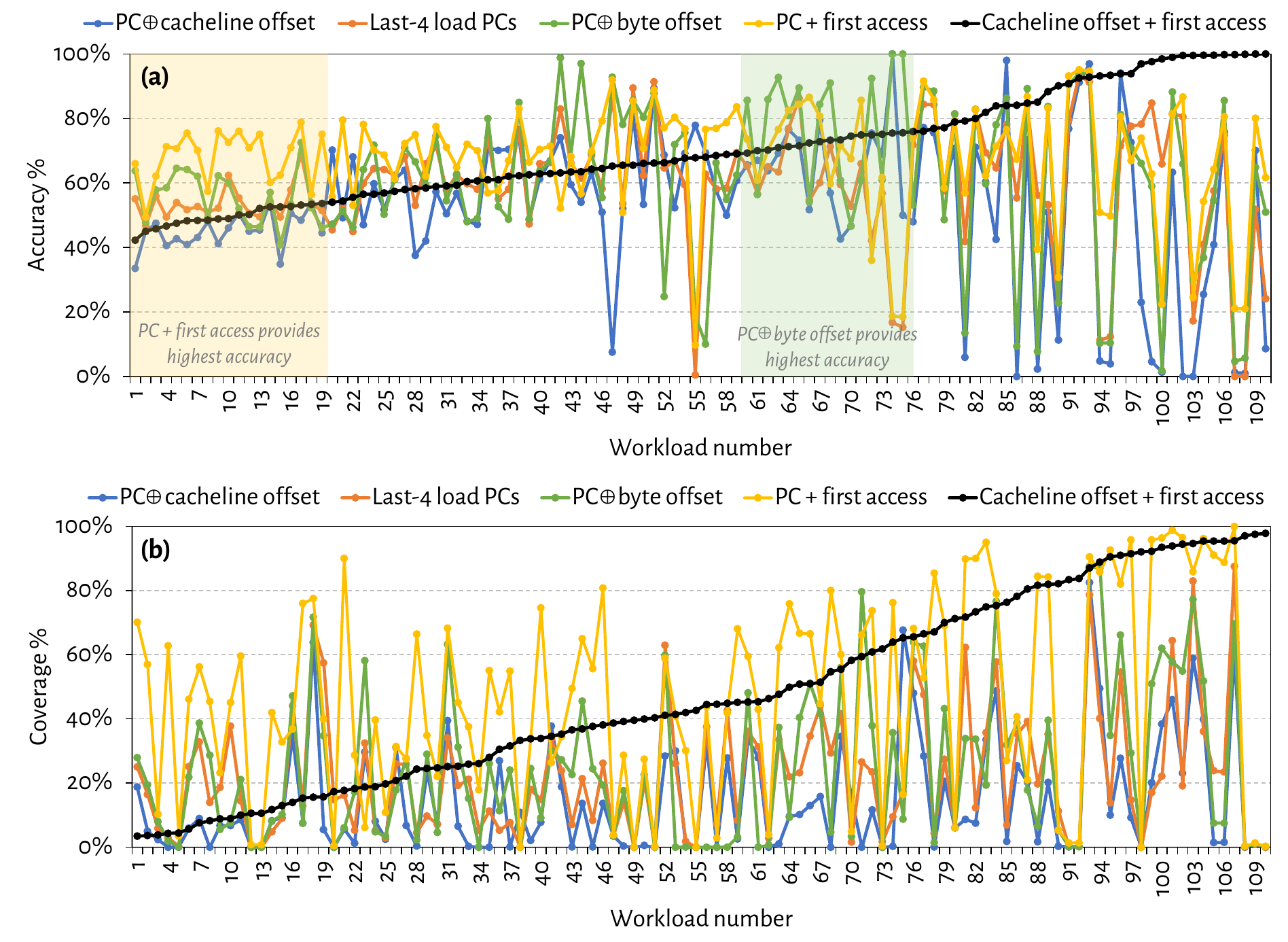}
\vspace{0.5em}
\caption{Line graph of \pred's \rbd{(a) accuracy and (b) coverage} using each \rbd{of the} five program features individually across all 110 single-core workloads. 
No single feature can provide the best accuracy \rbd{or coverage} across all workloads.
}
\label{fig:acc_line_chart}
\end{figure}

To understand the \rbd{usefulness} of multi-feature learning, we analyze per-trace accuracy and coverage of \pred using each individual program feature. Fig.~\mbox{\ref{fig:acc_line_chart}(a)} shows the line graph of \pred's prediction accuracy with each \rbd{of the} five program features individually for all single-core workload traces. The traces are sorted in ascending order of \pred accuracy using the feature cacheline offset + first access, since this feature individually has the highest average accuracy (as shown in Fig.~\mbox{\ref{fig:acc_cov_contri}(a)}). The key takeaway from Fig.~\mbox{\ref{fig:acc_line_chart}(a)} is that there is no \emph{single} program feature that individually provides the highest prediction accuracy across \emph{all} workloads. Out of $110$ workload traces, the features PC + first access, cacheline offset + first access, PC $\oplus$ byte offset, PC $\oplus$ cacheline offset, and last-4 load PCs provide the highest prediction accuracy in $47$, $29$, $20$, $9$, and $5$ workload traces, respectively. We observe similar variability in \pred's coverage, \rbd{as shown in Fig.~\ref{fig:acc_line_chart}(b)}, where no single program feature individually provides the highest coverage across \emph{all} workloads. This large variability of accuracy/coverage with different features in different workloads warrants learning using all features \emph{in unison} to provide higher accuracy \emph{and} coverage than any individual program feature across a \emph{wide range} of workloads.

\subsection{Single-core Performance Analysis} \label{sec:perf_1c_main}

\subsubsection{Performance Improvement} \label{sec:perf_1c}

\begin{figure}[!h]
\centering
\includegraphics[width=3.3in]{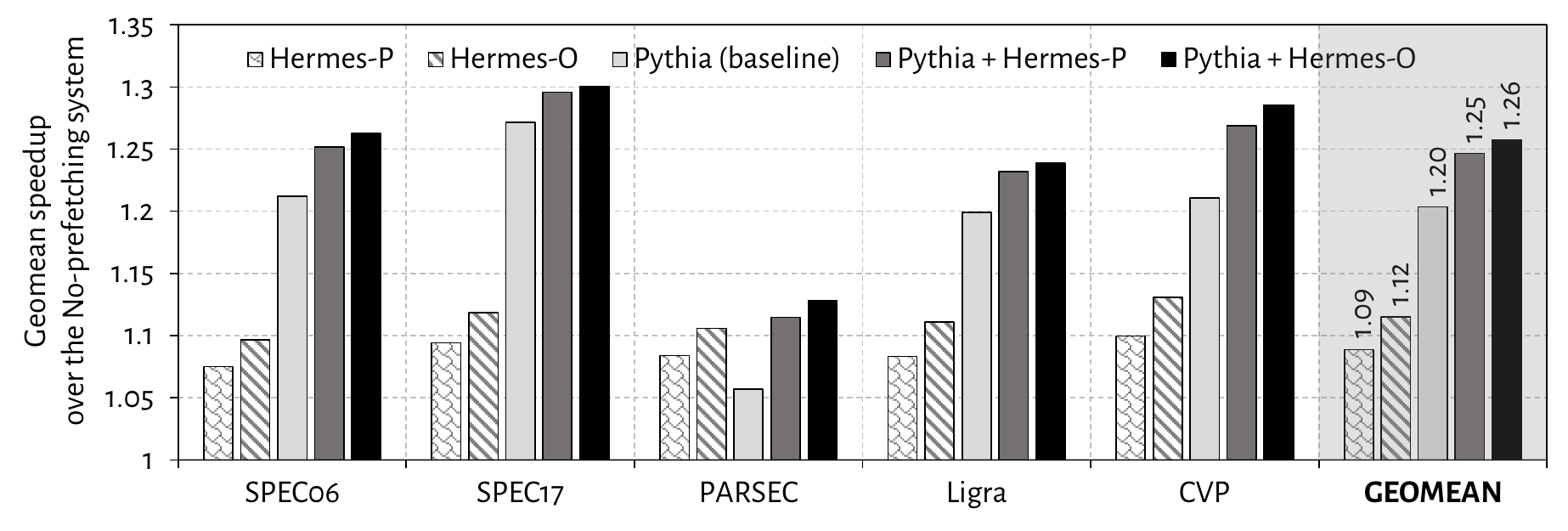}
\vspace{0.5em}
\caption{\rbd{Speedup in single-core workloads}}
\label{fig:perf_1c}
\end{figure}

Fig.~\ref{fig:perf_1c} shows \rbd{performance of} Hermes \rbd{(O and P)}, Pythia, and Hermes \rbd{combined with} Pythia \rbd{normalized to} the no-prefetching system in single-core workloads. We make three key observations. 
First, Hermes provides nearly half of the performance \rbd{benefit of} Pythia with only $\frac{1}{5}\times$ the storage overhead. On average, Hermes-O improves performance by $11.5\%$ over a no-prefetching system, whereas Pythia improves performance by $20.3\%$.
Second, Hermes-O \rbd{(Hermes-P) combined with} Pythia outperforms Pythia by $5.4\%$ ($4.3\%$). Third, Hermes \rbd{combined with} Pythia \emph{consistently} outperforms Pythia in \emph{every} workload category.

\begin{figure}[!h]
\centering
\includegraphics[width=3.3in]{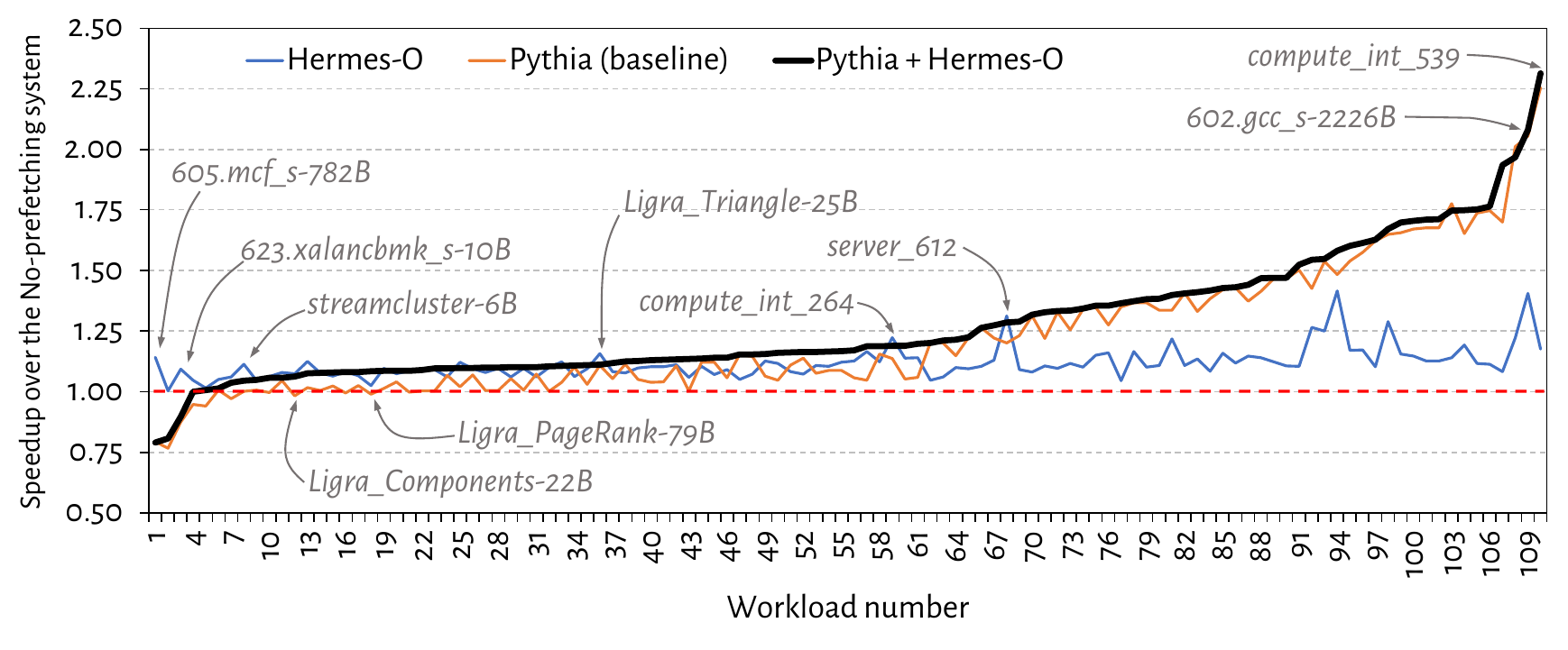}
\caption{Single-core performance of all 110 workloads}
\label{fig:perf_1c_line}
\end{figure}

\begin{sloppypar}
To better understand Hermes's performance improvement, Fig.~\ref{fig:perf_1c_line} shows the performance line graph of Hermes, Pythia, and Hermes \rbd{combined with} Pythia for every single-core workload trace. The traces are sorted in ascending order of performance gains by Hermes \rbd{combined with Pythia} over the no-prefetching system. We make four key observations from Fig.~\ref{fig:perf_1c_line}.
First, \rbd{Hermes combined with Pythia} outperforms the no-prefetching system in all but three single-core workload traces. The \texttt{compute\_int\_539} and \texttt{605.mcf\_s-782B} \rbd{traces} \rbd{experience} the highest and the lowest \rbd{speedup} \rbd{($2.3\times$ and $0.8\times$, respectively)}.
Second, unlike Pythia, Hermes \emph{always} improves performance over the no-prefetching system in \emph{every} workload trace. 
Third, Hermes outperforms Pythia by $7.9\%$ on average in $51$ traces (e.g., \texttt{streamcluster-6B}, \texttt{Ligra\_PageRank-79B}). In the remaining $59$ traces, Pythia outperforms Hermes by $26\%$ on average.
Fourth, \rbd{Hermes combined with Pythia} consistently outperforms \emph{both} \rbd{Hermes and Pythia alone} in almost every workload trace.
\end{sloppypar}
Based on our performance results, we conclude that, Hermes provides significant and consistent performance \rbd{improvements} over a wide range of workloads both \rbd{by itself and when combined with} the state-of-the-art prefetcher \rbd{Pythia}.

\subsubsection{\rbd{Effect of the Off-chip Load Prediction Mechanism}} \label{sec:dmp_with_hmp_perf}

\begin{figure}[!h]
\centering
\includegraphics[width=3.3in]{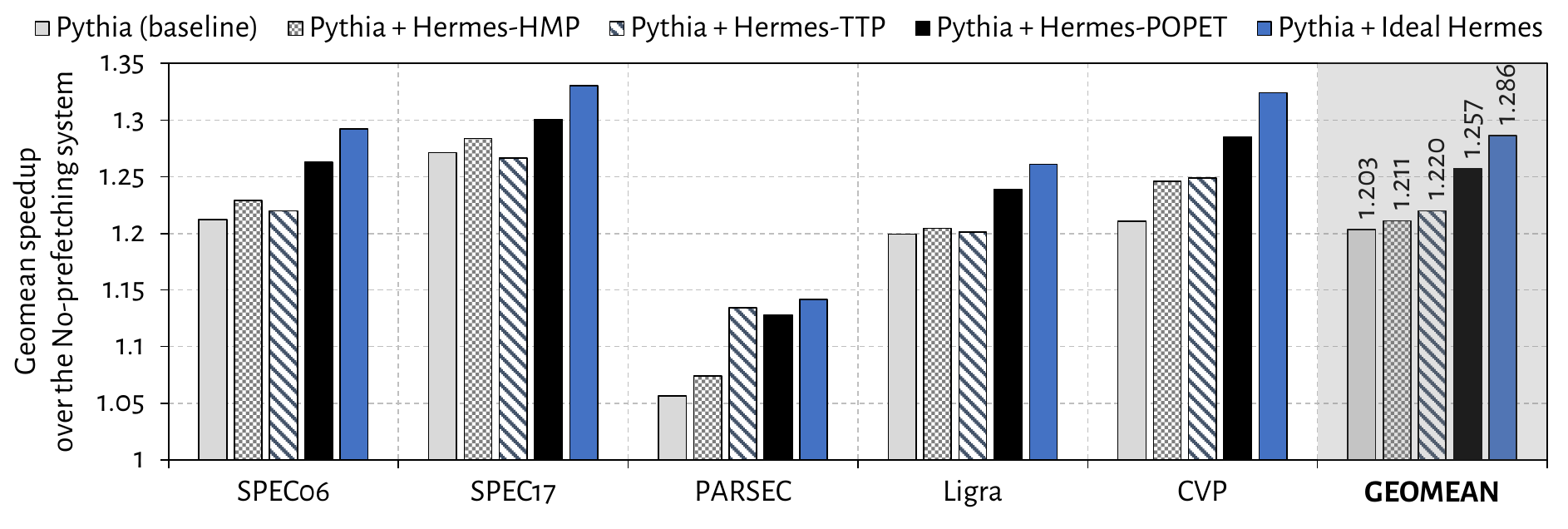}
\vspace{0.5em}
\caption{\rbd{Speedup of Hermes with three off-chip load predictors (HMP, \rbg{TTP}, and \pred) and the Ideal Hermes.}}
\label{fig:perf_1c_with_hmp}
\end{figure}

Fig.~\ref{fig:perf_1c_with_hmp} shows the performance of Hermes with \pred, Hermes-HMP, \rbg{Hermes-TTP}, and the Ideal Hermes (see~\cref{sec:headroom_study}) combined with Pythia normalized to the no-prefetching system in single-core workloads.
\rbe{We make two key observations.}
\rbe{First,} \rbd{Hermes with \pred} outperforms both Hermes-HMP and \rbg{Hermes-TTP}. 
\rbg{On average, \rbh{Hermes-HMP, Hermes-TTP}, and \rbd{Hermes with \pred} \rbd{combined with} Pythia provide \rbh{$0.8\%$, $1.7\%$}, and $5.4\%$ performance \rbh{improvement} \rbd{over} Pythia, respectively.}
\rbe{Second, Hermes-\pred provides nearly $90\%$ of the performance improvement provided by the Ideal Hermes that employs an ideal off-chip load predictor with $100\%$ accuracy and coverage.}
\rbd{We conclude that Hermes provides performance gains due to \rbg{both} \rbe{the} high \rbe{off-chip load prediction} accuracy and coverage of \pred. \rbe{Thus,} designing a good off-chip predictor is critical for Hermes to improve performance.}

\subsubsection{\rbd{Effect on Stall Cycles}}

\begin{figure}[!h]
\centering
\includegraphics[width=3.3in]{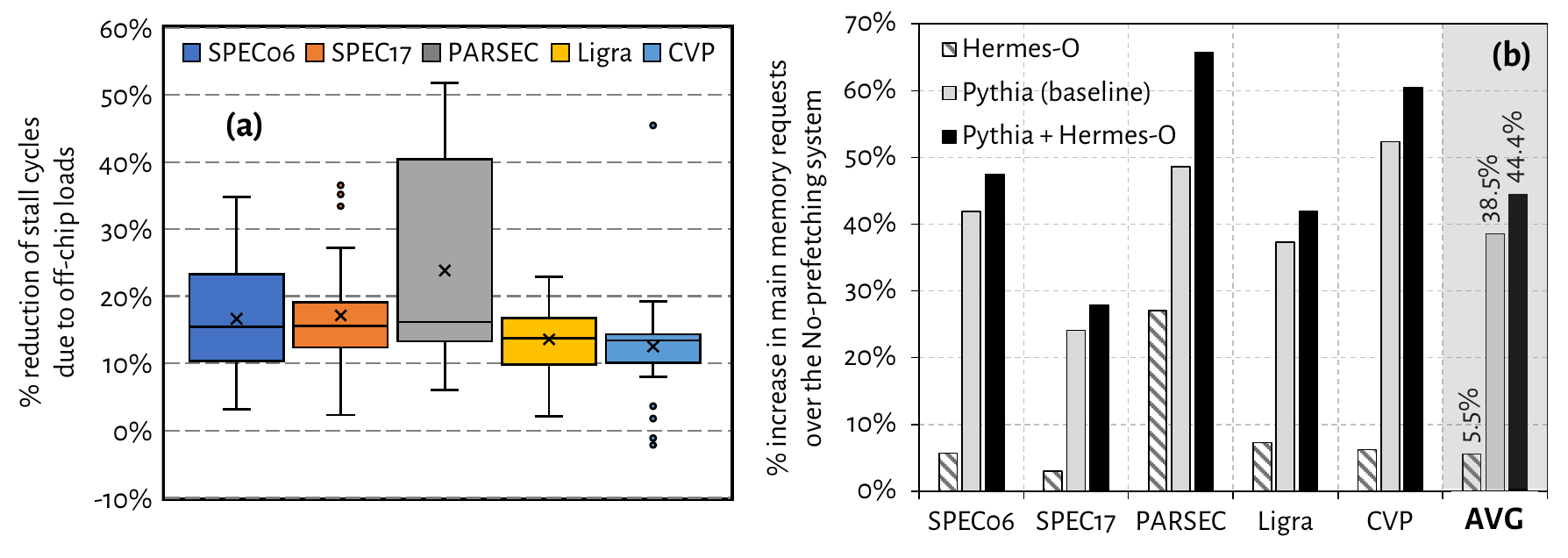}
\vspace{0.5em}
\caption{(a) Reduction in stall cycles caused by off-chip loads. (b) Overhead in the main memory requests.}
\label{fig:system_impact}
\end{figure}

Fig.~\ref{fig:system_impact}(a) \rbd{plots the distribution of} the percentage reduction \rbd{in} stall cycles due to off-chip load requests in a system with Hermes over the baseline system \rbe{in single-core workloads} as a \rbe{box-and-whiskers} plot.\footnote{\rbd{Each box is lower-bounded by the first quartile (i.e., the middle value between the lowest value and the median value of the data points) and upper-bounded by the third quartile (i.e., the middle value between the median and the highest value of the data points). The inter-quartile range ($IQR$) is the distance between the first and the third quartile (i.e., the length of the box). Whiskers extend an additional $1.5\times IQR$ on the either side of the box. Any outlier values that falls outside the range of whiskers are marked by dots.}
The cross marked value within each box represents the mean.
}
The key observation is that Hermes reduces the stall cycles caused by off-chip loads by $16.2\%$ on average (up to $51.8\%$) across all workloads. \texttt{PARSEC} workloads experience the highest \rbd{average} stall cycle reduction of $23.8\%$. $90$ out of $110$ \rbd{workloads experience} at least $10\%$ stall cycle reduction. 
We conclude that Hermes considerably reduces the stall cycles due to off-chip load requests, which \rbd{leads to} performance \rbd{improvement}.

\subsubsection{Overhead in Main Memory Requests} \label{sec:bw_1c}

Fig.~\ref{fig:system_impact}(b) shows the percentage increase in main memory requests in Hermes, Pythia, and \rbd{Hermes combined with Pythia} over the no-prefetching system in all single-core workloads. We make two key observations. First, Hermes increases main memory requests by only $5.5\%$ \rbd{(on average)} over the no-prefetching system, whereas Pythia by $38.5\%$. This means that, every $1\%$ performance gain (see Fig.~\ref{fig:perf_1c}) comes at a cost of only $0.5\%$ increase in main memory requests in Hermes, whereas nearly $2\%$ increase in main memory requests in Pythia.
\rbd{We attribute this result} to the highly-accurate predictions made by \pred, as compared to less-accurate prefetch decisions made by Pythia. 
Second, Hermes \rbd{combined with} Pythia further increases main memory requests by only $5.9\%$ over Pythia. This means that, every $1\%$ performance benefit by Hermes on top of Pythia comes at a cost of only $1\%$ overhead in main memory requests.
We conclude that, Hermes, due to its underlying high-accuracy prediction mechanism, adds considerably lower overhead in main memory requests while providing significant performance improvement both \rbd{by itself and when combined with Pythia}.

\subsection{Eight-core Performance Analysis} \label{sec:perf_mc}

\begin{figure}[!h]
\centering
\includegraphics[width=3.3in]{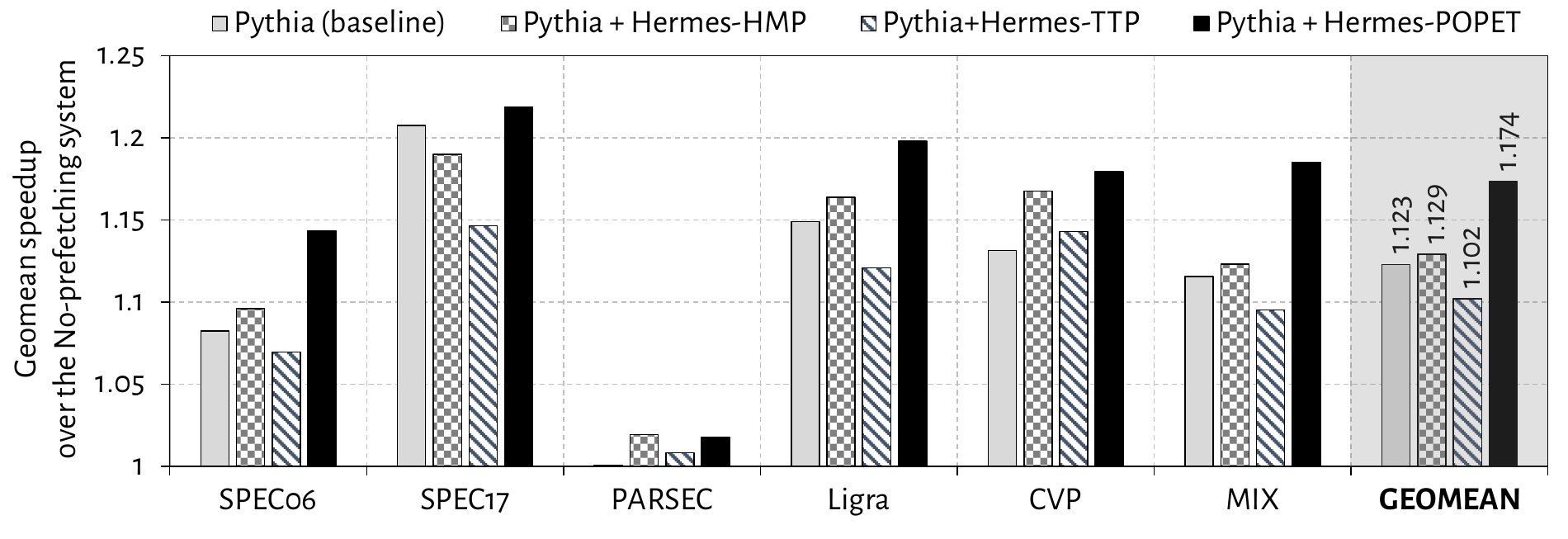}
\vspace{0.5em}
\caption{\rbd{Speedup in} eight-core workloads}
\label{fig:perf_8c}
\end{figure}

\rbd{Fig.~\ref{fig:perf_8c} shows \rbe{the} performance of Pythia, Hermes-HMP, \rbh{Hermes-TTP}, and Hermes-\pred combined with Pythia normalized to the no-prefetching system in all eight-core workloads.}
The key takeaway is that due to the highly-accurate predictions by \pred, Hermes-\pred combined with Pythia consistently outperforms Pythia in \emph{every} workload category. On average, Hermes-HMP, \rbg{Hermes-TTP}, and Hermes-\pred combined with Pythia \rbe{provide} $0.6\%$, \rbg{$-2.1\%$}, and $5.1\%$ \rbe{higher} performance on top of Pythia, respectively.
\rbh{Due to its inaccurate predictions, TTP generates many unnecessary main memory requests, which reduce the performance of Hermes-TTP combined with Pythia as compared to Pythia alone in the bandwidth-constrained eight-core configuration.}
\rbe{We conclude that Hermes provides significant and consistent performance improvement in the bandwidth-constrained eight-core system due to its highly-accurate off-chip load prediction.}

\begin{figure*}[!h]
\centering
\includegraphics[width=7.1in]{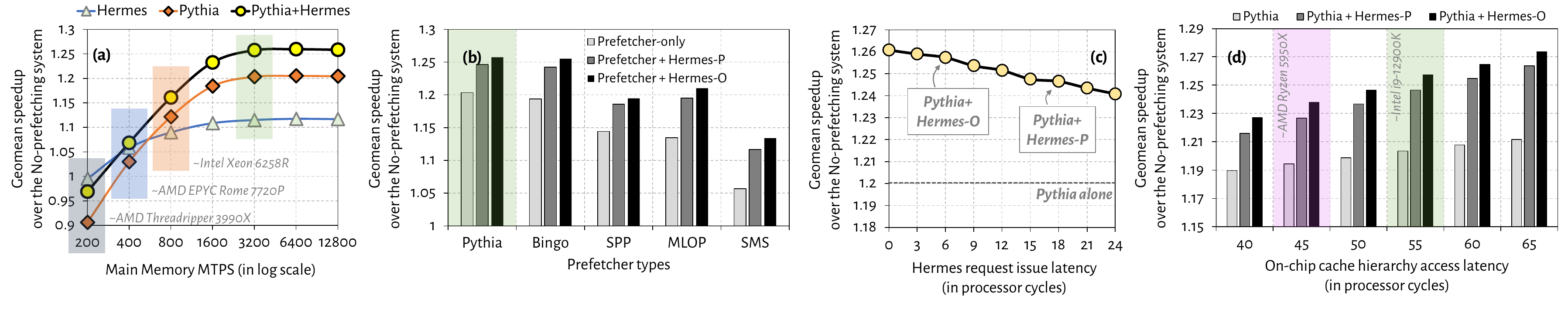}
\caption{Performance sensitivity to (a) main memory bandwidth, (b) different prefetching techniques, (c) \rbe{Hermes request issue latency}, and (d) \rbe{on-chip cache hierarchy access latency}. The baseline system configuration is highlighted in green. Other highlighted configurations closely match with various commercial processors~\cite{intel_xeon_gold,zen_epyc,zen_threadripper}.}
\label{fig:sensitivity_master}
\end{figure*}

\subsection{Performance Sensitivity Analysis} \label{sec:1c_sensitivity}

\subsubsection{\rbd{Effect of} Main Memory Bandwidth} \label{sec:dram_bw}

\rbd{Fig.~\ref{fig:sensitivity_master}(a) shows the speedup of Hermes, Pythia, and Hermes combined with Pythia over the no-prefetching system in single-core workloads by scaling the main memory bandwidth.}
We make two key observations. 
First, Hermes \rbd{combined with} Pythia \emph{consistently} outperforms Pythia in \emph{every} main memory bandwidth configuration from $\frac{1}{16}\times$ to $4\times$ of the baseline system. 
\rbd{Hermes combined with Pythia} outperforms Pythia \rbe{alone} by $6.2\%$ and $5.5\%$ in \rbe{the} main memory bandwidth configuration \rbd{with} $200$ and $12800$ million transfers per second (MTPS), \rbe{respectively}.
Second, Hermes \rbe{by itself} \emph{outperforms} Pythia in highly-bandwidth-constrained configurations. This is \rbd{due} to the highly-accurate off-chip load predictions made by \pred, which incurs less main memory \rbd{bandwidth} overhead than the aggressive, less-accurate prefetching decisions made by Pythia.  
Hermes outperforms Pythia by $2.8\%$ and $8.9\%$ in $400$ and $200$ MTPS configurations, respectively.

\subsubsection{\rbd{Effect of the Baseline} Prefetcher} \label{sec:prefetchers}

We evaluate Hermes \rbd{combined with} four recently-proposed data prefetchers: Bingo~\cite{bingo}, SPP~\cite{spp} (with perceptron filter~\cite{ppf}), MLOP~\cite{mlop}, and SMS~\cite{sms}. For each experiment, we replace the baseline LLC prefetcher Pythia with a new prefetcher and measure the performance improvement of the prefetcher \rbd{by itself} and Hermes \rbd{combined with} the prefetcher. 
Fig.~\ref{fig:sensitivity_master}(b) \rbe{shows the performance of the baseline prefetcher, and Hermes-P/O combined with the baseline prefetcher, normalized to the no-prefetching system in single-core workloads}. The key takeaway is that Hermes \rbd{combined with} \emph{any} \rbd{baseline} prefetcher consistently outperforms \rbd{the baseline prefetcher by itself} for \rbd{all four evaluated} prefetching techniques. Hermes+prefetcher outperforms the prefetcher \rbd{alone} by $6.2\%$, $5.1\%$, $7.6\%$, and $7.7\%$, \rbd{for} Bingo, SPP, MLOP, and SMS as the \rbd{baseline} prefetcher.

\subsubsection{\rbd{Effect of \rbe{the} Hermes Request} Issue Latency} \label{sec:load_issue_latency}

To analyze the performance benefit of Hermes over a wide range of processor \rbd{designs} with \rbd{simple or} complex on-chip datapath, we perform a performance sensitivity study by varying the Hermes request issue latency. 
\rbd{Fig.\ref{fig:sensitivity_master}(c) shows the \rbd{performance} of Hermes \rbe{combined with} Pythia \rbe{normalized to the no-prefetching system} in single-core workloads \rbe{as} Hermes \rbe{request} issue latency \rbe{varies from $0$ cycles to $24$ cycles}.}
\rbd{The dashed-line represents the performance of Pythia \rbe{alone}}.
We make two key observations. 
First, the \rbd{speedup} \rbe{of} Hermes \rbd{combined with} Pythia decreases \rbe{as} the \rbd{Hermes request} issue latency \rbe{increases}.
Second, even with a pessimistic \rbd{Hermes request} issue latency \rbe{of $24$ cycles}, Hermes \rbd{combined with} Pythia outperforms  Pythia.
Pythia+Hermes outperforms Pythia by $5.7\%$ and $3.6\%$ with $0$-cycle and $24$-cycle \rbd{Hermes request} issue latency, respectively.

\subsubsection{\rbd{Effect of} \rbe{the} On-chip Cache Hierarchy Access Latency} \label{sec:llc_latency}

We evaluate Hermes \rbe{by} varying the on-chip cache hierarchy access latency. 
For each experiment, we keep the L1 and L2 cache access latencies unchanged and vary the LLC access latency from $25$-cycles to $50$-cycles, to mimic the access latencies of a wide range of sliced LLC designs with \rbd{simple or complex} on-chip networks. 
\rbd{Fig.~\ref{fig:sensitivity_master}(d) shows the \rbe{performance} of Pythia, and Hermes (O and P) combined with Pythia, normalized to the no-prefetching system in single-core workloads.}
\rbd{We make two key observations.} 
\rbd{First,} Hermes \rbd{combined with} Pythia consistently outperforms Pythia \rbe{for} \emph{every} on-chip cache hierarchy latency. Hermes-O combined with Pythia outperforms Pythia alone by $3.6\%$ and $6.2\%$ in system with $40$-cycle and $65$-cycle on-chip cache hierarchy access latency, respectively.
\rbd{Second, the performance improvement by Hermes combined with Pythia increases \rbe{as the on-chip cache hierarchy access latency increases}. Thus, we \rbe{posit} that Hermes can provide even higher performance benefit in future processors with longer \rbe{on-chip} cache access \rbe{latencies}.}

\subsubsection{\rbd{Effect of} \rbe{the} Activation Threshold} \label{sec:act_thresh_sensi}

\begin{figure}[!h]
\centering
\includegraphics[width=3.3in]{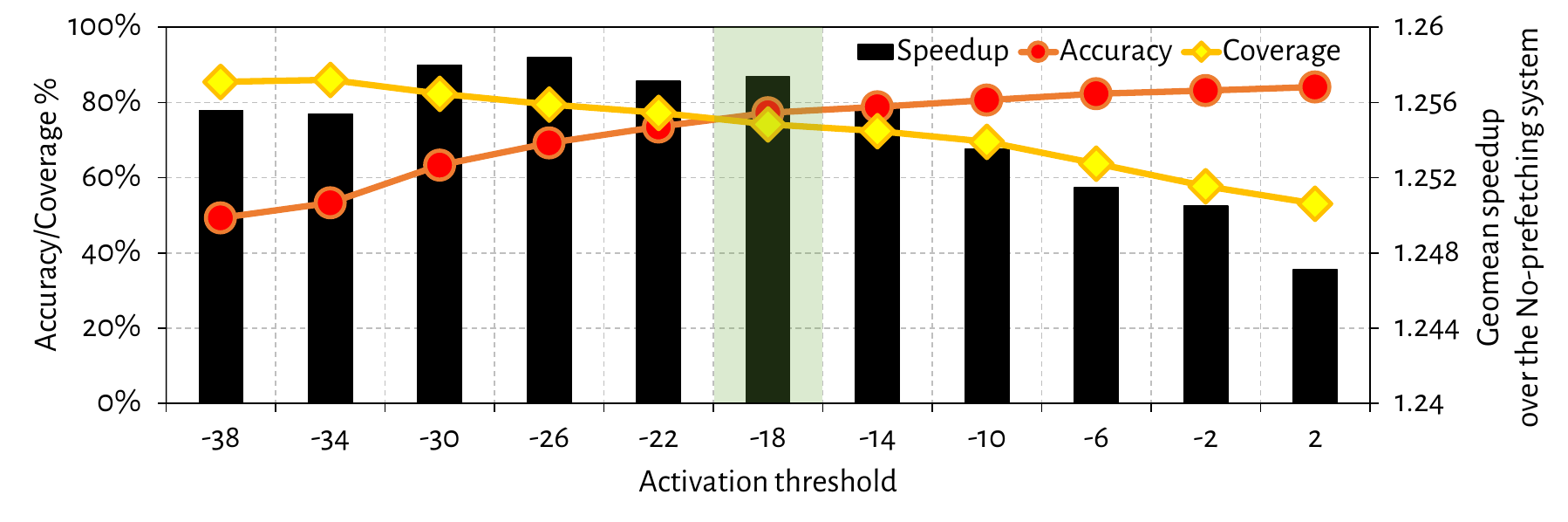}
\caption{\rbe{Effect of the activation threshold on \pred's accuracy and coverage (on the left y-axis) and Hermes's speedup (on the right y-axis) for all single-core workloads.}}
\label{fig:act_thresh_sensi}
\end{figure}

\rbe{We evaluate} the impact of the activation threshold ($\tau_{act}$) on \rbe{Hermes's performance} by varying $\tau_{act}$. Fig.~\mbox{\ref{fig:act_thresh_sensi}} shows \pred's accuracy and coverage (as line graphs on the \rbd{left} y-axis) and the \rbd{performance} of Hermes combined with Pythia over the no-prefetching system (as a bar graph on the \rbd{right} y-axis) across all single-core workloads \rbe{as $\tau_{act}$ varies from $-38$ to $2$}. 
The key takeaway from Fig.~\mbox{\ref{fig:act_thresh_sensi}} is that \pred's accuracy (coverage) increases (decreases) \rbe{as $\tau_{act}$ increases}. However, Hermes's performance \rbe{gain} peaks near $\tau_{act} = -26$, which favors higher coverage by trading off accuracy. As \pred's accuracy directly impacts Hermes's main memory request overhead (\rbe{and} hence its performance in bandwidth-constrained configurations), we set $\tau_{act} = -18$ in \pred. \rbd{Doing so} simultaneously optimizes both \pred's accuracy \emph{and} coverage.

\subsection{Power Overhead} \label{sec:power_overhead}

\begin{figure}[!h]
\centering
\includegraphics[width=3.3in]{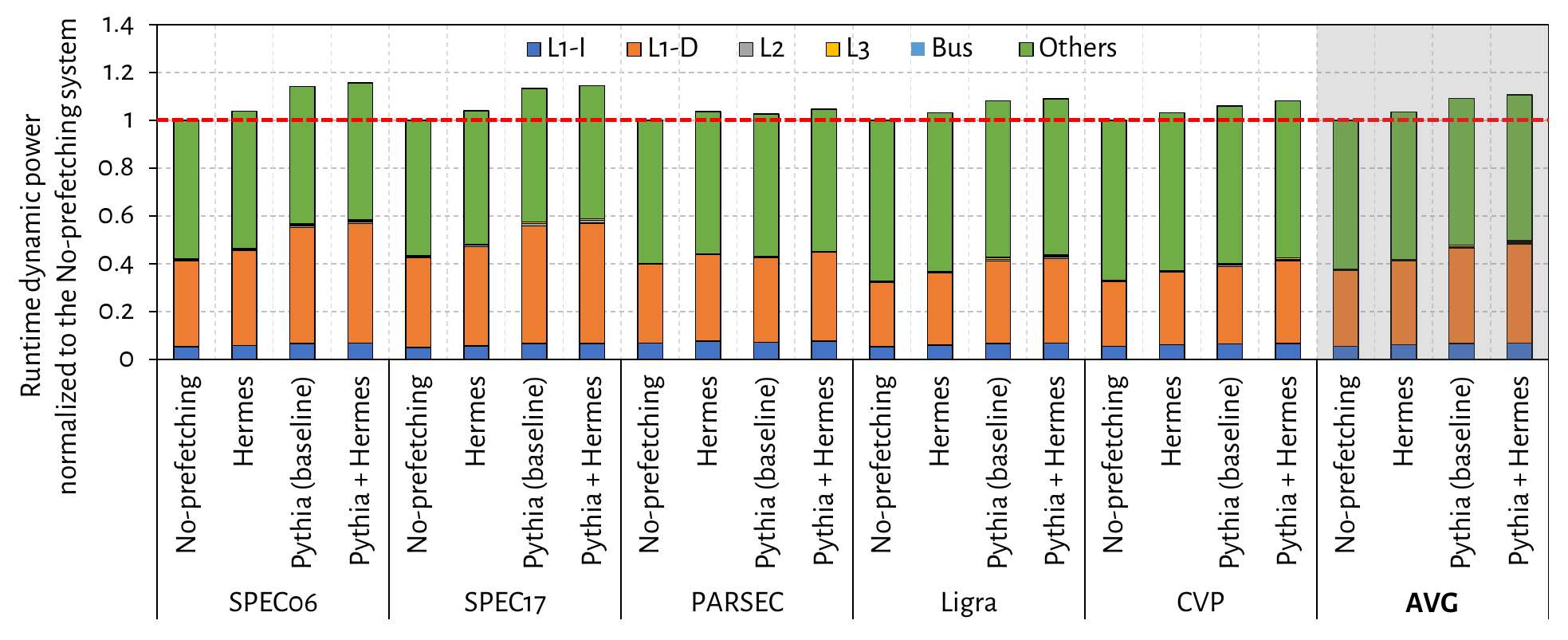}
\vspace{0.5em}
\caption{\rbe{Processor power consumption of Hermes, Pythia, and Hermes combined with Pythia.}}
\label{fig:power_overhead}
\end{figure}

\begin{sloppypar}
To accurately estimate Hermes's dynamic power consumption, we model our single-core configuration in McPAT~\mbox{\cite{mcpat}} and \rbd{compute \rbe{processor} power consumption} using statistics from performance simulations. Fig.~\mbox{\ref{fig:power_overhead}} shows the runtime dynamic power consumed by Hermes, Pythia, and \rbd{Hermes combined with Pythia}, normalized to the no-prefetching system \rbd{for all single-core workloads}. We make two key observations. First,  Hermes increases processor power consumption by only $3.6\%$ on average over the no-prefetching system, whereas Pythia increases power \rbe{consumption} by $8.7\%$. 
Second, Hermes \rbd{combined with} Pythia incurs only $1.5\%$ additional power overhead on top of Pythia. \rbe{We conclude} that Hermes incurs \rbd{only a} modest power overhead \rbd{and is more efficient than Pythia alone}.
\end{sloppypar}

%% file: 07related_works.tex
\vspace{-0.3em}
\section{Other Related Work} \label{sec:other_related_works}

To our knowledge, this is the first work that proposes \rbd{(1)} a lightweight perceptron-based off-chip \rbd{load} predictor \rbe{(\pred)} that makes accurate predictions without having large metadata overhead, \rbd{and (2) a mechanism \rbe{(Hermes)} that takes advantage of such a predictor to improve processor performance 
\rbe{by eliminating on-chip cache access latency from the critical path of a correctly-predicted off-chip load.}
}
We have already quantitatively and \rbd{qualitatively} compared (1) \rbd{\pred with} HMP~\cite{yoaz1999speculation} \rbd{and (2) Hermes with multiple state-of-the-art hardware data prefetchers: \rbi{Pythia~\cite{pythia}, Bingo~\cite{bingo}, SPP~\cite{spp,ppf}, MLOP~\cite{mlop}, and SMS~\cite{sms}}}. In this section, we \rbe{discuss} other related works.

\textbf{Hit/miss prediction.} 
Peir et al.~\cite{peir_bloom_filter} \rbe{propose} a \rbd{Bloom}-filter-based hit/miss predictor to optimize instruction scheduling \rbd{in} an out-of-order processor.
Memik et al.~\cite{mnm} \rbe{propose} five different heuristics to keep track of cache contents using simple tables. 
Prior works~\cite{loh_miss_map,alloy_cache} also explore hit/miss prediction in DRAM caches. Loh and Hill~\cite{loh_miss_map} propose \textit{MissMap}, \rbd{which} uses bitvectors to track the residency of cachelines in a large DRAM cache. 
\rbe{Adapting} MissMap \rbe{to} off-chip load prediction \rbe{poses} two key challenges. First, MissMap can have high false-positive prediction rate \rbd{(as discussed in~\cite{lp})}. Second, the size of MissMap can grow \rbe{very} large, \rbd{leading to \rbe{large} latency and storage overheads}.
\pred, on the other hand, requires \rbd{only} $4$~KB of storage overhead, while producing highly-accurate off-chip \rbd{load} predictions.

\textbf{Cache bypassing.}
High-performance cache management policies (e.g., ~\cite{teresa_cache_bypassing,teresa_cache_bypassing2,tyson1995modified,tyson1997managing,adaptive_cache_hierarchy,jayesh_bypass,ship,jain2016back,jimenez2010dead,jimenez2017multiperspective,sdbp,rivers1996reducing,rivers1998utilizing,hu2002timekeeping,kharbutli2008counter,liu2008cache,piquet2007exploiting,chi1989improving,dybdahl2006enhancing,wong2000modified}) dynamically bypass cache levels during a cache fill operation if the incoming cacheline is \rbe{not expected} 
\rbe{to be used by the program in the future. This expectation (i.e., reuse prediction) can be provided by either a hardware-based reuse prediction mechanism~\cite{rivers1996reducing,rivers1998utilizing,jain2016back,jimenez2010dead,jimenez2017multiperspective,sdbp,kharbutli2008counter,liu2008cache,piquet2007exploiting,dybdahl2006enhancing,wong2000modified} or \rbf{a software hint~\cite{chi1989improving,tyson1995modified,tyson1997managing,nt_load_intel_manual,movnti}, e.g., non-temporal load instructions employed by modern processors~\cite{movnti,nt_load_intel_manual}}.}
The goal of these \rbe{cache bypassing} techniques is to better utilize the cache space \rbe{by avoiding the insertion of useless cachelines into the cache}. 
\rbe{Hermes's goal is different and orthogonal to these techniques: to reduce the latency of a long-latency off-chip load by eliminating the on-chip cache hierarchy access latency from its critical path.}
\rbe{As such,} Hermes can be \rbe{combined} with \rbe{any} cache bypassing technique.

\textbf{Data prefetching.} 
Prior prefetching techniques can be broadly categorized into three classes:
(1) precomputation-based prefetchers that pre-execute \rbe{program code} to generate future loads~\cite{dundas,mutlu2003runahead,mutlu2003runahead2, mutlu2005techniques,mutlu2006efficient,hashemi2016continuous,mutlu2005address,hashemi2015filtered,vector_runahead,mutlu2005reusing,iacobovici2004effective},
(2) temporal prefetchers that predict future load addresses by memorizing long sequence of demanded cacheline addresses~\cite{markov,stems,somogyi_stems,wenisch2010making,domino,isb,misb,triage,wenisch2005temporal,chilimbi2002dynamic,chou2007low,ferdman2007last,hu2003tcp,bekerman1999correlated,cooksey2002stateless,karlsson2000prefetching}, and 
(3) spatial prefetchers that predict future load addresses by learning program access patterns over different memory regions~\cite{stride,streamer,baer2,jouppi_prefetch,ampm,fdp,footprint,sms,spp,vldp,sandbox,bop,dol,dspatch,bingo,mlop,ppf,ipcp,pythia}. 
As we show in \cref{sec:prefetchers}, Hermes can be \rbe{combined} with any baseline prefetcher to provide higher performance than just the prefetcher \rbd{alone}.

%% file: 08conclusion.tex
\section{Conclusion}

\rbe{We introduce Hermes, a technique that accelerates long-latency off-chip load requests by eliminating the on-chip cache hierarchy access latency from their critical path. To enable Hermes, we propose a perceptron-learning based off-chip load predictor (\pred) that accurately predicts which load requests might go off-chip.}
Our extensive evaluations using a wide range of workloads \rbd{and system configurations} show that Hermes provides significant performance benefits over a baseline system with \rbe{a} state-of-the-art prefetcher. 
\rbf{As on-chip cache hierarchy continues to grow in size and complexity in future processors, we believe and hope that Hermes's key observation and off-chip load prediction mechanism would inspire future works to explore a multitude of other memory system optimizations.}

%% file: 09ack.tex
\section*{Acknowledgments}
We thank Anant Nori and Sreenivas Subramoney for their valuable feedback on this work.
We thank the anonymous reviewers of MICRO 2022 for their encouraging feedback. We thank the SAFARI Research Group members for providing \rbd{a} stimulating intellectual environment.
We acknowledge the generous gifts \rbe{from} our industrial partners: Google, Huawei, Intel, Microsoft, and VMware.
This work is supported \rbe{in part} by \rbe{the Semiconductor Research Corporation} \rbd{and the ETH Future Computing Laboratory}.
The first author thanks his departed father, whom he lost in COVID-19 pandemic.

%% file: 10appendix.tex
\section{Artifact Appendix} \label{sec:artifact}

\subsection{Abstract}

We implement Hermes using \rbe{the} ChampSim simulator~\cite{champsim}. In this artifact, we provide the source code of Hermes and necessary instructions to reproduce its key performance results.\footnote{\rbe{This appendix focuses on reproducing four key results mentioned here. Nonetheless, the artifact contains files and necessary scripts to reproduce all results mentioned in the paper}.} We identify four key results to demonstrate Hermes's novelty: 
\begin{itemize}
    \item Comparison of accuracy and coverage of \pred against HMP and \rbg{TTP} (Fig.~\ref{fig:ocp_acc_cov}).
    \item Workload category-wise performance comparison in single-core workloads (Fig.~\ref{fig:perf_1c}).
    \item Workload category-wise performance comparison of Hermes with \pred with Hermes-HMP and \rbg{Hermes-TTP} in single-core workloads (Fig.~\ref{fig:perf_1c_with_hmp}).
    \item Performance sensitivity to different prefetching techniques (Fig.~\ref{fig:sensitivity_master}(b)).
\end{itemize}

The artifact can be executed in any machine with a general-purpose CPU and $36$ GB disk space. However, we strongly recommend running the artifact on a compute cluster with \texttt{slurm}~\cite{slurm} support for bulk experimentation.

\subsection{Artifact Check-list (Meta-information)}

\small
\begin{itemize}
  \item {\bf Compilation: } GCC 6.3.0 or above.
  \item {\bf Data set: } Download traces using the supplied script.
  \item {\bf Run-time environment: } Perl v5.24.1
  \item {\bf Metrics: } IPC, predictor's coverage, and accuracy.
  \item {\bf Experiments: } Generate experiments using supplied scripts.
  \item {\bf How much disk space required (approximately)?: } $36$ GB
  \item {\bf How much time is needed to prepare workflow (approximately)?: } $\sim2$ hours. Mostly depends on downloading bandwidth.
  \item {\bf How much time is needed to complete experiments (approximately)?: } $\sim12$ hours using a compute cluster with $640$ cores.
  \item {\bf Publicly available?: } Yes.
  \item {\bf Code licenses (if publicly available)?: } MIT
  \item {\bf Archived (provide DOI)?: } \url{https://doi.org/10.5281/zenodo.6909799}
\end{itemize}

\subsection{Description}

\subsubsection{How to Access}

The source code can be downloaded from either GitHub~\cite{hermes_github} or Zenodo~\cite{hermes_zenodo}.

\subsubsection{Hardware Dependencies}

Hermes can be run on any system with a general-purpose CPU and at least $36$ GB of free disk space.

\subsubsection{Software Dependencies}
\begin{itemize}
    \item \texttt{cmake >= 3.20.2}
    \item \texttt{gcc >= v6.3.0}
    \item \texttt{perl >= v5.24.1}
    \item \texttt{xz >= v5.2.5}
    \item \texttt{gzip >= v1.6}
    \item \texttt{megatools >= v1.11.0}
    \item \texttt{md5sum >= v8.26}
    \item \texttt{wget >= v1.18}
    \item \texttt{Microsoft Excel >= v16.51}
\end{itemize}

\subsubsection{Data Sets}

The ChampSim traces required to evaluate Hermes can be downloaded using the supplied script. Our implementation of Hermes is fully compatible with prior ChampSim traces that are used in previous cache replacement (CRC-2~\cite{crc2}), data prefetching (DPC-3~\cite{dpc3}) and value-prediction (CVP-2~\cite{cvp2}) championships.

\subsection{Installation}

\begin{enumerate}
    \item Clone Hermes from GitHub repository:
        \vspace{0.4em}
        \shellcmd{git clone https://github.com/CMU-SAFARI/ \\ Hermes.git}
        \vspace{0.4em}
        
    \item Please make sure to set environment variables as:
        \vspace{0.4em}
        \shellcmd{source setvars.sh}
        \vspace{0.4em}
        
    \item Clone Bloomfilter library inside Hermes home and build:
        \vspace{0.4em}
        \shellcmd{cd \$HERMES\_HOME}
        \shellcmd{git clone https://github.com/mavam/libbf.git libbf/}
        \shellcmd{cd libbf/}
        \shellcmd{mkdir build \&\& cd build/ \&\& cmake ../}
        \shellcmd{make clean \&\& make}
        \vspace{0.4em}
        
    \item Build Hermes for single-core Intel Goldencove configuration as:
        \vspace{0.4em}
        \shellcmd{cd \$HERMES\_HOME}
        \shellcmd{./build\_champsim.sh glc multi multi multi multi 1 1 0}
\end{enumerate}

\subsection{Preparing Traces} \label{sec:preparing_traces}
This section describes \rbe{the} steps to download and verify \rbe{the} necessary ChampSim traces. We recommend the reader to follow the README in the GitHub repository to get up-to-date information.

\begin{enumerate}
    \item Install the Megatools executable as:
        \vspace{0.4em}
        \shellcmd{cd \$HERMES\_HOME/scripts}
        \shellcmd{wget --no-check-certificate \\ https://megatools.megous.com/builds/builds/\\megatools-1.11.0.20220519-linux-x86\_64.tar.gz}
        \shellcmd{tar -xvf megatools-1.11.0.20220519-linux-x86\_64.tar.gz}
        \vspace{0.4em}
    
    \item Use \texttt{download\_traces.pl} script to download traces:
        \vspace{0.4em}
        \shellcmd{cd \$HERMES\_HOME/traces}
        \shellcmd{perl \$HERMES\_HOME/scripts/download\_traces.pl \\ --csv artifact\_traces.csv --dir ./}
        \vspace{0.4em}
        
    \item Once the script finishes downloading $110$ traces, please verify the checksum as follows. Make sure all traces pass the checksum test.
        \vspace{0.4em}
        \shellcmd{cd \$HERMES\_HOME/traces}
        \shellcmd{md5sum -c artifact\_traces.md5}
        \vspace{0.4em}
    \item If the traces are downloaded in other path, please update the full path in \texttt{MICRO22\_AE.tlist} file inside \texttt{\$HERMES\_HOME/experiments} directory appropriately.
    
\end{enumerate}

\subsection{Experimental Workflow}
This section describes \rbe{the} steps to generate and execute necessary experiments. We recommend the reader to follow \texttt{script/README.md} to know more about each script used in this section.

\subsubsection{Launching Experiments} \label{sec:artifact_launching_experiments}
The following instructions \rbe{enable launching} all experiments required to reproduce key results in a local machine. We \emph{\textbf{strongly}} recommend using a compute cluster with \texttt{slurm} support to efficiently launch experiments in bulk. To launch experiments using \texttt{slurm}, please provide \texttt{--local 0} (tested using \texttt{slurm v16.05.9}). 
\begin{enumerate}
    \setlength\itemsep{0.5em}
    \item Create the jobfile for the experiments as follows:
        \vspace{0.4em}
        \shellcmd{cd \$HERMES\_HOME/experiments}
        \shellcmd{perl \$HERMES\_HOME/scripts/create\_jobfile.pl --exe \$HERMES\_HOME/bin/glc-perceptron-no-\\multi-multi-multi-multi-1core-1ch --tlist \\ MICRO22\_AE.tlist --exp MICRO22\_AE.exp --local 1 > jobfile.sh}
        \vspace{0.4em}
        
    \item Please make sure the paths used in \texttt{tlist} and \texttt{exp} files are appropriately changed before creating the jobfile.
    
    \item If you are creating jobs for \texttt{slurm}, please set the \texttt{slurm} partition name appropriately (set as \texttt{slurm\_part} by default). You might also want to set some key parameters for \texttt{slurm} configuration as follows: (a) max memory per node: $2$ GB, (b) timeout: $12$ hours.
    
    \item Finally, launch the experiments as follows:
        \shellcmd{cd \$HERMES\_HOME/outputs/}
        \shellcmd{source ../jobfile.sh}
        \vspace{0.4em}
\end{enumerate}

\subsubsection{Rolling-up Statistics} \label{sec:rollup}
The \texttt{rollup.pl} script parses a set of output files of a given experiment and dumps all statistics in a comma-separated-value (CSV) format. To automate the roll-up process, we use the following set of instructions which \rbe{enable the creation of} four CSV files in experiments directory. These CSV files are later used for comparison in Appendix~\ref{sec:artifact_evaluation}.
\vspace{0.4em}
\shellcmd{cd \$HERMES\_HOME/experiments}
\shellcmd{bash automate\_rollup.sh}

\subsection{Evaluation}\label{sec:artifact_evaluation}
We use three metrics for comparing Hermes with previous works: (1) performance, (2) accuracy and (3) coverage of the off-chip predictor. The performance gain of a simulation configuration is measured with respect to the no-prefetching system using Eq.~\ref{eq:pref}.
\begin{equation}
    Perf_X = \frac{IPC_X}{IPC_{nopref}} \label{eq:pref}
\end{equation}
\vspace{1pt}

We measure the accuracy and coverage of each evaluated off-chip prediction mechanism using Eq.~\ref{eq:acc} and~\ref{eq:cov}.

\begin{equation} \label{eq:acc}
    Accuracy_X = \frac{TP}{TP+FP}
\end{equation}

\begin{equation} \label{eq:cov}
    Coverage_X = \frac{TP}{TP+FN}
\end{equation}
\vspace{1pt}

Here, \texttt{TP} represents the number of predicted off-chip requests that \emph{actually go} off-chip, \texttt{FP} represents the number of predicted off-chip requests that \emph{do not go} off-chip, and \texttt{FN} represents the number of requests that are \emph{not predicted} by the off-chip predictor \emph{but go} off-chip.

To easily calculate the metrics, we provide a Microsoft Excel template to post-process the rolled-up CSV files generated in Appendix~\ref{sec:rollup}. The template has five sheets. The first sheet (named \emph{metadata}) contains the metadata to identify the category of each workload trace and experiment. Each of the remaining sheets shares the same name of the rolled-up CSV file.
Each of these sheets is already populated with our collected results, necessary formulas, pivot tables, and charts to reproduce the results presented in the paper.
\emph{Only the blue-highlighted columns} in each sheet need to be populated by \emph{your own} CSV files.
Please follow \rbe{these} instructions to reproduce the results from \emph{your own} CSV statistics files:
\begin{enumerate}
    \item Copy and paste each CSV file into its corresponding sheet's top left corner (i.e., cell \texttt{A1}).
    \item If you have copied the CSV file using an CSV application that already automatically separates the columns, then go to Step $4$.
    \item Immediately after pasting, convert the comma-separated rows into columns by going to Data $\rightarrow$ Text-to-Columns $\rightarrow$ Select comma as a delimiter. This \rbe{replaces} the already existing data in the sheet with the newly collected data.
    \item \emph{Refresh each pivot table in each sheet} by clicking on them and then clicking Pivot-Table-Analyse $\rightarrow$ Refresh.
\end{enumerate}
The reader can also use any other data processor (e.g., Python pandas) to reproduce the same result. 

\subsection{Expected Results}
\begin{itemize}
    \item \textbf{Accuracy and coverage comparison.} \pred should show $77.1\%$ accuracy, whereas HMP and \rbg{TTP} should show $46.8\%$ and \rbg{$16.6\%$} accuracy, respectively. \pred should show $74.3\%$ coverage, whereas HMP and \rbg{TTP} should show $22.3\%$ and \rbg{$94.8\%$} coverage, respectively.
    \item \textbf{Performance comparison with Pythia.} Hermes-P, Hermes-O, Pythia, Pythia+Hermes-P, and Pythia+Hermes-O should provide $8.9\%$, $11.5\%$, $20.5\%$, $24.7\%$, and $25.6\%$ geomean performance improvement, respectively, across all single-core workloads.
    \item 
    \begin{sloppypar}
        \textbf{Performance comparison with prior predictors.} Pythia, Pythia+Hermes-HMP, \rbg{Pythia+Hermes-TTP}, and Pythia+Hermes-\pred should provide $20.5\%$, $21.1\%$, \rbg{$22.09\%$}, and $25.6\%$ geomean performance improvement, respectively, across all single-core workload traces. 
    \end{sloppypar}
    \item 
    \begin{sloppypar}
        \textbf{Performance comparison with varying prefetchers.} Both Hermes-P and Hermes-O improves performance than the prefetcher \rbe{by itself} for \emph{every} \rbe{baseline} prefetcher type: Pythia, Bingo, SPP, MLOP, and SMS.
    \end{sloppypar}
\end{itemize}

\subsection{Experiment Customization}
\begin{itemize}
    \item The configuration of \rbe{each} prefetcher can be customized by changing the \texttt{ini} files inside the \texttt{config} directory.
    \item The \texttt{exp} files can be customized to run new experiments with different prefetcher combinations. More experiment files can be found inside \texttt{experiments/extra} directory. One can use the same instructions mentioned in Appendix~\ref{sec:artifact_launching_experiments} to launch experiments.
\end{itemize}

\subsection{Quick Troubleshooting}
Please check the FAQ section in GitHub (\url{https://github.com/CMU-SAFARI/Hermes#frequently-asked-questions}) for quick troubleshooting tips.

\subsection{Methodology}

Submission, reviewing and badging methodology:

\begin{itemize}
  \item \url{https://www.acm.org/publications/policies/artifact-review-badging}
  \item \url{http://cTuning.org/ae/submission-20201122.html}
  \item \url{http://cTuning.org/ae/reviewing-20201122.html}
\end{itemize}

%% file: 11appendix2.tex
\section{Extended Results} \label{sec:extended_results}

\subsection{Performance Sensitivity to Reorder Buffer Size} \label{sec:perf_rob}

\begin{figure}[!h]
\centering
\includegraphics[width=3.3in]{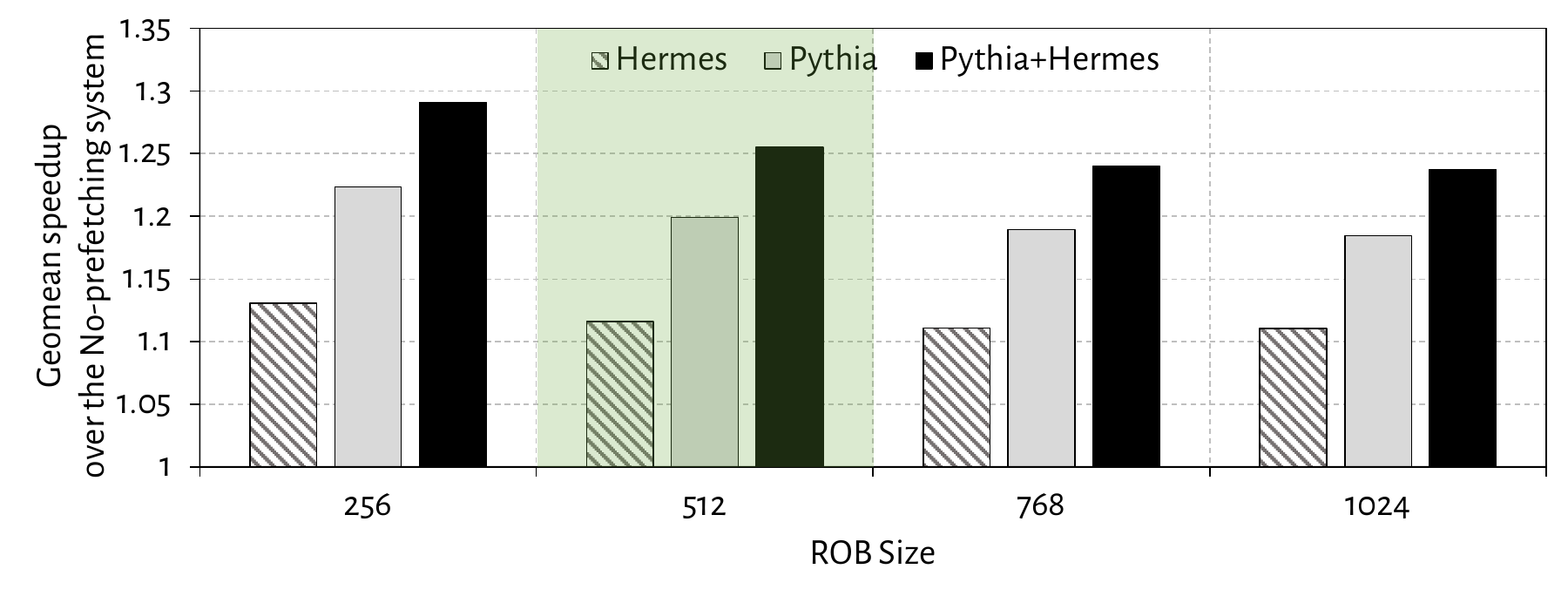}
\vspace{0.2em}
\caption{Performance sensitivity to reorder buffer size. The baseline configuration is marked in green.}
\label{fig:perf_rob_size}
\end{figure}

Fig.~\ref{fig:perf_rob_size} shows the performance of Hermes, Pythia, and Hermes combined with Pythia normalized to the no-prefetching system in single-core workloads as the size of reorder buffer (ROB) varies from $256$ entries to $1024$ entries. The key takeaway is that Hermes combined with Pythia 
outperforms Pythia alone in every ROB size configuration. Pythia+Hermes outperforms Pythia by $6.7\%$ and $5.3\%$ in a system with $256$-entry and $1024$-entry ROB.

\subsection{Performance Sensitivity to LLC Size} \label{sec:perf_llc}

\begin{figure}[!h]
\centering
\includegraphics[width=3.3in]{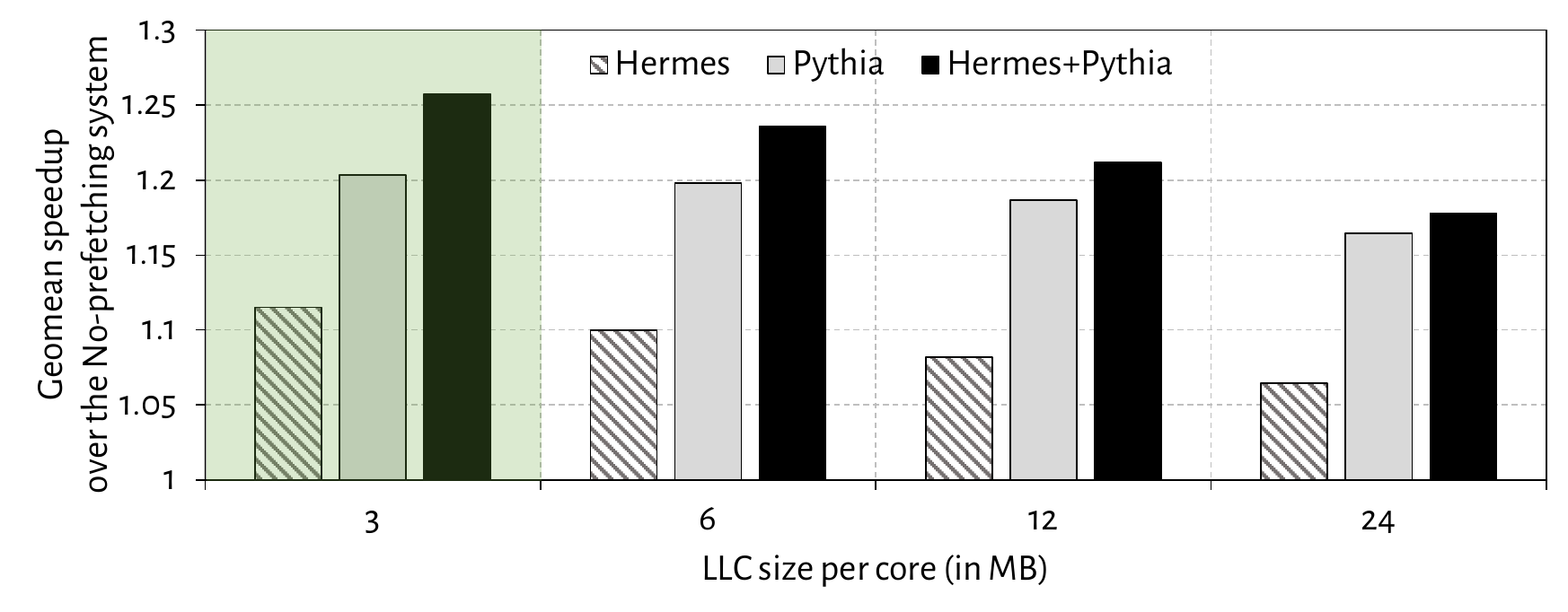}
\vspace{0.2em}
\caption{Performance sensitivity to LLC size. The baseline configuration is marked in green.}
\label{fig:perf_llc_size}
\end{figure}

Fig.~\ref{fig:perf_llc_size} shows the performance of Hermes, Pythia, and Hermes combined with Pythia normalized to the no-prefetching system in single-core workloads as the per-core last-level cache (LLC) size varies from $3$~MB to $24$~MB. The key takeaway is that Hermes combined with Pythia outperforms Pythia alone in every LLC size configuration. 
Even in a system with a $12$~MB and $24$~MB LLC per core, Pythia+Hermes provides $2.5\%$ and $1.3\%$ performance benefit over Pythia alone.

\subsection{Variation of Prediction Accuracy and Coverage with Different Prefetchers} \label{sec:acc_cov_variation_pref}

\begin{figure}[!h]
\centering
\includegraphics[width=3.3in]{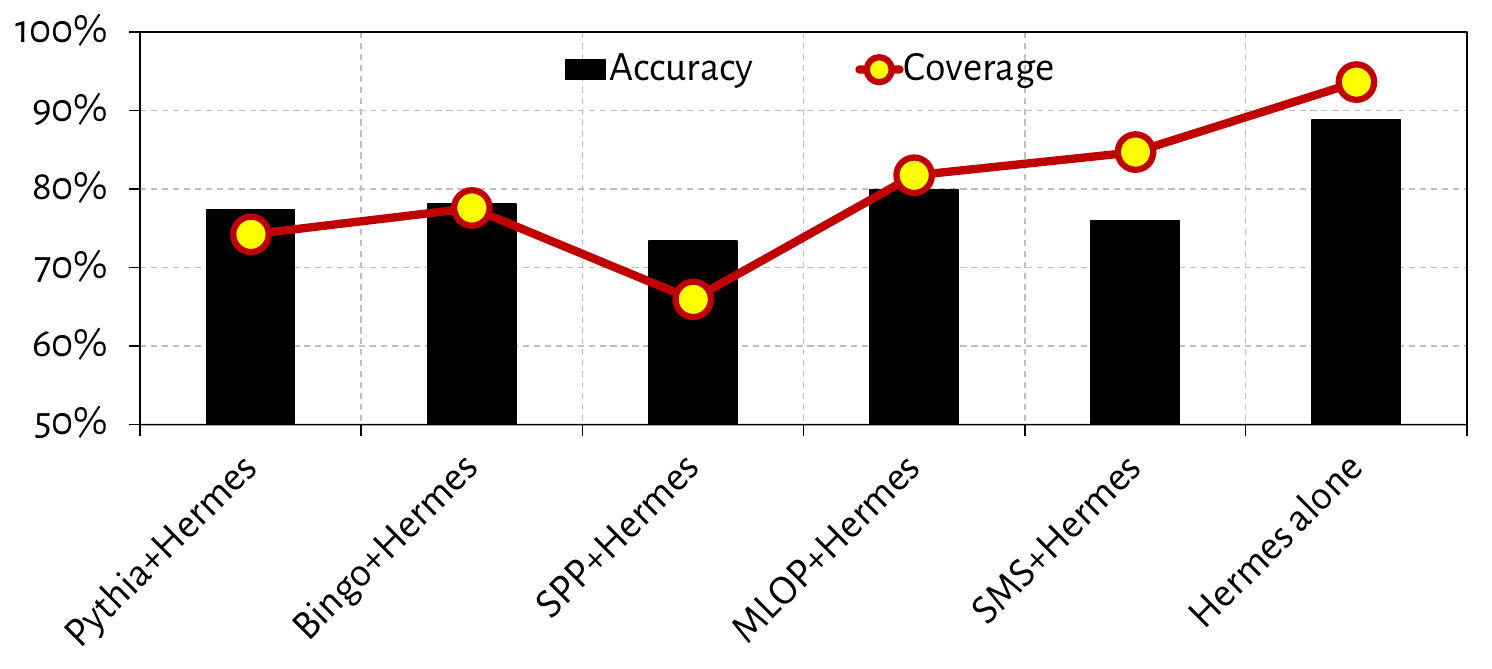}
\vspace{0.2em}
\caption{Variation of off-chip load prediction accuracy and coverage with different data prefetchers.
}
\label{fig:acc_cov_var}
\end{figure}

Fig.~\ref{fig:acc_cov_var} shows the off-chip load prediction accuracy and coverage when Hermes is combined with different baseline data prefetchers. We make two key observations. First, \pred's accuracy and coverage vary widely based on the baseline data prefetcher. When combined with Pythia, Bingo, SPP, MLOP, and SMS, \pred provides accuracy of $77.3\%$, $78.1\%$, $73.4\%$, $79.9\%$, and $76.0\%$, while providing coverage of $74.2\%$, $77.6\%$, $65.9\%$, $81.7\%$, and $84.7\%$, respectively. Second, in a system without any baseline data prefetcher, \pred provides significantly higher accuracy ($88.9\%$) and coverage ($93.6\%$) than any configuration with a baseline prefetcher. This shows that, the prefetch requests generated by a sophisticated data prefetcher interfere with the off-chip load prediction. This is why \pred's accuracy and coverage increases in absence of a data prefetcher.

\subsection{Overhead in Main Memory Requests with Different Prefetchers} \label{sec:main_mem_overhead_pref}

\begin{figure}[!h]
\centering
\includegraphics[width=3.3in]{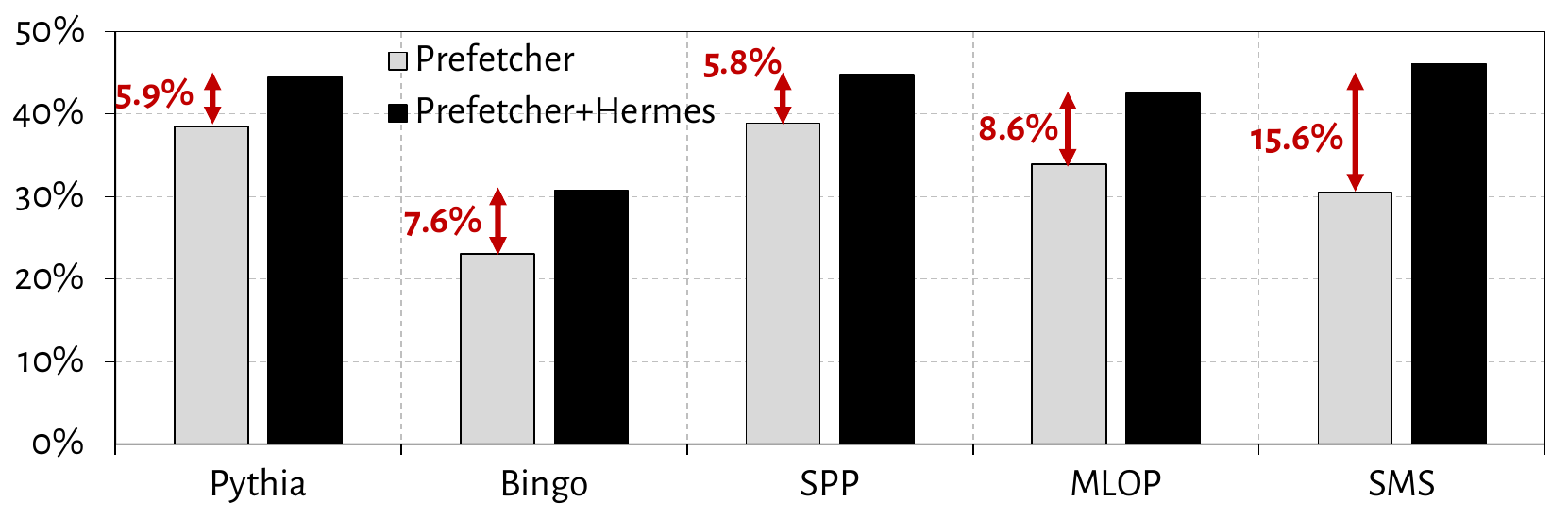}
\vspace{0.2em}
\caption{Overhead in main memory requests with different data prefetchers.}
\label{fig:main_mem_overhead_pref}
\end{figure}

Fig.~\ref{fig:main_mem_overhead_pref} shows the percentage increase in the main memory requests over the no-prefetching system by different types of data prefetchers alone, and in combination with Hermes in all single-core workloads. Combining Hermes with the baseline prefetcher increases the main memory request overhead by $5.9\%$, $7.6\%$, $5.8\%$, $8.6\%$, and $15.6\%$ for the baseline prefetchers Pythia, Bingo, SPP, MLOP, and SMS, respectively.

\subsection{Limit Study by Varying Hermes Request Issue Latency}
\label{sec:her_hermes_limit_study}

As we discuss earlier in~\Cref{sec:spec_load_issue}, Hermes's performance gain significantly depends on the Hermes request issue latency. 
We already conduct a bounded performance sensitivity study with Hermes request latency varying from $0$ cycle to $24$ cycles in~\Cref{sec:load_issue_latency}.
In this section, we conduct a performance limit study by pushing the range of the Hermes request issue latency even further to understand two key aspects: 
(1) under what range of Hermes request issue latency, Hermes's performance benefits become negligible, and 
(2) how much performance benefit we can expect from Hermes in a real system under a realistic on-chip network latency.

To accurately estimate the on-chip network latency in a real commercial processor, we first estimate the latency to access a banked-SRAM array of equal size to the L2/LLC of our baseline processor (which is modeled after the Intel Alder Lake processor; see~\Cref{sec:methodology}) using PCACTI~\cite{pcacti}, and then subtract the SRAM array access latency from the publicly-reported cache access latency (which inherently includes the on-chip network access latency).
Our evaluation yields an optimistic estimate of $2.01$ns (equivalent to $8$ processor cycles for our baseline processor clocked at 4GHz) and $2.65$ns (equivalent to $11$ processor cycles) latency for the L2- and LLC-sized SRAM arrays, respectively. This gives us a pessimistic estimate of $31$ cycles\footnote{Estimated latency spent on on-chip network for accessing L2 = reported L2 access latency (i.e., $10$ cycles) - estimated L2-sized SRAM access latency (i.e., $8$ cycles) = $2$ cycles; Estimated latency spent on on-chip network for accessing LLC = reported LLC access latency (i.e., $40$ cycles) - estimated LLC-sized SRAM access latency (i.e., $11$ cycles) = $29$ cycles;} for the total on-chip network latency for accessing L2 and LLC.

\begin{figure}[!ht]
\centering
\includegraphics[width=\columnwidth]{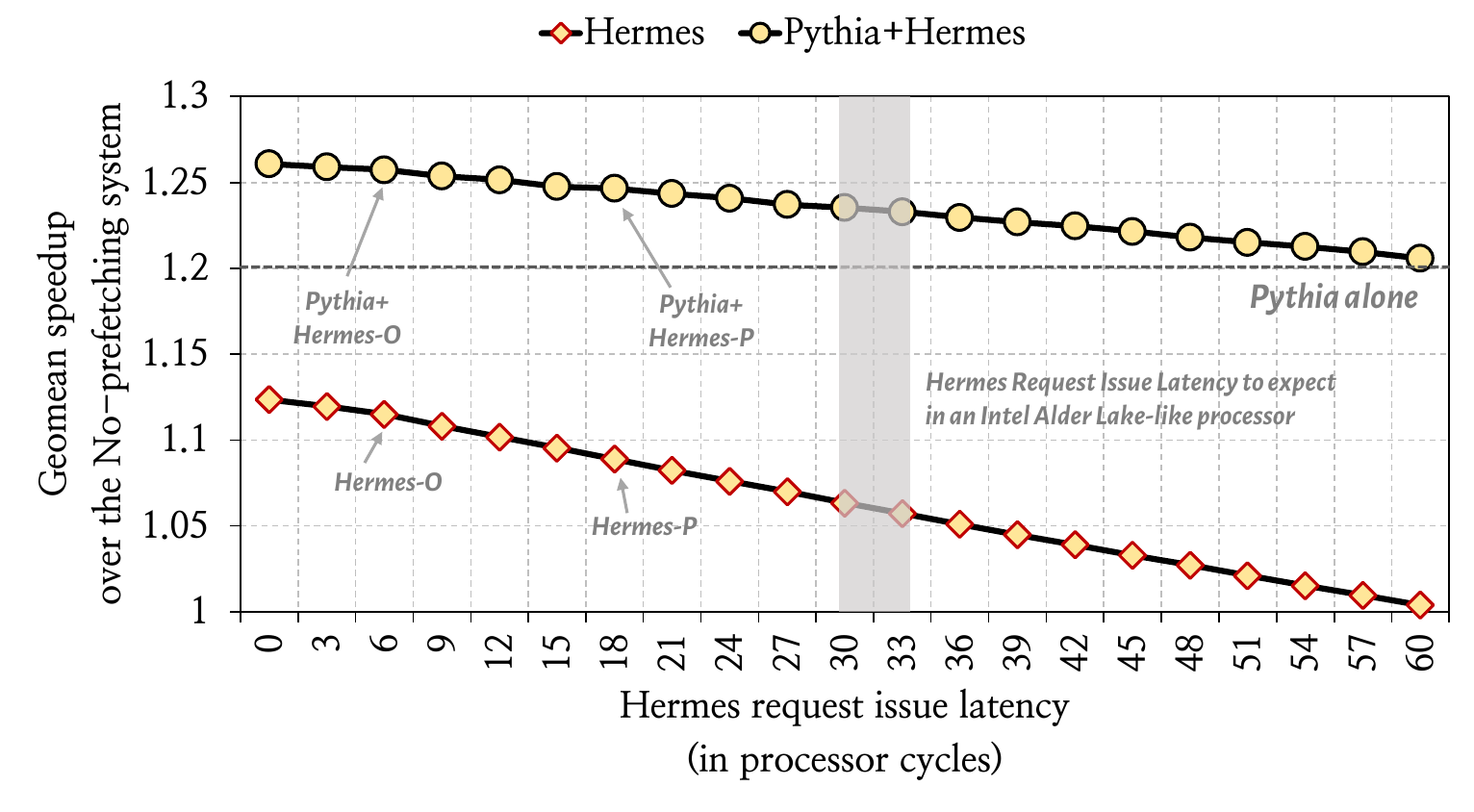}
\caption{Performance of Hermes and Pythia+Hermes while varying Hermes request issue latency.}
\label{fig:her_limit_study}
\end{figure}

\Cref{fig:her_limit_study} shows the geomean performance of Hermes and Pythia+Hermes under varying Hermes request issue latency. We make two key observations. First, Hermes's performance gain, both in standalone and in presence of Pythia, reduces for Hermes request issue latency of $51$ cycles or more. 
This is expected, as with such latency, a speculative Hermes request would arrive at the memory controller nearly at the same time as its corresponding regular load request,\footnote{The minor performance gains we observe even with Hermes request issue latencies higher than $51$ cycles is primarily stemming from hiding any latency caused by queuing delays during cache hierarchy traversal.} thus losing the latency hiding opportunity. With $51$-cycle Hermes request issue latency, Hermes improves performance by only $2.1\%$ on average over a no-prefetching system, and Hermes combined with Pythia improves performance by only $1.5\%$ over Pythia-alone.
Second, even with our realistic $31$-cycle on-chip network latency, Hermes-alone provides a significant $6.3\%$ performance gain over no-prefetching system, and Hermes with Pythia provides $3.3\%$ performance over Pythia-alone.
Thus we conclude that Hermes would provide tangible performance benefit even on a real-system with complex on-chip network topology and latency.

\subsection{Performance Evaluation using DPC4 Traces}
\label{subsec:hermes_dpc4}

\paraheading{Methodology.}
We leverage the DPC4 infrastructure~\cite{dpc4} to establish a rigorous evaluation methodology to validate the generalization capability of Hermes. 
More specifically, we use the corpus of $483$ previously-unseen single-core DPC4 workload traces that span across three application domains of AI, graph mining, and Google datacenter, to evaluate Hermes  . 
As Hermes has been designed, implemented, and finalized long \emph{before} the collection of these DPC4 workload traces, this trace corpus never contributed to the design-space exploration (e.g., feature selection, hyperparameter optimization) of Hermes.
Moreover, we do not perform any additional hyperparameter optimization, feature engineering, or architectural adjustment for these traces. 
Instead, we evaluate Hermes with its baseline configuration.
Thus, this methodology enables us to empirically answer a central research question: does Hermes merely \emph{memorize} the characteristics of familiar benchmark suites, or does it \emph{autonomously learn} a prediction strategy that \emph{generalizes} across fundamentally different and previously-unseen workloads?

\begin{figure}[!ht]
\centering
\includegraphics[width=\columnwidth]{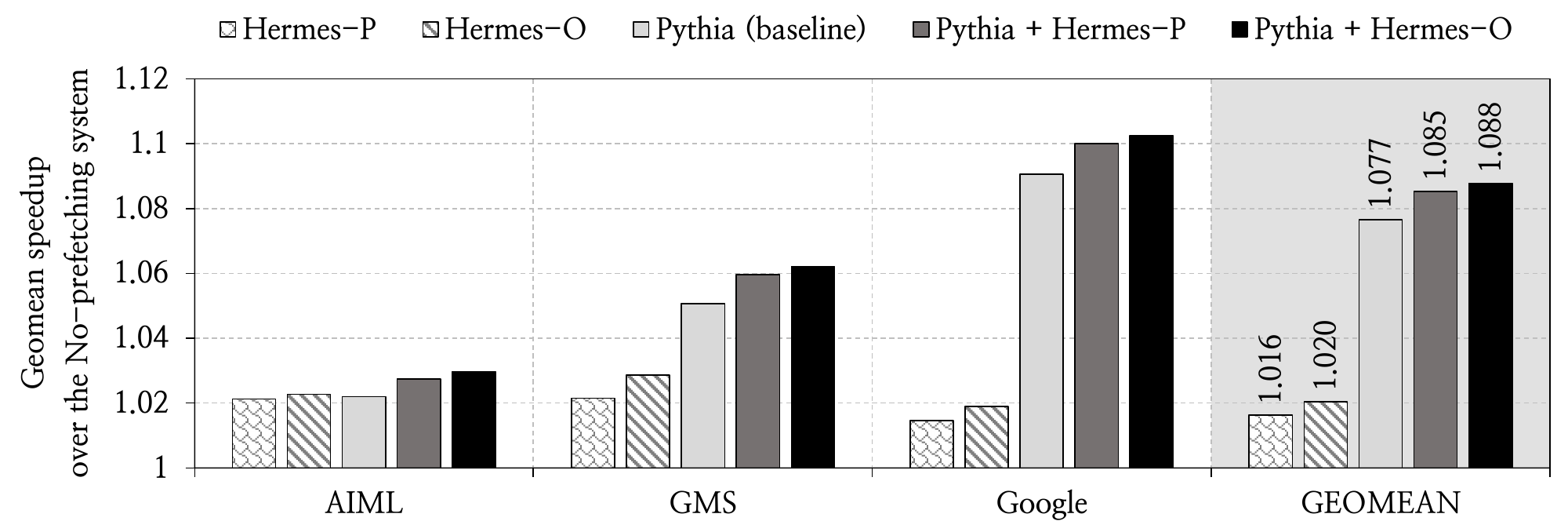}
\caption{Speedup in 483 single-core DPC4 workload traces.}
\label{fig:her_dpc4_overall}
\end{figure}

\paraheading{Speedup in Single-Core DPC4 Workload Traces.}
\Cref{fig:her_dpc4_overall} shows performance of Hermes (P and O), Pythia, and Hermes combined with Pythia normalized to the no-prefetching system in $483$ single-core DPC4 workloads.
The key observation is that, without any additional tuning for these traces, Hermes-O alone and combined with Pythia improve performance over a no-prefetching system by $2.0\%$ and $8.8\%$ on average, respectively, whereas Pythia alone provides performance gain of $7.7\%$.
In every workload category, Hermes provides performance benefit when applied alone and when combined with Pythia.

\begin{figure}[!ht]
\centering
\includegraphics[width=\columnwidth]{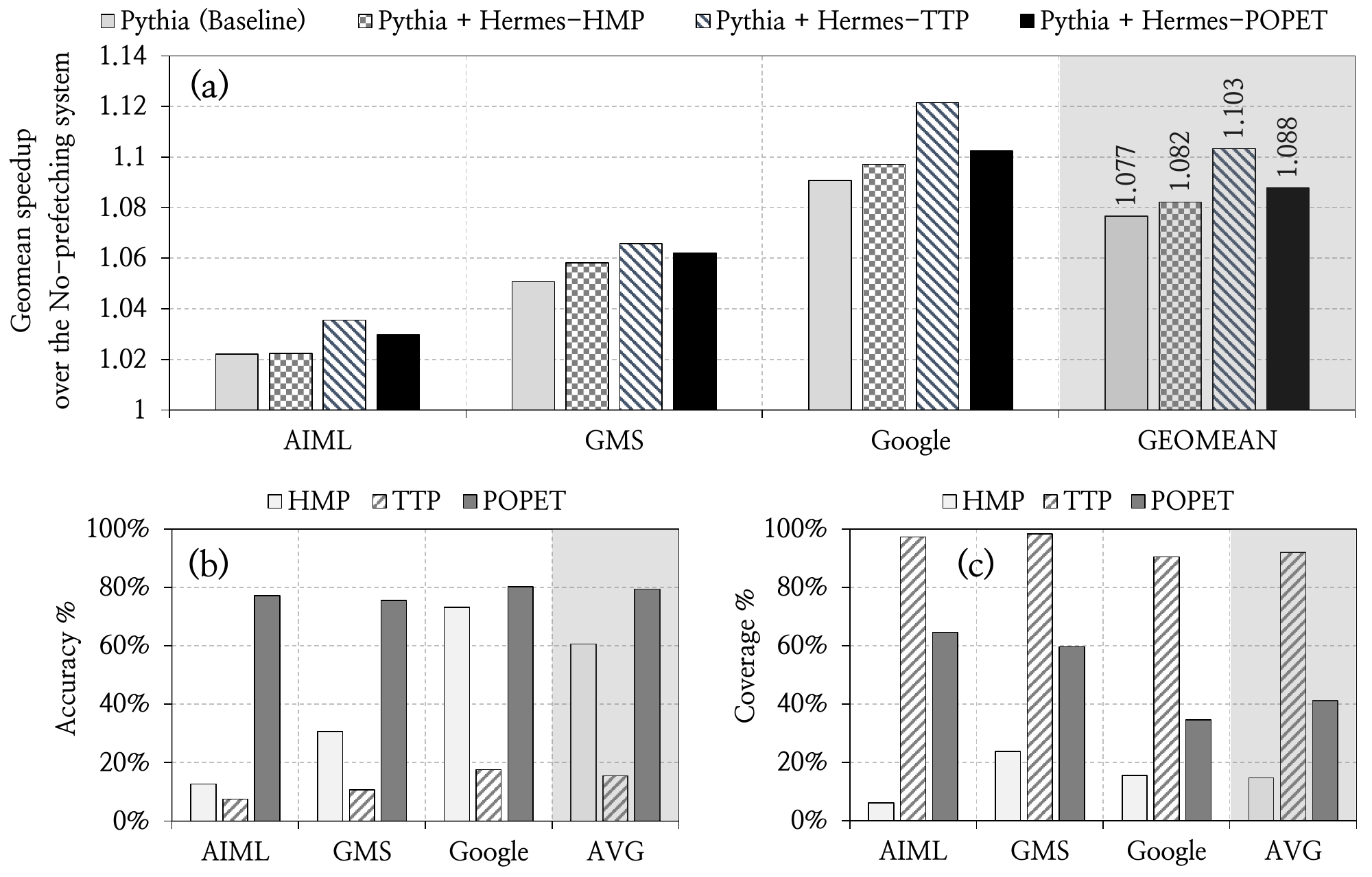}
\caption{(a) Speedup of Hermes with three different off-chip predictors, HMP, TTP, and POPET in single-core DPC4 workloads. Comparison of (b) accuracy and (c) coverage of POPET against those of HMP and TTP in single-core DPC4 workloads.
}
\label{fig:her_dpc4_ocp}
\end{figure}

\paraheading{Effect of the Off-Chip Load Prediction Mechanism}.
\Cref{fig:her_dpc4_ocp}(a) shows the geomean performance of Pythia alone and Pythia combined with Hermes with three different off-chip prediction mechanisms: POPET, HMP, and TTP, across all $483$ DPC4 workloads.
We make two key observations.
First, Hermes with POPET consistently outperforms both Pythia alone and Pythia with Hermes-HMP in all workload categories.
Second, when combined with Pythia, Hermes-POPET underperforms Hermes-TTP.
To understand the reason behind POPET's under-performance,~\Cref{fig:her_dpc4_ocp}(b) and~\Cref{fig:her_dpc4_ocp}(c) report the off-chip prediction accuracy and coverage of HMP, TTP, and POPET, respectively.
Although POPET achieves a substantially higher prediction accuracy ($79.5\%$) than TTP ($15.5\%$), its coverage is significantly lower than that of TTP.
POPET only covers $41.2\%$ of the true off-chip load requests, whereas TTP achieves a coverage of $92.1\%$.
This limited coverage constrains POPET’s overall effectiveness and ultimately results in lower performance compared to TTP when combined with Pythia.

\begin{figure}[!ht]
\centering
\includegraphics[width=\columnwidth]{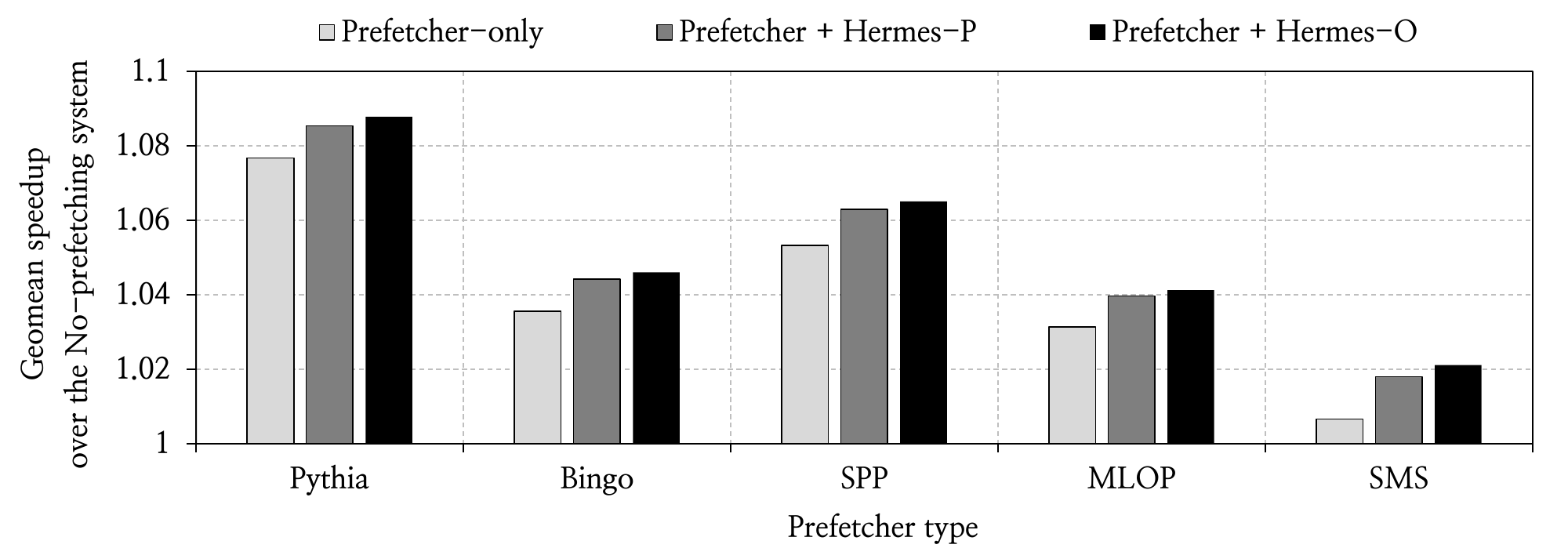}
\caption{Performance sensitivity to baseline prefetcher in single-core DPC4 workloads.}
\label{fig:her_dpc4_pref}
\end{figure}

\paraheading{Effect of the Baseline Prefetcher}.
\Cref{fig:her_dpc4_pref} shows the geomean performance of the baseline prefetcher, and Hermes-P/O combined with the baseline prefetcher, normalized to the no-prefetching system across all $483$ single-core DPC4 workload traces.
The key takeaway is that Hermes combined with any baseline prefetcher consistently outperforms the baseline prefetcher by itself for all four evaluated prefetching techniques. Hermes+prefetcher outperforms the prefetcher alone by $1.1\%$, $1.0\%$, $1.2\%$, $1.0\%$, $1.5\%$, for Pythia, Bingo, SPP, MLOP, and SMS as the baseline prefetcher, respectively.

Overall, the performance results with DPC4 traces provide empirical evidence that Hermes generalizes beyond its design-time workloads. Despite never observing these traces during development and despite operating in a plug-and-play configuration without additional tuning, Hermes provides considerable, and often the state-of-the-art, performance gains in emerging workloads.

\subsection{Performance Evaluation using SPEC CPU 2026 Benchmarks}
\label{subsec:hermes_spec26}

\paraheading{Methodology.}
To further demonstrate Hermes' effectiveness, we evaluate it using memory-intensive workload traces from the recently-released SPEC CPU 2026 benchmark suite~\cite{spec2026}. 
We collect the workload traces using a three step methodology. First, we run each workload in the benchmark suite using the ref input size\footnote{Some workloads (e.g., \texttt{821.gcc\_s}, \texttt{823.llvm\_s}) get executed multiple times with different command line parameters when run with the ref input. For such workloads, we extract at most ten execution commands lines and run them individually for the rest of this methodology.} and capture at most five representative regions of interest (RoI), each with 1 billion instructions, using the SimPoint methodology~\cite{simpoint}. 
Second, we capture the workload trace from each SimPoint RoI using ChampSim's PIN-based tracing tool. Third, we filter out any trace from this study that has less than two LLC misses per kilo instructions (MPKI) in the no-prefetching system.
Overall, we consider $149$ workload traces (abbreviated as \texttt{SPEC26}) in this study, which are summarized in~\Cref{tab:spec26}.
Similar to~\ref{subsec:hermes_dpc4}, these traces neither contributed to the design-space exploration of Hermes, nor we perform any additional hyperparameter optimization, feature engineering, or architectural adjustment for these traces. Instead, we evaluate Hermes as is with its baseline configuration.

\begin{table}[htbp]
  \centering
  \caption{SPEC CPU 2026 benchmark traces used for evaluation.}
  \small
    \begin{tabular}{l||L{12em}||r}
    \thickhline
          & \textbf{Workload type} & \multicolumn{1}{l}{\textbf{\# traces}} \\
    \thickhline
    \multirow{3}[1]{*}{SPEC rate} & \Tabval{Floating point (rate-fp)} & \Tabval{32} \\
        \cline{2-3}          & \Tabval{Integer (rate-int)} & \Tabval{28} \\
        \cline{2-3}          & \Tabval{\textbf{\textit{Total}}} & \Tabval{\textbf{60}} \\
    \hline
    \hline
    \multirow{3}[1]{*}{SPEC speed} & \Tabval{Floating point (speed-fp)} & \Tabval{33} \\
        \cline{2-3}          & \Tabval{Integer (speed-int)} & \Tabval{56} \\
        \cline{2-3}          & \Tabval{\textbf{\textit{Total}}} & \Tabval{\textbf{89}} \\
    \thickhline
    \end{tabular}%
  \label{tab:spec26}%
\end{table}%

\begin{figure}[!ht]
\centering
\includegraphics[width=\columnwidth]{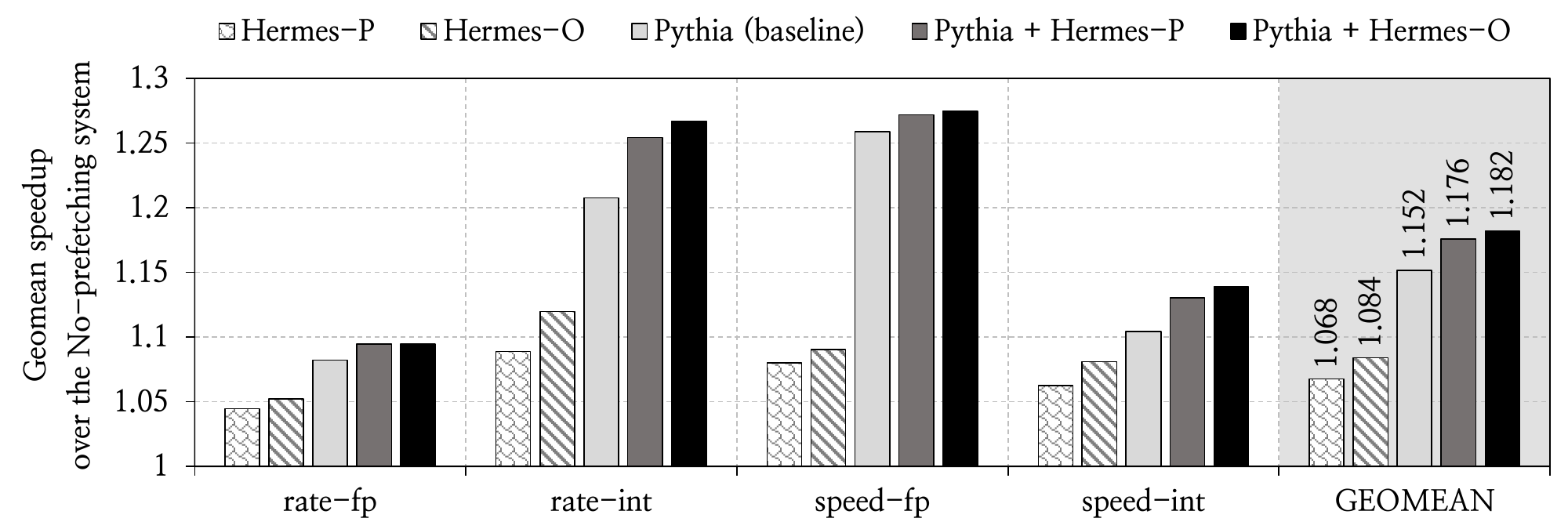}
\caption{Speedup in 149 single-core SPEC26 traces.}
\label{fig:her_spec26_perf}
\end{figure}

\paraheading{Speedup in Single-Core Workloads.}
\Cref{fig:her_spec26_perf} shows performance of Hermes (P and O), Pythia, and Hermes combined with Pythia normalized to the no-prefetching system in $149$ single-core \texttt{SPEC26} traces.
The key observation is that, without any additional tuning, Hermes-O alone and Hermes-O combined with Pythia improve performance over a no-prefetching system by $8.4\%$ and $18.2\%$ on average, respectively, whereas Pythia alone provides performance gain of $15.2\%$. In every workload category, Hermes provides performance benefit when applied alone and when combined with Pythia.

\begin{figure}[!ht]
\centering
\includegraphics[width=\columnwidth]{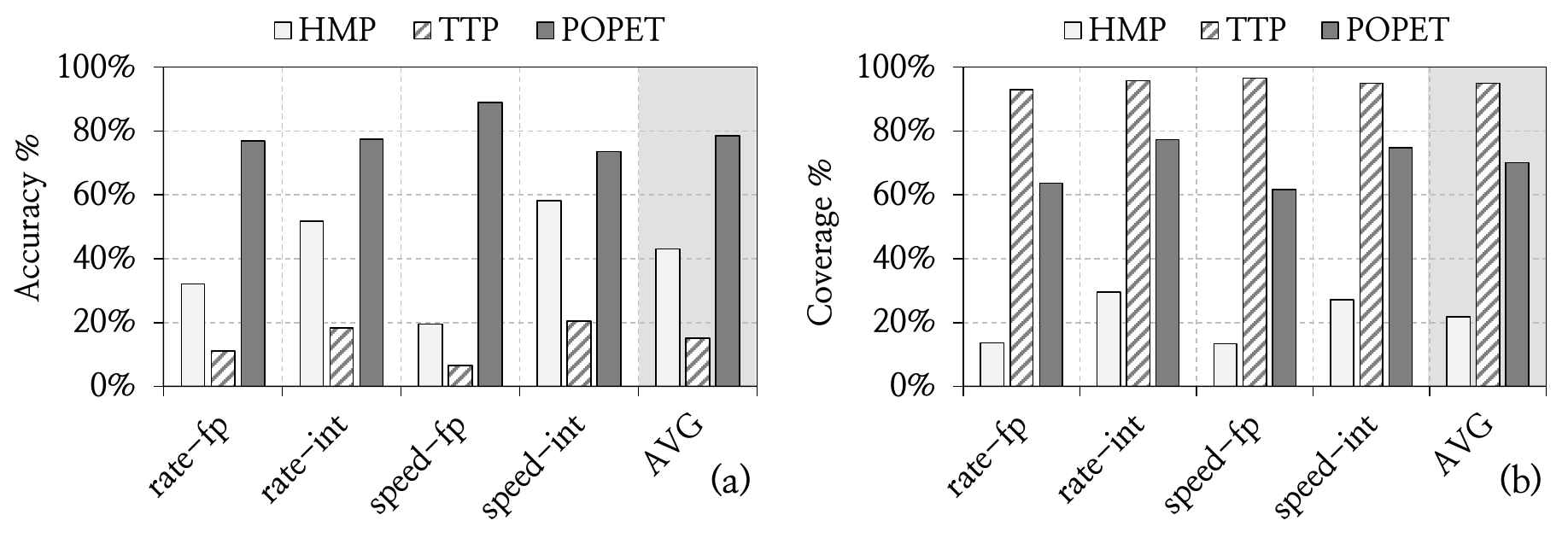}
\caption{Comparison of (a) accuracy and (b) coverage of POPET against those of HMP and TTP in SPEC26 traces.}
\label{fig:her_spec26_ocp_cov_acc}
\end{figure}

\paraheading{Effect of the Off-Chip Load Prediction Mechanism}.
Fig.~\ref{fig:her_spec26_ocp_cov_acc} shows the comparison of \rbh{\pred's} \rbe{off-chip load prediction} accuracy and coverage  against \rbh{those of} HMP and TTP. The key observation is that \pred has significantly higher accuracy \emph{and} coverage than HMP. \rbh{\pred provides} $78.5\%$ accuracy with $70.1\%$ coverage \rbd{on average across all \texttt{SPEC26} traces}, whereas HMP \rbh{provides} $43.1\%$ accuracy with $21.8\%$ coverage. 
\rbg{TTP, with a metadata budget of $1.5$~MB, \rbh{provides the highest} coverage \rbh{($95.1\%$)} but with a significantly lower accuracy \rbh{($15.1\%$)}.} 

\begin{figure}[!ht]
\centering
\includegraphics[width=\columnwidth]{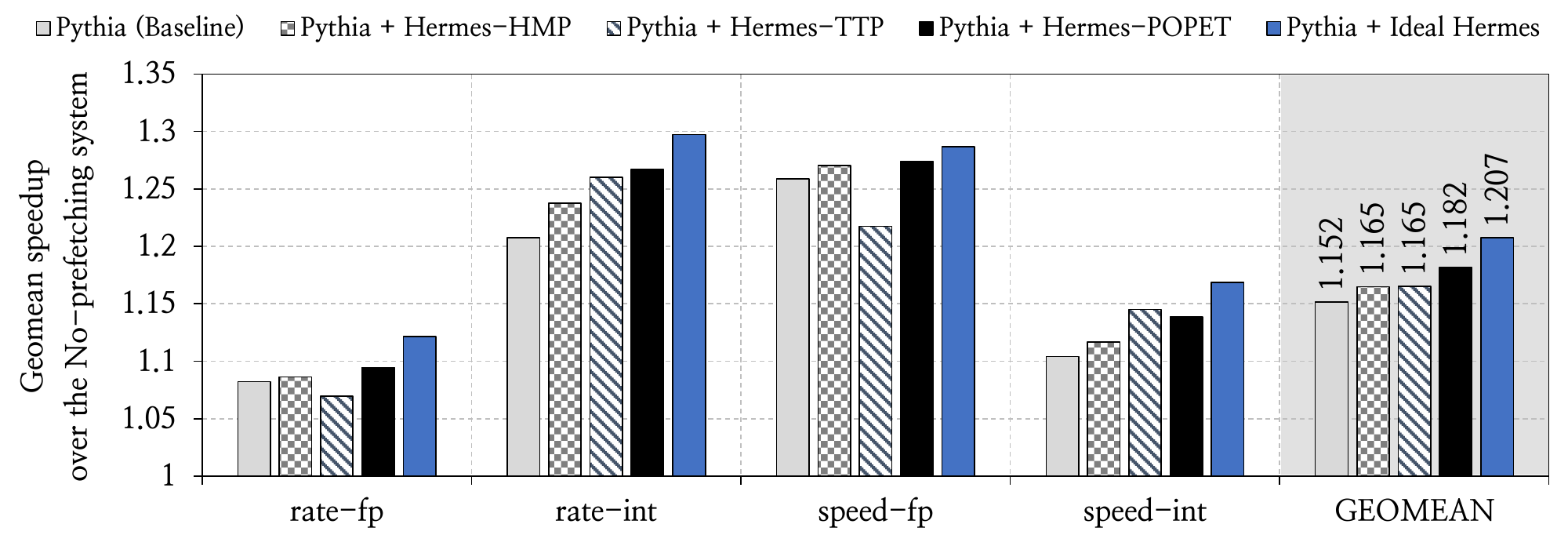}
\caption{Speedup of Hermes with three different off-chip predictors, HMP, TTP, and POPET in SPEC26 traces.}
\label{fig:her_spec26_ocp}
\end{figure}

\pred's superior accuracy \emph{and} coverage directly translates to performance benefits.
Fig.~\ref{fig:her_spec26_ocp} shows the performance of Hermes with \pred, Hermes-HMP, \rbg{Hermes-TTP}, and the Ideal Hermes (see~\cref{sec:headroom_study}) combined with Pythia normalized to the no-prefetching system in \texttt{SPEC26} traces.
\rbe{We make two key observations.}
\rbe{First,} \rbd{Hermes with \pred} outperforms both Hermes-HMP and \rbg{Hermes-TTP}. 
\rbg{On average, \rbh{Hermes-HMP, Hermes-TTP}, and \rbd{Hermes with \pred} \rbd{combined with} Pythia provide \rbh{$1.4\%$, $1.4\%$}, and $3.0\%$ performance \rbh{improvement} \rbd{over} Pythia, respectively.}
\rbe{Second, Hermes-\pred provides $88\%$ of the performance improvement provided by the Ideal Hermes that employs an ideal off-chip load predictor with $100\%$ accuracy and coverage.}
\rbd{We conclude that Hermes provides performance gains due to \rbg{both} \rbe{the} high \rbe{off-chip load prediction} accuracy and coverage of \pred.} 

\begin{figure}[!ht]
\centering
\includegraphics[width=\columnwidth]{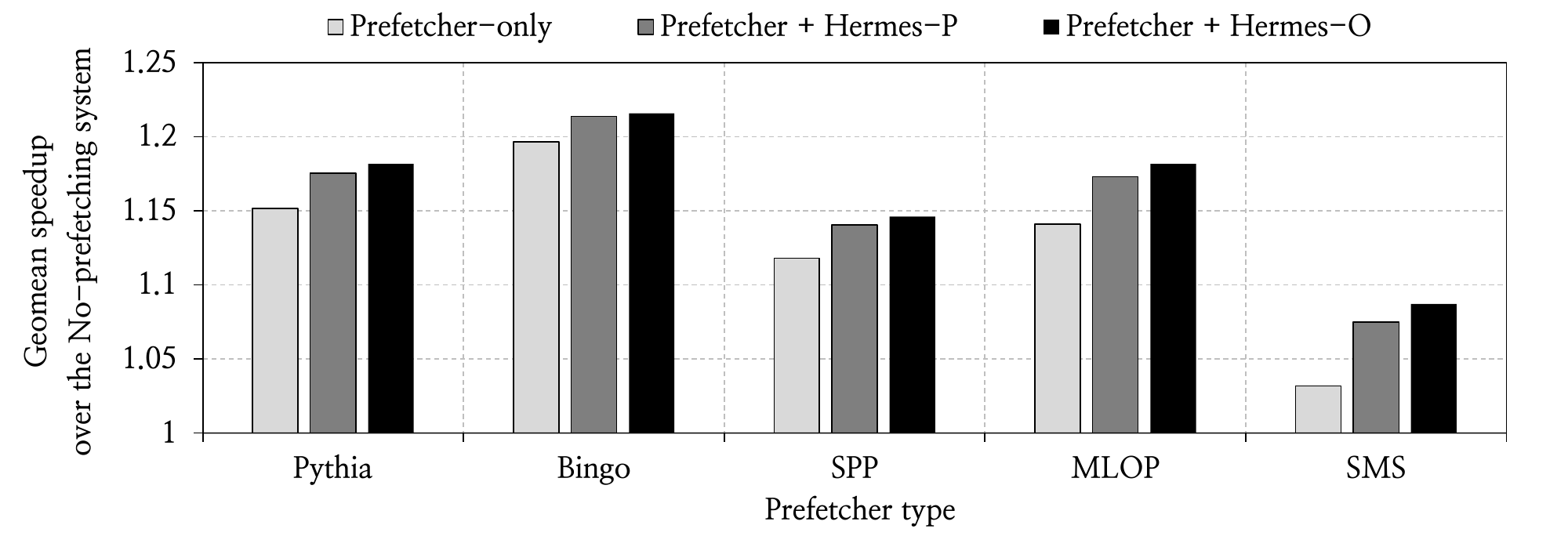}
\caption{Performance sensitivity to baseline prefetcher in SPEC26 traces.}
\label{fig:her_spec26_pref}
\end{figure}

\paraheading{Effect of the Baseline Prefetcher}.
\Cref{fig:her_spec26_pref} shows the geomean performance of the baseline prefetcher, and Hermes-P/O combined with the baseline prefetcher, normalized to the no-prefetching system across all \texttt{SPEC26} workload traces.
The key takeaway is that Hermes combined with any baseline prefetcher consistently outperforms the baseline prefetcher by itself for all four evaluated prefetching techniques. Hermes+prefetcher outperforms the prefetcher alone by $3.0\%$, $1.3\%$, $2.8\%$, $4.1\%$, $5.6\%$, for Pythia, Bingo, SPP, MLOP, and SMS as the baseline prefetcher, respectively.

